\documentclass[twoside,11pt]{article}

\usepackage{subfig,tikz}
\usepackage{tkz-graph} 
\usetikzlibrary{arrows.meta} 
\usetikzlibrary{positioning}
\usetikzlibrary{arrows,shapes.arrows,shapes.geometric,shapes.multipart,decorations.pathmorphing, positioning,shapes.swigs,}
\tikzset{
    -Latex,auto,node distance =1 cm and 1 cm,semithick,
    state/.style ={ellipse, draw, minimum width = 0.7 cm},
    point/.style = {circle, draw, inner sep=0.04cm,fill,node contents={}},
    bidirected/.style={Latex-Latex,dashed},
    el/.style = {inner sep=2pt, align=left, sloped}
}

\usepackage{xr-hyper} 
\usepackage{jmlr2e}
\usepackage{enumitem} 
\usepackage{amsmath} 
\usepackage{bbm} 
\usepackage{float} 
\usepackage{bm} 
\usepackage{comment} 
\usepackage{nopageno} 
\usepackage{csquotes} 
\usepackage{afterpage} 
\usepackage{xurl} 
\usepackage{multirow} 
\usepackage{array} 

\usepackage{hyperref}
\usepackage{cleveref}
\usepackage{layouts}
\usepackage{threeparttable}
\makeatletter

\newcommand*{\addFileDependency}[1]{
\typeout{(#1)}
\@addtofilelist{#1}
\IfFileExists{#1}{}{\typeout{No file #1.}}
}\makeatother

\newcommand*{\myexternaldocument}[1]{%
\externaldocument{#1}%
\addFileDependency{#1.tex}%
\addFileDependency{#1.aux}%
}
\myexternaldocument{supplementary}

\newcommand{\indep}{\perp\hskip -7pt \perp } 

\DeclareMathOperator{\expit}{expit}
\DeclareMathOperator{\Bernoulli}{Bern}

\newtheorem{assumption}{Assumption}

\allowdisplaybreaks 
\makeatletter
\newcommand{\labeltext}[2]{%
  \@bsphack
  \csname phantomsection\endcsname 
  \def\@currentlabel{#1}{\label{#2}}%
  \@esphack
}
\mathchardef\m="2D

\ShortHeadings{Contrasting Identifying Assumptions}{Gorbach, de Luna, Karvanen, and Waernbaum}
\firstpageno{1}

\begin{document}

\title{Contrasting Identifying Assumptions of Average Causal Effects: Robustness and Semiparametric Efficiency}

    \author{\name Tetiana Gorbach \email tetiana.gorbach@umu.se \\
        \name Xavier de Luna \email xavier.de.luna@umu.se\\
       \addr Department of Statistics\\
       Umeå University\\
       901 87 Umeå, Sweden\\
       \AND
       \name  Juha Karvanen \email juha.t.karvanen@jyu.fi\\
       \addr Department of Mathematics and Statistics\\
       P.O.Box 35 (MaD) FI-40014 University of Jyvaskyla, Jyv\"askyl\"a, Finland\\
       \AND
       \name  Ingeborg Waernbaum  \email ingeborg.waernbaum@statistik.uu.se\\
       \addr Department of Statistics\\
       Box 513, 751 20 Uppsala University, Uppsala, Sweden\\}

\editor{}

\maketitle

\begin{abstract}
Semiparametric inference on average causal effects from observational data is based on assumptions yielding identification of the effects. In practice, several distinct identifying assumptions may be plausible; an analyst has to make a delicate choice between these models. In this paper, we study three identifying assumptions based on the potential outcome framework:  the back-door assumption, which uses pre-treatment covariates, the front-door assumption, which uses mediators, and the two-door assumption using pre-treatment covariates and mediators simultaneously. We provide the efficient influence functions and the corresponding semiparametric efficiency bounds that hold under these assumptions, and their combinations. We demonstrate that neither of the identification models provides uniformly the most efficient estimation and give conditions under which some bounds are lower than others. We show when semiparametric estimating equation estimators based on influence functions  attain the bounds, and study the robustness of the estimators to misspecification of the nuisance models. The theory is complemented with simulation experiments on the finite sample behavior of the estimators. The results obtained are relevant for an analyst facing a choice between several plausible identifying assumptions and corresponding estimators. Our results show that this choice implies a trade-off between efficiency and robustness to misspecification of the nuisance models.   
\end{abstract}


\section{Introduction}

    This paper deals with observational studies, where the treatment assignment is not randomized. In such studies, inference on a causal parameter of interest relies on assumptions that allow identification of the parameter. Depending on the scientific context and observed data, the analyst may consider several such assumptions, hereafter also called identification models. We investigate how the choice of the identification model affects the efficiency and robustness of the semiparametric estimation of a causal parameter.   
    
    In this paper, we use the potential outcome framework \citep{neyman1923applications, rubin1974estimating} to define causal estimands, and consider three semiparametric models used to identify the average causal effect (ACE) of a treatment on an outcome of interest. The first model considers observed data on the treatment, pre-treatment covariates and outcome and is defined through a back-door assumption (also called ignorability of treatment assignment, \citealp{rosenbaum1983central}), where the treatment assignment is assumed randomized-like given a set of observed pre-treatment covariates. The second identification model  considers  observed data on the treatment, mediators and outcome and uses a front-door assumption to identify the ACE. The third model considers observed data on the treatment, outcome, pre-treatment covariates and mediators to identify the ACE using both pre-treatment covariates and mediators \citep{fulcher2020robust}. We say that this model is defined via the two-door assumption because it combines pre-treatment covariates from the back-door assumption and mediators from the front-door assumption.  Other types of identification models  not covered in this paper do exist, for example,  \citet{bowden1984instrumental,  shpitser2010validity, helske2021estimation}. In particular, instrumental variables are often used in empirical economics, although in that case a different target causal parameter (local average causal effect, \citealp{angrist1996identification}) is identified. 
    
    The back-door assumption has been fundamental for the study of efficient estimation of semiparametric regular asymptotically linear (RAL) estimators of the ACE. \citet{robins1994estimation} and  \citet{hahn1998role} independently derived the semiparametric efficiency bound \citep{vaart1998}, that is, the lower bound for the asymptotic variance of semiparametric RAL estimators under the back-door assumption. This bound has served as a benchmark for a wide range of proposed estimators; see, for example, \citet{scharfstein1999adjusting, abadie2006large, chan2016globally, tan2020model}.
    
    Recently, \citet{fulcher2020robust}, also using the potential outcome framework, provided the efficient influence function, the  variance of which is equal to the semiparametric efficiency bound, under the two-door assumption. Related results have been obtained using the graphical framework \citep{pearl2009causality}. In particular,  \citet{rotnizky2020efficient} derived  efficient  influence functions  for  structural  causal  models  described through directed acyclic graphs, and   \citet{bhattacharya2020semiparametric} extended  Rotnizky’s  results  to  a specific type of acyclic directed mixed graphs allowing for hidden/unobserved variables.

    In this context, we contribute in several directions. First, we show that none of the considered identification strategies is uniformly the most efficient for ACE estimation.   Second, we deduce the efficient influence functions and the corresponding semiparametric efficiency bounds under pairwise combinations of the back-, the front-, and the two-door assumptions. Several of the derived efficient influence functions are identical to those obtained within a graphical model framework in \citet{rotnizky2020efficient} and \citet{bhattacharya2020semiparametric}. 
    This happens when the causal parameters and the identifying models with two frameworks yield the same observed data distribution and statistical parameter (functional of the observed data distribution) on which the inference is performed. 
    
    Furthermore, we contrast the efficiency and robustness of semiparametric ACE estimators constructed under different identification models. For this purpose, we introduce estimators based on the derived efficient influence functions. These estimators use different sets of nuisance models of the data generating mechanisms. In particular, to achieve the bounds, estimators fit different sets of nuisance models, which need to be estimated at a sufficiently high rate of convergence. We also show that the proposed estimators have different robustness properties. Firstly, they are consistent even if some of the fitted nuisance models are misspecified. Secondly, they reach the semiparametric efficiency bound even when some of the nuisance models are estimated at a slow convergence rate, provided that the other nuisance models are estimated at a high enough rate (a product rate condition is used; e.g., \citealt{farrell2015robust}). In addition to the theoretical properties, the finite sample behaviors of the proposed estimators are compared via simulation experiments. Our results make it clear that there is a trade-off between efficiency and bias (robustness) of the estimation using different identifying assumptions. Thus, an analyst needs to consider a trade-off between the plausibility of the identifying assumptions and the difficulty in specifying (or non-parametrically estimating) nuisance models.
    
    The rest of the paper is organized as follows. Section~\ref{sec:backgroud} introduces notation and the potential outcome framework, presents the back-, front-, and two-door identifying assumptions studied, and provides the corresponding efficient influence functions. Semiparametric efficiency bounds based on different models are derived and compared in  Section~\ref{sec:comparison}. Section~\ref{sec:semipar_est} introduces semiparametric estimators, including their robustness  and efficiency properties. A simulation study on finite sample properties is included in Section~\ref{sec:simulation studies}. The findings are discussed in Section~\ref{sec:discussion}.

\section{Background} \label{sec:backgroud}
We start this section by providing definitions and notation. We then describe the identification models and present the corresponding efficient influence functions.

\subsection{Definitions and Notation}\label{sec:notation}
We consider a scalar treatment variable $A.$ Following the Neyman–Rubin causal model 
\citep{neyman1923applications, rubin1974estimating, holland1986statistics} and the mediation framework \citep{robins1992identifiability, pearl2001direct}, let $Y(a)$ and $Z(a)$ denote, respectively,  the values of a univariate outcome and a possibly multivariate mediator that is observed if the treatment $A$ is set to value $a.$ 
The parameter of interest is the average causal effect of a treatment, $\theta = EY(a^*)-EY(a)$, where $EY(a)$ denotes the expectation of $Y(a).$ Further, let $Y(a,z)$ denote the value of the outcome that would be observed if the treatment $A$ were set to $a$ and the mediating variable were set to $z.$  Also, let $Y(a, Z(a^*))$ denote the value of $Y$ that would occur if $A$ were set to $a$ and $Z$ were set to what it would have been if $A$ were set to $a^*.$
The observed outcome and the observed mediator are denoted by  $Y $ and $Z,$ respectively. Let $C$ and $U$ denote a set of, respectively, observed and unobserved, pre-treatment covariates that may confound the \mbox{$A$--$Z$}, \mbox{$Z$--$Y$} or \mbox{$A$--$Y$} relationships. Thus, $C$ is not allowed to be affected by $A, Z,$ and $Y$; $A$ is not allowed to be affected by $Z$ and $Y$; $Z$ is not allowed to be affected by $Y.$  The sets of possible values of the considered random variables are correspondingly denoted as $\mathcal{A, Y}, \mathcal{Z}, \mathcal{C}$ and $\mathcal{U}.$  We consider an i.i.d. sample of size $n$ and  use index $i$ to represent subjects in the sample. 

We use $p(x)=p(X=x)$ to denote the probability density function of a continuous random variable $X$ at point $x$ and, correspondingly, the probability mass function if $X$ is a discrete random variable. We use $\mathbb{P}f(X) $ to denote expectations of $f(X)$ for a new observation $X$ (treating the function $f$ as fixed); $\mathbb{P}\hat{f}(X) $ is random since it depends on the sample used to estimate $f$. $||f||^2 = \mathbb{P}f^2$.   In what follows, we assume discrete random variables; accordingly, we use summations of probabilities over the space of values of a random variable. If the random variables were continuous, the summations would be replaced by integrals.

The indicator function $I(A=a^*)$ is equal to 1 if $A=a^*$ and 0 otherwise. 
When $\sqrt{n}(\hat{\theta} - \theta) \rightarrow_d N(0, V),$  $V$ will be called the asymptotic variance of $\hat{\theta}.$ 
We make the following assumption throughout.
\begin{assumption}
\begin{enumerate}[label=(\roman*)]
    \item If $A=a$, then $ Y = Y(a) $ and $ Z = Z(a)$ with probability 1 for any $a \in \mathcal{A}$ (consistency assumption, e.g., \citealp[p.229]{pearl2009causality}, \citealp{goetghebeur2020formulating}),
    \item If $A=a$ and $Z=z$, then $Y = Y(a,z) $ with probability 1 for any $a \in \mathcal{A},$ $z \in \mathcal{Z}$ (consistency assumption)
    \item For any $a \in \mathcal{A},$ $Y(a) = Y(a,Z(a))$ (composition assumption, e.g., \citealp[p.229]{pearl2009causality}, \citealp[p.462]{vanderweele2015explanation}). 
\end{enumerate}
\end{assumption}

From the definitions, it follows that if $Z(a^*)=z,$ then $Y\big(a, Z(a^*)\big)=Y(a, z)$ for any $z \in \mathcal{Z}$ and $a, a^* \in \mathcal{A}.$  This means that the outcome is the same regardless of the mediator being assigned or occurring as a response to some treatment. 
We also assume in the sequel that $p(Y, A, Z, C)>0$  for any  $a \in \mathcal{A}$ (positivity assumption), and that the variances of the observed variables are bounded.

\subsection{Identification Models} \label{sec:identification}
When the treatment is not randomized, as in observational studies, $\theta$ can be  estimated if it is identified in the considered model. Below we present back-, front-, and two-door assumptions under which $\theta$  is identified. We also provide the corresponding identification expressions.

In the first identification model, consider observed data on $C, A$, and  $Y.$ Under the following back-door assumption (also called  ignorability assumption,  \citet{rosenbaum1983central}),
\begin{enumerate}[label=\textbf{BD}]
\itemsep0em
\item \label{ass:bd} $Y(a)  \indep  A|C, \ \forall a   \in \mathcal{A},$
\end{enumerate}
$\theta$ is identified through the back-door adjustment
$ p(Y(a^*)=y) = \sum\limits_{c} p(c)p(y|a^*,c). $

The second identification model considers observed data on $A, Z$, and $Y.$ Under the following front-door assumptions
\begin{enumerate}[label=\textbf{FD\arabic*}] \label{ass:fd}
\itemsep0em
\item \label{ass:fd1} $Y(a,z) = Y(a^*,z)=$ $Y(z)$ for any $a, a^* \in \mathcal{A}$ and $z \in \mathcal{Z},$
\item \label{ass:fd2} $Z(a)$ $\indep A$ for any $a$ $\in \mathcal{A},$
\item \label{ass:fd3}$Y(a,z) \indep Z(a^*) |A=a$ for any $a, a^* \in \mathcal{A}$ and $z \in \mathcal{Z},$
\end{enumerate}
$\theta$ is identified via the front-door adjustment $p(Y(a^*)=y)=\sum\limits_{z} p(z|a^*)\sum\limits_{\bar{a}}p(y|\bar{a},z)p(\bar{a}). $
 The adjustment follows from Lemma 1 in \citet{fulcher2020robust}, by treating  $C$ as unobserved variables.
In the sequel,  assumptions \ref{ass:fd1}--\ref{ass:fd3} will collectively be denoted by \hyperref[ass:fd]{\hyperref[ass:fd]{\textbf{FD}}}.

Finally, in the third model, consider observed data on $C, A, Z,$ and $Y$ and the two-door assumptions    
\begin{enumerate}[label=\textbf{TD\arabic*}] \label{ass:td}
\item \label{ass:td1} $Y(a,z) = Y(a^*,z)=$ $Y(z)$ for any $a, a^* \in \mathcal{A}$ and $z \in \mathcal{Z},$
\item \label{ass:td2} $Z(a)$ $\indep A|C$ for any $a \in \mathcal{A},$
\item \label{ass:td3} $Y(a,z) \indep Z(a^*) | A= a, C$ for any $a, a^* \in \mathcal{A},$ $z \in \mathcal{Z},$
\end{enumerate}

If Assumption \hyperref[ass:td]{\textbf{TD}} holds, $\theta$ is identified via the two-door adjustment
$ p(Y(a^*)=y)=\sum\limits_{z,c} p(z|a^*,c)\sum\limits_{\bar{a}}p(y|\bar{a},z,c)p(\bar{a},c).$ The two-door adjustment follows from Lemma 1 in \citet{fulcher2020robust}, because, under Assumption \ref{ass:td1}, $Y(A, Z(a^*)) =  Y(a^*, Z(a^*)) = Y(a^*),$ that is,
there is no direct effect of the treatment $A$ on the outcome $Y.$ 

\begin{figure}[t] 
\vspace{-1cm}
\centering
\subfloat[]{\label{fig:bd}
    \centering
    \begin{tikzpicture}
    \tikzset{line width=0.6pt, outer sep=0pt,
    ell/.style={draw,fill=white, inner sep=2pt,
    line width=0.6pt},
    swig vsplit={gap=2pt},
    node distance=0.5cm}
        \node[name=A, shape=swig vsplit]{
            \nodepart{left}{$A$}
            \nodepart{right}{$a$} };
        \node(Z)       [right=of A] {};
        \node[name=Y,  right=of Z, ell, shape=ellipse]{$Y(a)$};
        \node(C)       [below=of Z] {$\bm{C}$};
        \draw[-{Latex[length=1mm, width=1mm]}] (A) -- (Y);
        \draw[->,-{Latex[length=2mm, width=2mm]}](C) edge[out=180,in=300] (A.south west);
        \draw[-{Latex[length=1mm, width=1mm]}] (C) -- (Y);
    \end{tikzpicture}
    }   
\subfloat[]{\label{fig:fd}
    \centering
    \begin{tikzpicture}
    \tikzset{line width=0.6pt, outer sep=0pt,
    ell/.style={draw,fill=white, inner sep=2pt,
    line width=0.6pt},
    swig vsplit={gap=2pt},
    node distance=0.5cm}
    \node[name=A, shape=swig vsplit]{
            \nodepart{left}{$A$}
            \nodepart{right}{$a$} };
        \node[name=Z, shape=swig vsplit, right=of A]{
            \nodepart{left}{$Z(a)$}
            \nodepart{right}{$z$} };
        \node[name=Y,  right=of Z, ell, shape=ellipse]{$Y(z)$};
        \node(C)       [below=of Z] {};
        \node(U)       [above=of Z] {$\bm{U}$};
        \draw[-{Latex[length=2mm, width=2mm]}] (A) -- (Z);
        \draw[-{Latex[length=2mm, width=2mm]}] (Z) -- (Y);     
        \draw[->,-{Latex[length=2mm, width=2mm]}](U) edge (A.north west);
        \draw[-{Latex[length=2mm, width=2mm]}] (U) -- (Y);
    \end{tikzpicture}
    }
\subfloat[]{\label{fig:td}
    \centering
    \begin{tikzpicture}
    \tikzset{line width=0.6pt, outer sep=0pt,
    ell/.style={draw,fill=white, inner sep=2pt,
    line width=0.6pt},
    swig vsplit={gap=2pt},
    node distance=0.5cm};
        \node[name=A, shape=swig vsplit]{
            \nodepart{left}{$A$}
            \nodepart{right}{$a$} };
        \node[name=Z, shape=swig vsplit, right=of A]{
            \nodepart{left}{$Z(a)$}
            \nodepart{right}{$z$} }; 
        \node[name=Y,  right=of Z, ell, shape=ellipse]{$Y(z)$};
        \node(C)       [below=of Z] {$\bm{C}$};
        \node(U)       [above=of Z] {$\bm{U}$};
        \draw[-{Latex[length=1mm, width=1mm]}] (A) -- (Z);
        \draw[-{Latex[length=1mm, width=1mm]}] (Z) -- (Y);
        \draw[->,-{Latex[length=1mm, width=1mm]}](C) edge[out=180,in=300] (A.south west);
        \draw[-{Latex[length=1mm, width=1mm]}] (C) -- (Z);
        \draw[-{Latex[length=1mm, width=1mm]}] (C) -- (Y.south west);
        \draw[->,-{Latex[length=1mm, width=1mm]}](U) edge (A.north west);
        \draw[-{Latex[length=1mm, width=1mm]}] (U) -- (Y);
    \end{tikzpicture}
    }
\caption{  
Examples of Single-World Intervention Graphs that represent the distributions that satisfy 
(a) Assumption \ref{ass:bd}; (b) Assumption \hyperref[ass:fd]{\textbf{FD}}; (c) Assumption \hyperref[ass:td]{\textbf{TD}}. } 
\label{fig:model}
\end{figure}
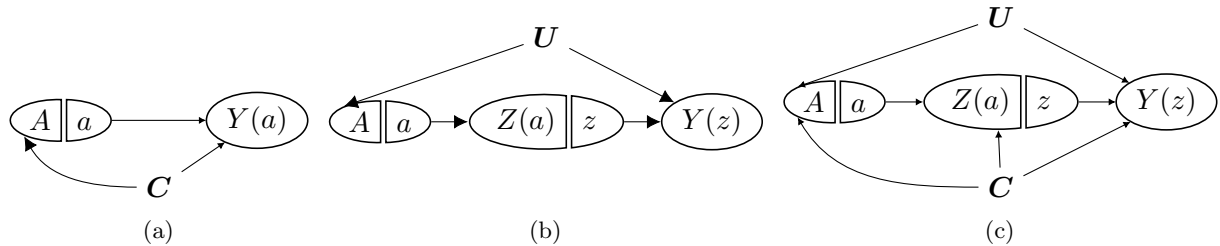

Assumption \ref{ass:td1} is the same as Assumption \ref{ass:fd1}, and
corresponds to Assumption 5 in \citet{fulcher2020robust}.
 Assumptions \ref{ass:td2}--\ref{ass:td3} correspond to Assumptions 2 and 3 in \citet{fulcher2020robust}, and are conditional independencies given $C,$ instead of the marginal independencies in \ref{ass:fd2}--\ref{ass:fd3}. Thus, assumptions \ref{ass:td2}--\ref{ass:td3} allow pre-treatment covariates $C$ to affect the mediator $Z$ (see Figure \ref{fig:td}, for an example).  
 In the sequel, we will denote the set of assumptions \ref{ass:td1}--\ref{ass:td3} by Assumption \hyperref[ass:td]{\textbf{TD}}.

We work within the potential outcome framework and the identification models above are potential outcome assumptions. An alternative framework to causal inference is based on directed acyclic graphs (DAGs) and do-calculus \citep{pearl2009causality}. These 
    two frameworks are used to define causal parameters and map those to statistical parameters (functional of the observed data distribution) through identification models. The frameworks are not equivalent. However, direct comparisons of identification assumptions within the two frameworks can be done by looking at what consequences the assumptions have on the observed data distribution and its mapping into a statistical parameter. The results derived under the potential outcome framework in Section \ref{sec:comparison} can thus be compared with results deduced elsewhere within the DAG framework to provide such comparisons. Furthermore, a recent third causal framework combines DAGs and potential outcome variables through Single-World Intervention Graphs \citep[SWIGs,][]{richardson2013single}. We use SWIGs informally here to describe examples of data generating mechanisms for which the above identification assumptions hold, and relate these scenarios to criteria used in the graphical literature. 


  In particular, when the probability distribution function of $(Y(a), A, C)$ is compatible with a SWIG on these variables, model \ref{ass:bd} holds if $C$ satisfies Pearl's back-door criterion \citep[see Def. 3.3.1,][]{pearl2009causality} relative to $(A,Y)$ in the corresponding DAG. This is the case in Figure \ref{fig:bd}.   
Furthermore, for a distribution for $(Y(z), Z(a), A, U)$ compatible with the SWIG in Figure \ref{fig:fd}, then model \hyperref[ass:fd]{\textbf{FD}} holds if Pearl's front-door criterion \citep[see Def. 3.3.3,][]{pearl2009causality} holds in the corresponding DAG.
Finally, consider a distribution function for $(Y(a,z)=Y(z), Z(a), A, C, U)$ compatible with the SWIG in Figure \ref{fig:td}. Then, Assumption \ref{ass:td1} (identical to Assumption \ref{ass:fd1}) corresponds to condition (i) of Pearl's front-door criterion and Assumption 5 in \citet{fulcher2020robust}, i.e. $Z$ should intercept all directed paths from  $A$ to $Y$.  Assumption \ref{ass:td2} corresponds to all back-door paths from $Z(a)$ to $A$ being blocked by $C$ (no unmeasured confounding of the exposure-mediator relationship conditional on $C$, i.e. (ii) in Pearl's front-door criterion, but here conditioning on $C$; see also \citealt[p.464]{vanderweele2015explanation}). Assumption \ref{ass:td3} corresponds to conditioning on $C,$ all back-door paths from $Z$ to $Y$ being blocked by $A$ (condition (iii) in Pearl's front-door criterion with blocking set including not only treatment $A$, but also covariates $C$).

\subsection{Efficient Influence Functions} \label{sec:eif}
In what follows, we present the efficient influence functions for $\theta$ under the three considered sets of identifying assumptions. The reader unfamiliar with semiparametric inference is referred to Appendix \ref{sec:appendix_semiparametric_efficiency}, \citet{newey1990semiparametric}, and \citet{tsiatis2007semiparametric}.
Under Assumption \ref{ass:bd}, \citet{robins1994estimation} and \citet{hahn1998role} derived the efficient influence function for $\theta$:   
\begin{align*}
    & \varphi_{bd}  =
    \frac{I(A=a^*)}{p(a^*|C)}(Y - E(Y|a^*, C))-  \frac{I(A=a)}{p(a|C)}(Y - E(Y|a, C)) \nonumber\\
    & +  E(Y|a^*C) - E(Y|a, C) - \theta, 
\end{align*}
and \citet{hahn1998role} provided  the corresponding semiparametric efficiency bound:
\begin{align}
    var& \varphi_{bd}  = E\left[\frac{var\big(Y|a^*, C\big)}{p(a^*|C)}+\frac{var\big(Y|a, C\big)}{p(a|C)}\right] + E\bigg(E\big(Y|a^*, C\big)-E\big(Y|a, C\big)-\theta\bigg)^2. \label{eq:var_eif_bd}
\end{align}

Theorem 1 in \citet{fulcher2020robust} provides the efficient influence function for parameter $E(Y(Z(a^*))),$ which, under Assumption \ref{ass:fd1} (equivalently \ref{ass:td1}), is equal to $EY(a^*).$ Since \ref{ass:td1} does not restrict the observed data distribution in any way, Theorem 1 from \citet{fulcher2020robust} can be used directly to derive efficient influence functions under Assumption \hyperref[ass:fd]{\textbf{FD}} and under Assumption \hyperref[ass:td]{\textbf{TD}}. Thus,
ignoring the fact that pre-treatment covariates $C$ are observed (i.e., treating $C$ as the unobserved $U$), and using the linearity of the differentiation operation, one can show using Theorem 1 in \citet{fulcher2020robust} that, under the \hyperref[ass:fd]{\textbf{FD}} assumption,  the efficient influence function for $\theta$ is    
\begin{align}
        &\varphi_{fd}= 
        \left(Y - E(Y|A,Z)\right) \frac{p(Z|a^*)-p(Z|a)}{p(Z|A)} +\frac{I(A=a^*)}{p(A)}\left(\sum_{\bar{a}} E(Y|\bar{a}, Z)p(\bar{a})-EY(a^*)\right) \label{eq:eif_ate_fd_def}\\
        &-\frac{I(A=a)}{p(A)}\left(\sum_{\bar{a}} E(Y|\bar{a}, Z)p(\bar{a})-EY(a)\right)+\sum_{z}E(Y|A, z)(p(z|a^*)-p(z|a))-\theta. \nonumber
\end{align}
Furthermore, the efficient influence function for $\theta$ under the \hyperref[ass:td]{\textbf{TD}} assumption follows directly from Theorem 1 in \citet{fulcher2020robust}: 
\begin{align}
    &\varphi_{td}  =
    \left(Y-E(Y|A,  Z, C)\right) \frac{p(Z|a^*,C)-p(Z|a,C)}{p(Z|A,C)} \nonumber \\
     & +\left(\sum\limits_{\bar{a}}E(Y|\bar{a}, Z, C)p(\bar{a}|C)-\sum_{\bar{a},z}E(Y|\bar{a}, z, C)p(z|A,C)p(\bar{a}|C)\right) \frac{I(A=a^*) - I(A=a)}{p(A|C)}\nonumber \\
    & + \sum\limits_{z} E(Y|A, z, C)\big(p(z|a^*, C)- p(z|a, C)\big)-\theta. \label{eq:eif_td}
\end{align}
The semiparametric efficiency bounds under Assumption \hyperref[ass:fd]{\textbf{FD}}, $var \varphi_{fd},$ and  under Assumption \hyperref[ass:td]{\textbf{TD}}, $var \varphi_{td},$ are presented in Appendix \ref{sec:appendix_var_eif}. The expressions for the bounds do not include any counterfactuals and can, therefore, be estimated from the observed data.

\section{Efficiency Comparisons}\label{sec:comparison}

We consider here models where pairs of \ref{ass:bd}, \hyperref[ass:fd]{\textbf{FD}}, and  \hyperref[ass:td]{\textbf{TD}} assumptions are fulfilled. We first compare the asymptotic variances of the semiparametric estimation based only on one of the assumptions within a pair respectively, showing that none is uniformly lower.  Note that the efficiency bound under the model where a pair of assumptions are fulfilled can be even lower than the bound under either individual identifying assumption.  We, therefore, deduce the semiparametric efficiency bounds valid under each pair of assumptions.

\subsection{Back-door versus Two-door Identification} \label{sec:two_door_vs_back_door}

The model that satisfies Assumptions \ref{ass:bd} and \hyperref[ass:td]{\textbf{TD}} considers the observed variables $C, A, Z, Y.$ 


\begin{proposition} \label{th:sufficient_conditions_td_vs_bd}
Suppose that Assumptions \ref{ass:bd} and \hyperref[ass:td]{\textbf{TD}} are fulfilled and $Y (a; z)  \indep A | C$ for any $a \in \mathcal{A}, z \in \mathcal{Z}.$ Consider semiparametric estimation based only on Assumption \ref{ass:bd} and  semiparametric estimation based only on Assumption  \hyperref[ass:td]{\textbf{TD}}. 
The difference between $var \varphi_{td}$ and $var \varphi_{bd}$ can be represented as follows: 
\begin{align}
    v&ar \varphi_{td}- var \varphi_{bd} \nonumber\\
    &= \sum_{z,c}p(c)var(Y|z, c) \Bigg( (p(z|a^*,c) - p(z|a,c))^2  
    \sum\limits_{\bar{a}}\frac{p(\bar{a}|c)}{p(z|\bar{a},c)}- \frac{p(z|a^*,c)}{p(a^*|c)}  - \frac{p(z|a,c)}{p(a|c)}  \Bigg) \label{eq:var_td_minus_var_bd}
\end{align}
Neither of the estimation strategies is uniformly more efficient than the other  (in terms of the lowest asymptotic variance) regardless of model parameters. For example, if
\begin{align}
(p(z|a^*,c)-p(z|a,c))^2  \sum\limits_{\bar{a}}\frac{p(\bar{a}|c)}{p(z|\bar{a},c)}- \frac{p(z|a^*,c)}{p(a^*|c)}  - \frac{p(z|a,c)}{p(a|c)}\leq 0  \text{ for all } z \in \mathcal{Z},c \in \mathcal{C}, \label{eq:suff_td_better_bd}
\end{align}
then $var \varphi_{td} \leq var \varphi_{bd}. $ 
Moreover, if
\begin{align}
(p(z|a^*,c)-p(z|a,c))^2  \sum\limits_{\bar{a}}\frac{p(\bar{a}|c)}{p(z|\bar{a},c)} - \frac{p(z|a^*,c)}{p(a^*|c)}  - \frac{p(z|a,c)}{p(a|c)} \geq 0\text{ for all } z \in \mathcal{Z}, c \in \mathcal{C},\label{eq:suff_td_worse_bd}
\end{align} then $var \varphi_{td} \geq var \varphi_{bd}. $
\end{proposition}
The proof of Proposition \ref{th:sufficient_conditions_td_vs_bd} is provided in Appendix \ref{sec:appendix_proof_sufficient_conditions_td_less_than_bd}. Note that the conditions \eqref{eq:suff_td_better_bd} and \eqref{eq:suff_td_worse_bd}  involve only the distribution of the treatment and the mediator. Conditional ignorability for $Y(a,z),$ $Y (a; z)  \indep A | C$ in Proposition \ref{th:sufficient_conditions_td_vs_bd}  follows from Assumption \ref{ass:bd} for a distribution compatible with a SWIG (see Proposition 4 in   \citealp{malinsky2019potential}). 

Conditions \eqref{eq:suff_td_better_bd} and \eqref{eq:suff_td_worse_bd} in Proposition \ref{th:sufficient_conditions_td_vs_bd} are strict since they should be fulfilled for possible combinations of $z$ and $c$. However, one can notice that if the relationship between the treatment $A$ and the mediator $Z$ is weak, that is, $p(z|a^*,c)-p(z|a,c)$ is small, then $ var \varphi_{td}- var \varphi_{bd}$ is expected  to be negative and the estimation based on Assumption \hyperref[ass:td]{\textbf{TD}} 
 to be more efficient than the estimation based on Assumption \ref{ass:bd}. How weak the relationship between $A$ and $Z$ has to be depends on the distribution of $A$:  the smaller  $p(a|c)$ or $p(a^*|c),$ the closer is  $ var \varphi_{td}-var \varphi_{bd}$ to be negative. Stronger $A$--$Z$ relationship requires smaller $p(a|c)$ or $p(a^*|c)$ for the estimation based on Assumption \hyperref[ass:td]{\textbf{TD}} to be more efficient than the one based on Assumption \ref{ass:bd}.

In the  specific case of a binary treatment  $A$, the sufficient conditions \eqref{eq:suff_td_better_bd} and \eqref{eq:suff_td_worse_bd} might be further simplified as follows (see Appendix \ref{sec:example}). If for all $ z \in \mathcal{Z},c \in \mathcal{C} $, we have:
 \begin{align*}
    \frac{p(z|a^*,c)}{p(z|a,c)} \in I_{td \leq bd}= \bigg[\frac{2u+1-\sqrt{4u+1}}{2u},  \frac{2u+1+\sqrt{4u+1}}{2u}\bigg], 
\end{align*}
 where $u=(1-p(a^*|c))p(a^*|c),$ then $var \varphi_{td} \leq var \varphi_{bd}.$ If for all $ z \in \mathcal{Z},c \in \mathcal{C} $, $\frac{p(z|a^*,c)}{p(z|a,c)} \in I_{td \leq bd}^C, $ where $I_{td \leq bd}^C $ is the complement set of $I_{td \leq bd},$
then $var \varphi_{td} > var \varphi_{bd}.$ One can note that  for any $p(a^*|c) \in (0,1),$ the interval $(3-2\sqrt{2}, 3+2\sqrt{2}) \in  I_{td \leq bd}.$    Therefore, if  for all $ z,c, \quad \frac{p(z|a^*,c)}{p(z|a,c)} \in (3-2\sqrt{2}, 3+2\sqrt{2})  $ and  $p(a^*|c) \in (0,1)$ for any  $c \in \mathcal{C},$ then $var \varphi_{td} \leq var \varphi_{bd}.$ Appendix \ref{sec:example} illustrates  further the sufficient conditions when all observed variables are binary.

The observed data distribution when Assumptions \hyperref[ass:td]{\textbf{TD}}, \ref{ass:bd},  and $Y (a; z)  \indep A | C$ are fulfilled differs from the distributions under the \hyperref[ass:td]{\textbf{TD}} or the \ref{ass:bd} assumption  alone.  When the mediator is observed, estimation based on the back-door assumption does not use the information about the mediator.
Under Assumption \hyperref[ass:td]{\textbf{TD}}, the observed data distribution is unrestricted since all observed variables $C, A, Z, Y$ may be related to each other. However, since unobserved confounding of the treatment-outcome relationship is not allowed by Assumption \ref{ass:bd} together with $Y (a; z)  \indep A | C$, such restriction implies conditional independence of $Y$ and $A$ given $Z,C$ (see Appendix \ref{sec:appendix_proof_sufficient_conditions_td_less_than_bd}). 
This implies that, when Assumptions \ref{ass:bd}, \hyperref[ass:td]{\textbf{TD}},  and $Y (a; z)  \indep A | C$ are fulfilled simultaneously, the semiparametric efficiency bound for $\theta$ differs from the bound under the models defined only by  Assumption \ref{ass:bd} or Assumption \hyperref[ass:td]{\textbf{TD}}.  Therefore, we conclude this section by providing the semiparametric efficiency bound under this new set of assumptions. The estimators with the influence function that corresponds to this bound have an asymptotic variance at least as low  as the estimators constructed using the efficient influence functions under either the two-door or the back-door assumption.

\begin{proposition} \label{th:td_bd_eif} 
    When Assumptions \ref{ass:bd}, \hyperref[ass:td]{\textbf{TD}}, and $Y (a; z)  \indep A | C$ are fulfilled simultaneously, the efficient influence function for $\theta$ is
    \begin{align}
        \varphi_{bd, td}&
        = \left(Y-E(Y|Z, C)\right) \frac{ p(Z|a^*,C)-p(Z|a,C)}{\sum_a p(Z|a, C)p(a|C)} \nonumber \\
        & +\left(E(Y|Z, C) - \sum_{z}E(Y|z,C) p(z|A,C)\right) \frac{I(A=a^*)-I(A=a)}{p(A|C)}\nonumber\\
        & + \sum\limits_z E(Y|z,C) \big( p(z|a^*, C)  - p(z|a, C) \big)  - \theta.\label{eq: EIF_td_bd}
    \end{align}
\end{proposition}
The proof of Proposition \ref{th:td_bd_eif} can be found in Appendix \ref{sec:proof of lemma EIF TD BD}. Additionally, Appendix \ref{sec:appendix_var_eif} provides the expression for $var \varphi_{bd,td}$ in terms of observed data distribution that allows for direct estimation of the bounds from the observed data.

Note that because the independencies in the observed
data distribution here are the same as for the Bayesian network compatible with a DAG
defined by the paths $C \rightarrow A \rightarrow Z \rightarrow Y,$ $C \rightarrow Y,$ $ C\rightarrow Z,$ the influence function $ \varphi_{bd,td}$  in Equation \eqref{eq: EIF_td_bd} is the same as the efficient influence function in Theorem 14 from  \citet{rotnizky2020efficient} for this Bayesian network.  

\subsection{Front-door versus Two-door Identification} \label{sec:front_door_vs_two_door}
In order to compare the bounds under the \hyperref[ass:fd]{\textbf{FD}} and the \hyperref[ass:fd]{\textbf{TD}} assumptions, we first derive the lowest asymptotic variance attainable by the semiparametric estimators that are constructed under Assumption \hyperref[ass:fd]{\textbf{FD}}, $ var\varphi_{fd}$ (see Equation \ref{eq:var_eif_fd} in Appendix \ref{sec:appendix_var_eif}). Note that the conditioning set in the outcome and mediator models in $ var\varphi_{fd}$ does not include the pre-treatment covariates, in contrast to $var \varphi_{td}$ given in Equation \eqref{eq:var_eif_td}. 

In the sequel, and in line with common models considered under the front-door criterion (see, for example, Figure 3 in \citealp{pearl1995causal}), when pre-treatment covariates $C$ are observed under Assumption \hyperref[ass:fd]{\textbf{FD}}, we will consider the case where they confound the \mbox{$A$--$Y$} relationship only. Then,  the pre-treatment covariates must be unrelated to the mediator for the \hyperref[ass:fd]{\textbf{FD}} assumption to be fulfilled, that is, $Z(a) \indep C|A.$

\begin{proposition}\label{th:fd_vs_td}
    Suppose that Assumptions \hyperref[ass:fd]{\textbf{FD}} and \hyperref[ass:td]{\textbf{TD}} are fulfilled, and $Z(a) \indep C|A$ for all $a \in \mathcal{A}$.  Consider semiparametric estimation based only on Assumption \hyperref[ass:fd]{\textbf{FD}} and  semiparametric estimation based only on Assumption  \hyperref[ass:td]{\textbf{TD}}. Neither of the estimation strategies is uniformly more efficient than the other  (in terms of the lowest asymptotic variance) regardless of model parameters. 
    For example,  consider 
    $E(Y|Z, C, A) = \gamma_0 + \gamma_1Z+ \gamma_2C + \gamma_3A$ for some $\gamma_0, \gamma_1, \gamma_2, \gamma_3.$ We have   
    \begin{align*}
    & var \varphi_{fd} - var \varphi_{td} =  \sum_{z} (p(z|a^*)-p(z|a))^2 \sum_{\bar{a}}\frac{\gamma_2^2 var(C|\bar{a})p(\bar{a})}{p(z|\bar{a})} \\ 
    & + \gamma_1^2 var(Z|a^*)\left( \frac{1}{p(a^*)} - \sum\limits_{c}\frac{p(c)}{p(a^*|c)}\right) + \gamma_1^2 var(Z|a)\left(\frac{1}{p(a)} - \sum\limits_{c}\frac{p(c)}{p(a|c)}\right). &
    \end{align*}
    If
    \begin{align}
        \frac{1}{p(a^*)} \geq \sum\limits_{c}\frac{p(c)}{p(a^*|c)} \text{ and }\frac{1}{p(a)} \geq \sum\limits_{c}\frac{p(c)}{p(a|c)}, \label{eq:suff_fd_worse_td}
    \end{align} 
    then $var \varphi_{fd} \geq var \varphi_{td}.$
\end{proposition}
See proof in \ref{sec:appendix_proof_proposition_fd_vs_td}. 
Note that if $A \indep C,$ then the inequalities \eqref{eq:suff_fd_worse_td} in Proposition \ref{th:fd_vs_td} are satisfied and $var \varphi_{fd} \geq var \varphi_{td}.$ This is true even if $C$ does not affect $Y$ ($\gamma_2 = 0$). Thus, even if the covariates are not related to the mediator or the outcome, using the information about the covariates might improve estimation efficiency. When $C$ and $A$ are binary the inequalities \eqref{eq:suff_fd_worse_td} hold only if $A$ is independent of $C.$
When $A$ depends on $C, $ the ordering of the variances depends on other model parameters. For example, when $\gamma_2$ is large (strong \mbox{$C$--$Y$})  and/or $A$--$Z$ relationship  is strong and  $\gamma_1$ is close to 0 (weak \mbox{$Z$--$Y$} relationship),  $var \varphi_{td}$ might be lower than $var \varphi_{fd}.$

The linear assumptions made here are not meant to be done in a given empirical study (where hopefully flexible machine learning models should be used), otherwise making such assumptions would obviously change the efficiency bound for the ACE. Instead, these linearity assumptions are used to exemplify the models and understand the implications for the comparison of the efficiency bounds if the assumptions would hold but would not be known nor assumed in the analysis.

Below, we provide the semiparametric efficiency bound for $\theta$ in the model which satisfies Assumptions  \hyperref[ass:fd]{\textbf{FD}} (together with $Z(a) \indep C|A$) and \hyperref[ass:td]{\textbf{TD}}.

\begin{proposition} \label{th:fd_td_eif}
   Suppose that Assumptions  \hyperref[ass:fd]{\textbf{FD}} and \hyperref[ass:td]{\textbf{TD}} are fulfilled, and $Z(a) \indep C|A.$ The efficient influence function for $\theta$ is
    \begin{align*}
        \varphi&_{fd, td} =  \left(Y-E(Y|A,  Z, C)\right) \frac{p(Z|a^*)-p(Z|a)}{p(Z|A)}   \\
            &  + \sum_{c}\left(\sum\limits_{\bar{a}}E(Y|\bar{a}, Z, c)p(\bar{a}|c)
            - \sum_{\bar{a},z}E(Y|\bar{a}, z, c)p(z|A)p(\bar{a}|c)\right) \times \\
            & \times \frac{(I(A=a^*)-I(A=a)) p(c)}{p(A)} + \sum\limits_{z} E(Y|A, z, C)\big(p(z|a^*)- p(z|a)\big)-\theta.  
    \end{align*}
\end{proposition}
See the expression for $var \varphi_{fd,td}$ in Appendix \ref{sec:appendix_var_eif}; the proof of Proposition \ref{th:fd_td_eif} is similar to the proof of Proposition \ref{th:td_bd_eif} and is provided in \ref{sec:appendix_proof_th_fd_td_eif}.
It is interesting to note that $\varphi_{fd, td}$ is equal to the efficient influence function under  the assumptions of Section~\ref{sec:notation}, \hyperref[ass:td]{\textbf{TD}}, if $Z(a) \indep C|A.$ This happens because the \hyperref[ass:fd]{\textbf{FD}} assumption imposes no additional independence assumptions on the joint distribution which can improve the efficiency in the estimation of the ACE. Additionally,
because conditional independencies in the observed data distribution implied by the assumptions in Proposition \ref{th:fd_td_eif} are the same as those for a distribution compatible with the SWIG in Figure \ref{fig:td} where the arrow from $C$ to $Z(a)$ is absent, $\varphi_{fd, td}$ obtained here is the same as the efficient influence function in Theorem 12 from \citet{bhattacharya2020semiparametric} for the corresponding DAG.

\subsection{Front-door versus Back-door Identification}\label{sec:front_door_vs_back_door}
The expressions $var \varphi_{fd}$  in Equation \eqref{eq:var_eif_fd} and $var \varphi_{bd}$ in Equation \eqref{eq:var_eif_bd} can be used to compare the bounds obtained under the front-door and the back-door assumptions as follows.

\begin{proposition}\label{th:fd_vs_bd}
    Suppose that Assumptions  \ref{ass:bd}, \hyperref[ass:fd]{\textbf{FD}}  are fulfilled and  $Y (a; z)  \indep A | C,$ \\${Z(a)\indep C|A}$ for any $a \in \mathcal{A}, z \in \mathcal{Z}.$ Consider semiparametric estimation based only on Assumption \ref{ass:bd} and  semiparametric estimation based only on Assumption  \hyperref[ass:td]{\textbf{FD}}.  Neither of the estimation strategies is uniformly more efficient than the other  (in terms of the lowest asymptotic variance) regardless of model parameters. In particular, if for the true data distribution,   $E(Y|Z, C) = \gamma_0 + \gamma_1Z+ \gamma_2C$ for some $\gamma_0, \gamma_1, \gamma_2,$ then

    \begin{flalign*}
        & var\varphi_{fd} - var\varphi_{bd} \\
        & = \sum_{z,c}var(Y|z,c) p(c) \left(\sum_{\bar{a}}\frac{p(\bar{a}|c)(p(z|a^*)-p(z|a))^2}{p(z|\bar{a})}-\frac{p(z|a^*)}{p(a^*|c)}-\frac{p(z|a)}{p(a|c)}\right)\\
        & + \sum_{z} (p(z|a^*)-p(z|a))^2 \sum_{\bar{a}}\frac{\gamma_2^2var(C|\bar{a})p(\bar{a})}{p(z|\bar{a})}\\
        & +\gamma_1^2 var(Z|a^*)\left( \frac{1}{p(a^*)} - \sum\limits_{c}\frac{p(c)}{p(a^*|c)}\right) + \gamma_1^2 var(Z|a)\left(\frac{1}{p(a)} - \sum\limits_{c}\frac{p(c)}{p(a|c)}\right).
\end{flalign*}
 If
    \begin{align*}
        & \frac{1}{p(a^*)} \geq  \sum\limits_{c}\frac{p(c)}{p(a^*|c)}, \frac{1}{p(a)} \geq \sum\limits_{c}\frac{p(c)}{p(a|c)} \text{ and}\\
        & (p(z|a^*)-p(z|a))^2  \sum\limits_{\bar{a}}\frac{p(\bar{a}|c)}{p(z|\bar{a})}- \frac{p(z|a^*)}{p(a^*|c)}  - \frac{p(z|a)}{p(a|c)} \geq 0 \text{ for any } z \in \mathcal{Z},\bm{c} \in \mathcal{C},
    \end{align*}
   then $var\varphi_{fd} \geq var\varphi_{bd}.$ 
\end{proposition}
See  \ref{sec:appendix_proof_proposition_fd_vs_bd} for a proof. Note that the first inequality ensures $var\varphi_{fd} \geq var\varphi_{td}$ according to Proposition~\ref{th:fd_vs_td}, while the second inequality is the same as that in Proposition~\ref{th:sufficient_conditions_td_vs_bd} which ensured $var\varphi_{td} \geq var\varphi_{bd}.$ 

When both the \hyperref[ass:fd]{\textbf{FD}} and the \ref{ass:bd} assumptions are satisfied,  there is no unmeasured confounding of the \mbox{$A$--$Y$}  relationship and the two-door door assumption is also fulfilled. These restrictions could be used to further improve the efficiency of ACE estimation.   

\begin{proposition} \label{th:bd_fd_td_eif}
    Suppose that Assumptions \ref{ass:bd},  \hyperref[ass:fd]{\textbf{FD}},  and \hyperref[ass:td]{\textbf{TD}} are fulfilled and  $Z(a) \indep C|A.$ The efficient influence function for $\theta$ is
    \begin{align*}
        \varphi_{bd, fd, td} & = \frac{\left(Y-E(Y|Z, C)\right) (p(Z|a^*)-p(Z|a))}{\sum_a p(a|C)p(Z|a)}  \\
        & +  \sum_{c}\left(E(Y|Z,c) - \sum_{z} E(Y|z,c)p(z|A)\right)\frac{(I(A=a^*) - I(A=a))p(c)}{\sum_{c}{p(c)p(A|c)}} \\
        & +\sum_{z} E(Y|z,C) (p(z|a^*)- p(z|a))  -\theta. &
    \end{align*}
\end{proposition}
See the corresponding expression for $var \varphi_{bd,fd,td}$ in Appendix \ref{sec:appendix_var_eif}. The proof of Proposition \ref{th:bd_fd_td_eif} is similar to the proof of Proposition \ref{th:td_bd_eif} and is provided in \ref{sec:appendix_proof_lemma_fd_td_bd}. Since the independencies in the observed data distribution here are the same as for the Bayesian network  compatible with a DAG defined by the paths $C\rightarrow A \rightarrow Z \rightarrow Y, C \rightarrow Y$, the influence function $\varphi_{bd, fd, td}$ is the same as the efficient influence function in Theorem 14 from  \citet{rotnizky2020efficient}  for the distribution compatible with such a DAG.

\section{Semiparametric Estimation via Estimating Equation Estimators} \label{sec:semipar_est}
Influence functions can be used to semiparametrically estimate $\theta$ via, e.g., one-step, targeted learning and estimating equation estimators \citep[see, e.g.,][]{hines2022demystifying}.
Here, we consider the semiparametric estimator $\hat{\theta}$ as a solution of the estimating equation 
$$\frac{1}{n}\sum_{i=1}^n \varphi(X_i, \hat{\eta}, \hat{\theta}) = 0, $$
where $\hat{\eta}$ is an estimator of unknown nuisance parameters, for example,  $\hat{p}(C),$ $\hat{p}(A|C),$ $\hat{p}(Z|A, C),$ or $\hat{E}(Y|A, Z, C).$  

Influence functions described above are all of the form $\varphi (X, \eta, \theta) = m(X, \eta) - \theta$, i.e. a difference between a function of the data and the nuisance parameters $m(X, \eta)$ (different for different influence functions) and $\theta.$  
Therefore, the solutions of the estimating equations take the form 

$$\hat{\theta} = \frac{1}{n}\sum_{i=1}^n m(X_i, \hat{\eta}).$$
Properties of these estimators depend on the properties of nuisance parameter estimators $\hat{\eta}.$  
Below we discuss conditions for consistency, asymptotic normality and efficiency of such plug-in estimators of $\theta$. 

It is widely known that the estimators based on $\varphi_ {bd}$ are doubly robust, that is, a solution $\hat{\theta}_{bd}$ of the estimating equation  $\frac{1}{n}\sum_{i=1}^n \varphi_{bd,i}(X_i, \hat{\eta}, \hat{\theta}_{bd}) = 0$  is a consistent  estimator of $\theta$ if at least one of  the estimators of $E(Y|A, C)$ or $p(A|C)$ is consistent (see, e.g., \citealp{kennedy2016semiparametric}).

From Theorem 2 in \citet{fulcher2020robust}, we have that a solution $\hat{\theta}_{td}$ of the equation  \\ 
$\frac{1}{n}\sum_{i=1}^n \varphi_{td,i}(X_i, \hat{\eta}, \hat{\theta}_{td}) = 0$ is  consistent if $p(Z|A,C)$ or  $E(Y|A, Z, C)$ and $p(A|C)$ are correctly specified. These results are also in line with the consistency conditions of Theorem 9 in \citet{bhattacharya2020semiparametric} for a nonparametric saturated model under the \hyperref[ass:td]{\textbf{TD}} assumption.
Additionally, from \citet{fulcher2020robust}, solutions $\hat{\theta}_{fd}$ of the estimating equation \\  
$\frac{1}{n}\sum_{i=1}^n \varphi_{fd,i}(X_i, \hat{\eta}, \hat{\theta}_{fd}) = 0$ are consistent if $p(Z|A)$ or $E(Y|A,Z)$ and $p(A)$  are correctly specified. \autoref{th:efficiency_of_fd_and_td_est} below  provides conditions for the efficiency of $\hat{\theta}_{fd}$ and $\hat{\theta}_{td}.$  We need the following assumption. 
\begin{assumption}\label{ass:regularity_asympt_normality}
    \mbox{}
        \begin{itemize}
            \item[(i)] $|| m(X_i, \hat{\eta}) -  m(X_i, \eta_0))||^2 = o_p(1),$ where $\eta_0$ is the true value of $\eta,$ 
            \item[(ii)]  $\hat{\eta}$ is constructed using sample splitting  or 
           $m (X, \hat{\eta})$ belongs to a Donsker class \citep[for more details on this assumption see, e.g.,][]{kennedy2016semiparametric}.
        \end{itemize} 
    \end{assumption}

\begin{theorem} \label{th:efficiency_of_fd_and_td_est}

Under Assumption \hyperref[ass:td]{\textbf{TD}}, \Cref{ass:regularity_asympt_normality} and \Cref{ass:an_td}, given in the appendix, a solution $\hat{\theta}_{td}$ of the equation  $\frac{1}{n}\sum_{i=1}^n \varphi_{td,i}(X_i, \hat{\eta}, \hat{\theta}_{td}) = 0$ is a regular asymptotically linear estimator of $\theta = EY(a^*)-EY(a)$ with influence function $\varphi_{td}$ and reaches the efficiency bound when 
\begin{enumerate}[label=(\roman*)]
    \item  $ || \hat{E}(Y|A, Z, C) - E (Y|A, Z, C)|| \cdot
            ||   \hat{p}(Z|A,C) - p(Z|A,C) ||= o_p(n^{-1/2}),$ 
    \item $ || \hat{E}(Y|A, Z, C) - E (Y|A, Z, C)
            || \cdot
            ||  \hat{p}(Z|\breve{a},C) - p(Z|\breve{a},C)
            ||= o_p(n^{-1/2}), $ $\forall \breve{a} \in \{a^*, a\},$
    \item $ ||  \hat{p}(Z|\breve{a},C) - p(Z|\breve{a},C) || \cdot
            ||  \hat{p}(\breve{a}|C) - p(\breve{a}|C) ||
            = o_p(n^{-1/2}), $ $\forall \breve{a} \in \{a^*, a\},$
    \item $ || \hat{p}(Z|\breve{a},C) - p(Z|\breve{a},C)|| 
    \cdot
            ||  \hat{p}(A|C) - p(A|C) ||
            = o_p(n^{-1/2}), $ $\forall \breve{a} \in \{a^*, a\}.$
\end{enumerate}

Under Assumption \hyperref[ass:fd]{\textbf{FD}}, \Cref{ass:regularity_asympt_normality} and \Cref{ass:an_fd}, given in the appendix,  a solution $\hat{\theta}_{fd}$ of the equation  
$\frac{1}{n}\sum_{i=1}^n \varphi_{fd,i}(X_i, \hat{\eta}, \hat{\theta}_{fd}) = 0$ is a regular asymptotically linear estimator of $\theta = EY(a^*)-EY(a)$ with influence function $\varphi_{fd}$   and reaches the efficiency bound when
\begin{enumerate}[label=(\roman*)]
\setcounter{enumi}{3}
    \item $ || \hat{E}(Y|A, Z) - E (Y|A, Z)|| \cdot
            ||   \hat{p}(Z|A) - p(Z|A) ||= o_p(n^{-1/2}),$ 
    \item $ || \hat{E}(Y|A, Z) - E (Y|A, Z)
            || \cdot
            ||  \hat{p}(Z|\breve{a}) - p(Z|\breve{a})
            ||= o_p(n^{-1/2}), $ $\forall \breve{a} \in \{a^*, a\},$
    \item $ ||  \hat{p}(Z|\breve{a}) - p(Z|\breve{a}) || \cdot
            ||  \hat{p}(\breve{a}) - p(\breve{a}) ||
            = o_p(n^{-1/2}), $ $\forall \breve{a} \in \{a^*, a\},$
    \item $ || \hat{p}(Z|\breve{a}) - p(Z|\breve{a})|| 
    \cdot
            ||  \hat{p}(A) - p(A) ||
            = o_p(n^{-1/2}), $ $\forall \breve{a} \in \{a^*, a\}.$ 
\end{enumerate}
\end{theorem}
The proof of \autoref{th:efficiency_of_fd_and_td_est} is similar to the proof of \autoref{th:consistency_efficiency_of_bd_td_est} and is provided in \ref{sec:appendix_proof_lemma_efficiency_of_td_and_fd}. \autoref{th:consistency_efficiency_of_bd_td_est} below provides conditions for consistency, asymptotic normality, and efficiency of $\hat{\theta}_{bd,td}.$

\begin{theorem} \label{th:consistency_efficiency_of_bd_td_est}
    Suppose that Assumptions \ref{ass:bd} and \hyperref[ass:td]{\textbf{TD}} are fulfilled and $Y (a; z)  \indep A | C$ for any $a \in \mathcal{A}, z \in \mathcal{Z}.$ Suppose that for some $\bar{E} (Y|Z, C),$ $\bar{p}(Z|A, C),$ and $\bar{p}(A|C),$ 
    $\hat{E}(Y|Z, C) = o_p \big( \bar{E} (Y|Z, C) \big),$ 
    $\hat{p} (Z|A, C) = o_p \big( \bar{p} (Z|A, C)\big),$ 
    $\hat{p} (A|C) = o_p \big( \bar{p} (A|C)\big).$
    Under \Cref{ass:regularity_consistency}, a solution $\hat{\theta}_{bd, td}$ of the equation 
    $\frac{1}{n}\sum_{i=1}^n \varphi_{bd, td, i}(X_i, \hat{\eta}, \hat{\theta}_{bd, td}) = 0$ is consistent  for $\theta = EY(a^*)-EY(a)$ if at least one of the following conditions holds:
    \begin{enumerate}[label=(\roman*)]
        \item $\bar{E} (Y|Z, C) = E(Y|Z, C)$ and $\bar{p}(A|C) = p(A|C),$ 
        \item $\bar{E} (Y|Z, C) = E(Y|Z, C)$ and $\bar{p}(Z|A, C) = p(Z|A,C),$
        \item $\bar{p}(Z|A, C) = p(Z|A,C)$ and $\bar{p}(A|C) = p(A|C).$
    \end{enumerate} 

Under \Cref{ass:regularity_asympt_normality} and \Cref{ass:an_bd_td}, given in the appendix, $\hat{\theta}_{bd, td}$ is a regular asymptotically linear estimator of $\theta $ with influence function $\varphi_{bd, td}$ and reaches the efficiency bound when additionally the following conditions hold
    \begin{enumerate}[label=(\roman*)]
    \setcounter{enumi}{3}
         \item $ || \hat{E}(Y|Z,C) - E(Y|Z,C) || \cdot
            || \hat{p}(A| C) - p(A| C)
            || = o_p(n^{-1/2}), 
            $ 
        \item $ || \hat{E}(Y|Z,C) - E(Y|Z,C) || \cdot
            || \hat{p}(Z|A, C) - p(Z|A, C)
            || = o_p(n^{-1/2}), 
            $
        \item $ || \hat{p}(Z|\breve{a}, C) - p(Z|\breve{a}, C) 
        || \cdot ||    \hat{p}(\breve{a}|C) - p(\breve{a}|C)
            || = o_p(n^{-1/2}), \quad \breve{a} \in \{ a^*, a\}.
        $
    \end{enumerate}
\end{theorem}

See Appendix \ref{sec:appendix_proof_lemma_consistency_efficiency_of_bd_td_est} for the proof.    Conditions for the consistency and the efficiency of $\hat{\theta}_{fd, td}$ are provided in \autoref{th:consistency_efficiency_of_fd_td_est} below.

\begin{theorem} 
\label{th:consistency_efficiency_of_fd_td_est}
Suppose that Assumptions \hyperref[ass:fd]{\textbf{FD}} and \hyperref[ass:td]{\textbf{TD}} are fulfilled, and $Z(a) \indep C|A$ for all $a \in \mathcal{A}$.
    Suppose that for some $\bar{E} (Y|A, Z, C),$ $\bar{p}(Z|A),$ $\bar{p}(A|C),$ and $\bar{p}(C), $ \linebreak
    $\hat{E}(Y|A, Z, C) = o_p \big( \bar{E} (Y|A, Z, C) \big),$ 
    $\hat{p} (Z|A) = o_p \big( \bar{p} (Z|A)\big),$ 
    $\hat{p} (A|C) = o_p \big( \bar{p} (A|C)\big),$
    $\hat{p} (C) = o_p \big( \bar{p} (C)\big).$
    Under \Cref{ass:regularity_consistency}, a  solution $\hat{\theta}_{fd, td}$ of the equation \linebreak 
    $\frac{1}{n}\sum_{i=1}^n \varphi_{fd, td,i}(X_i, \hat{\eta}, \hat{\theta}_{fd, td}) = 0$ is consistent  for $\theta = EY(a^*)-EY(a)$ if  at least one of the following conditions holds:
    \begin{enumerate}[label=(\roman*)]
        \item $\bar{E}(Y|A, Z, C) = E (Y|A, Z, C),$  $\bar{p}(A|C) = p(A|C),$ and $\bar{p}(C) = p(C),$
        \item $\bar{p}(Z|A) = p(Z|A).$
    \end{enumerate}
    Under \Cref{ass:regularity_asympt_normality} and \Cref{ass:an_fd_td}, given in the appendix, $\hat{\theta}_{fd, td}$ is a regular asymptotically linear estimator of $\theta $ with influence function $\varphi_{fd, td}$ and reaches the efficiency bound when additionally the following conditions hold
    \begin{enumerate}[label=(\roman*)]
    \setcounter{enumi}{2}
    \item $|| \hat{E}(Y|A,Z,C) - E(Y|A,Z,C) || \cdot
            || \hat{p}(Z| A) - p(Z|A) ||  = o_p(n^{-1/2}),$  
    \item $ ||  \hat{p}(Z|\breve{a}) - p(Z|\breve{a}) || \cdot
            || \hat{p}(\breve{a}) - p(\breve{a})||  = o_p(n^{-1/2}), \quad \breve{a} \in \{ a^*, a\}, $
    \item $||\hat{p}(Z|\breve{a}) - p(Z|\breve{a}) || \cdot
         || \hat{p}(C) - p(C)|| = o_p(n^{-1/2}), \quad \breve{a} \in \{ a^*, a\},$
    \item $ || \hat{p}(Z|\breve{a}) - p(Z|\breve{a}) || \cdot 
            || \hat{p}(A|C) - p(A|C)|| = o_p(n^{-1/2}), \quad \breve{a} \in \{ a^*, a\},$
    \item $ || \hat{E}(Y|A,Z,C) - E(Y|A,Z,C) || \cdot
            ||  \hat{p}(Z|\breve{a} ) - p(Z|\breve{a} )|| = o_p(n^{-1/2}), \quad \breve{a} \in \{ a^*, a\}.$ 
\end{enumerate}
\end{theorem}
The proof of Theorem \ref{th:consistency_efficiency_of_fd_td_est} is similar to the proof of \autoref{th:consistency_efficiency_of_bd_td_est}  and can be found in \ref{sec:appendix_proof_lemma_consistency_of_fd_td_est}. Finally, the following  \autoref{th:consistency_efficiency_of_bd_fd_td_est} provides conditions for the consistency and the efficiency of $\hat{\theta}_{bd, fd, td}.$
\begin{theorem}\label{th:consistency_efficiency_of_bd_fd_td_est}
    Suppose that Assumptions  \ref{ass:bd}, \hyperref[ass:fd]{\textbf{FD}}  are fulfilled and  $Y (a; z)  \indep A | C,$ \\${Z(a)\indep C|A}$ for any $a \in \mathcal{A}, z \in \mathcal{Z}.$ Suppose that for some $\bar{E} (Y| Z, C),$ $\bar{p}(Z|A),$ $\bar{p}(A|C),$ and $\bar{p}(C), $ 
    $\hat{E}(Y|Z, C) = o_p \big( \bar{E} (Y| Z, C) \big),$ 
    $\hat{p} (Z|A) = o_p \big( \bar{p} (Z|A)\big),$ 
    $\hat{p} (A|C) = o_p \big( \bar{p} (A|C)\big),$
    $\hat{p} (C) = o_p \big( \bar{p} (C)\big).$
    Under \Cref{ass:regularity_consistency},  a solution $\hat{\theta}_{bd, fd, td}$ of the equation \\   $\frac{1}{n}\sum_{i=1}^n \varphi_{bd, fd, td,i}(X_i, \hat{\eta}, \hat{\theta}_{bd, fd, td}) = 0$ is consistent  for $\theta = EY(a^*)-EY(a)$ if at least one of the following conditions holds:
    \begin{enumerate}[label=(\roman*)]
        \item $\bar{E}(Y| Z, C) = E (Y| Z, C),$  $\bar{p}(A|C) = p(A|C),$ and $\bar{p}(C) = p(C),$
        \item $\bar{E}(Y|Z,  C)= E(Y|Z,  C),$ and $\bar{p}(Z|A) = p(Z|A),$ 
        \item $\bar{p}(Z|A) = p(Z|A)$ and $\bar{p}(A|C) = p(A|C).$ 
    \end{enumerate} 
    Under  \Cref{ass:regularity_asympt_normality} and \Cref{ass:an_bd_fd_td}, given in the appendix, $\hat{\theta}_{bd, fd, td}$ is a regular asymptotically linear estimator of $\theta$ with influence function $ \varphi_{bd, fd, td}$ and reaches the efficiency bound when additionally the following conditions hold
    \begin{enumerate}[label=(\roman*)]
    \setcounter{enumi}{3}
    \item $|| \hat{E}(Y|Z,C) - E(Y|Z,C) || \cdot
           || \hat{p}(A|C) - p (A|C) || = o_p(n^{-1/2}),$
    \item $|| \hat{E}(Y|Z,C) - E(Y|Z,C) || \cdot
           || \hat{p}(Z|A) - p (Z|A) || = o_p(n^{-1/2}),$
    \item $|| \hat{p}(Z|\breve{a}) - p (Z|\breve{a}) || \cdot
           || \hat{p}(\breve{a}) - p(\breve{a})||  = o_p(n^{-1/2}),$ $\breve{a} \in \{a^*, a\},$
     \item $|| \hat{p}(Z|\breve{a}) - p (Z|\breve{a}) || \cdot
           || \hat{p}(C) - p(C)||  = o_p(n^{-1/2})
           $ $\forall \breve{a} \in \{a^*, a\}.$
\end{enumerate}
\end{theorem}
The proof of Theorem \ref{th:consistency_efficiency_of_bd_fd_td_est} is provided in \ref{sec:appendix_proof_lemma_consistency_efficiency_of_bd_fd_td_est}.

The efficiency assumptions of Theorems \ref{th:consistency_efficiency_of_bd_td_est} - \ref{th:consistency_efficiency_of_bd_fd_td_est} (assumptions \textit{(iv) - (vi)} in \autoref{th:consistency_efficiency_of_bd_td_est}, \textit{(iii) - (vii)} in \autoref{th:consistency_efficiency_of_fd_td_est}, and \textit{(iv) - (vii)} in \autoref{th:consistency_efficiency_of_bd_fd_td_est}) are \mbox{$o_p(n^{-1/2})$-rate} of convergence for a product of estimation errors of two nuisance models; so called product rate conditions; see, e.g., \cite{chernozhukov2017double, farrell2015robust,moosavi2021costs}. This allows flexibility in the estimation. For example, if one of the models in the product is estimated using the correct parametric model, the other estimator needs only be consistent. The required convergence rates can also be achieved when both errors in the products are $o_p(n^{-1/4})$. 

\section{Simulation Studies} \label{sec:simulation studies}
This section presents two simulation studies. The first simulation study compares the asymptotic behaviors of semiparametric plug-in estimators introduced in Section~\ref{sec:semipar_est}.  The considered ACE estimators are consistent and reach the respective efficiency bounds because they are  based on $\sqrt{n}$-consistent  estimators of the nuisance parameters. Therefore, we draw attention to the comparison of estimators' empirical variances.  In the second simulation study, we investigate the robustness of the estimators under model misspecification. In both studies, we consider  binary pre-treatment covariate $C$ and treatment $A$.  According to assumptions \hyperref[ass:fd]{\textbf{FD2}} and \hyperref[ass:td]{\textbf{TD2}}, the potential mediator is assumed to be independent of $A$ or $C,$ that is, $Z(a) | A, C   \sim N(\beta a,1), \ a \in \mathcal{A}.$ To ensure that the remaining of \hyperref[ass:fd]{\textbf{FD}} and \hyperref[ass:td]{\textbf{TD}} assumptions hold,  we consider $Y(a,z)$ equal to $Y(z)$ for any $a \in \mathcal{A}$ and independent of $A$ or $Z(a),$ $Y(a^*,z) | Z(a), A, C  \sim N(\gamma_1z+\gamma_2C,1)$ for any $a, a^* \in \mathcal{A}.$  
We also consider $Y(a)|Z(a), A, C \sim N(\gamma_1 Z(a) + \gamma_2 C, 1)$ in order for Assumption \ref{ass:bd} to hold. 

The observed data is generated according to consistency assumptions as follows:
\begin{align*}
    C &\sim \text{Bernoulli}(0.5) \\
    A |C & \sim \text{Bernoulli}(\expit (C))\\
    Z | A, C  & \sim N(\beta A,1), \\ 
     Y| Z, A, C & \sim N(\gamma_1Z+\gamma_2C,1).
\end{align*}

All computations were performed in R \citep{rcite};  the code is available at \url{https://github.com/tetianagorbach/semiparametric_inference_ACE_BD_FD_TD_efficiency_robustness}.

\subsection{Simulation Study 1}\label{sec:sim_study1}
Here, we vary the effect of the mediator $Z$ and covariate $C$ on the outcome by considering eight data generating mechanisms corresponding to all combinations of $\beta, \gamma_1, \gamma_2 \in \{0.5, 1.5\}$. 

To show that the (scaled) empirical variance of the considered ACE estimators tends to the respective bounds with increasing sample size, we calculate the bounds  using the true conditional distributions. The variance bounds under the \ref{ass:bd} and the \hyperref[ass:fd]{\textbf{FD}} assumptions were calculated using Equations \eqref{eq:var_eif_bd} and \eqref{eq:var_eif_fd}, respectively.  To calculate the variance bound under the \hyperref[ass:td]{\textbf{TD}} assumption, we used the simplified expression from Equation \eqref{eq:var_td_minus_var_bd}.  

For the data generating mechanism considered, $E(Y|A, Z, C) = E(Y|Z,  C)$ are linear functions of the variables in the corresponding conditioning set.  \ref{sec:appendix_var_phi_fd_in_the_simulation_study} shows that $E(Y|A, Z) $  can also be represented as a linear function of $A$ and $Z$ for the considered distribution. Therefore,  the parameters of all outcome models, $E(Y|A, Z, C), E(Y|Z,  C),$ and $E(Y|A, Z),$ were estimated using ordinary least squares. Parameter $\beta$ in the mediator model was also estimated using ordinary least squares. The density of a normal distribution was used in the estimation of $p(Z|A).$ Further, $p(A|C)$ was estimated using iteratively reweighted least squares method in  the corresponding logistic regression, and $p(A=1)$ and $p(C=1)$ were consistently estimated using the proportion of  $A=1$ and $C=1,$ respectively. For the considered distribution, $var(Y | a, z, c) = var(Y |  z, c) = var(Z|a) = 1, $ see also \autoref{sec:appendix_expressions_simulation_study} for the expression for $var(Y|a,z), var(C|a).$  
\begin{center}
    \begin{figure}[t]
        \centering
        \includegraphics{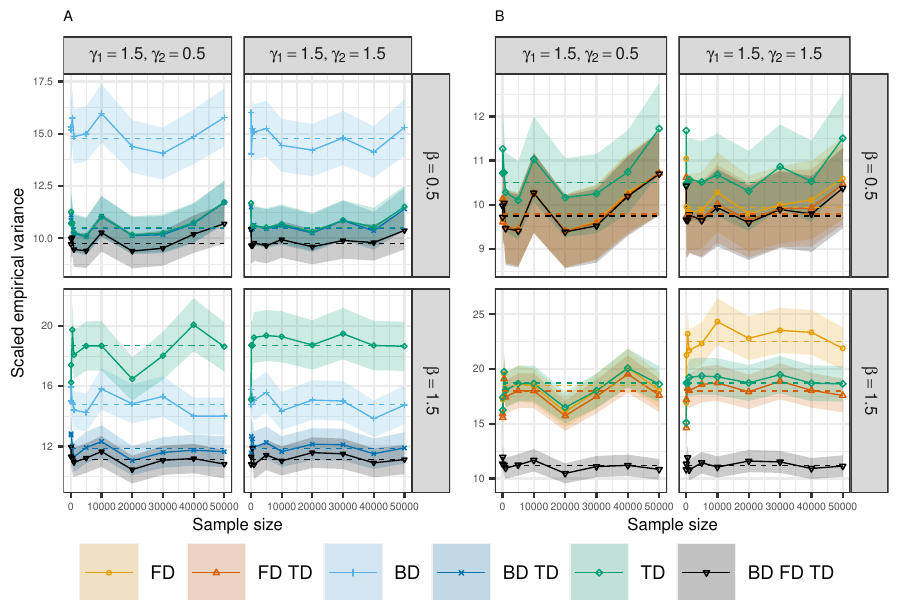}
        \caption{Scaled empirical variance of semiparametric estimators based on influence functions. Horizontal dashed lines represent the  lowest variance of regular and asymptotically linear estimators. Panel A represents estimators under Assumption \ref{ass:bd};  \hyperref[ass:td]{\textbf{TD}}; \ref{ass:bd} and \hyperref[ass:td]{\textbf{TD}}; and  \ref{ass:bd},  \hyperref[ass:fd]{\textbf{FD}} and  \hyperref[ass:td]{\textbf{TD}}. Panel B represents estimators based on Assumption \hyperref[ass:fd]{\textbf{FD}}; \hyperref[ass:td]{\textbf{TD}}; \hyperref[ass:fd]{\textbf{FD}} and \hyperref[ass:td]{\textbf{TD}};  and  \ref{ass:bd},  \hyperref[ass:fd]{\textbf{FD}}, \hyperref[ass:td]{\textbf{TD}}.  Shaded areas represent 95\% confidence intervals for the estimated scaled variance, based on the normal approximation.  Note the difference in scales between the panels, and that estimators based on Assumption \hyperref[ass:td]{\textbf{TD}} and Assumptions \ref{ass:bd},  \hyperref[ass:fd]{\textbf{FD}}, \hyperref[ass:td]{\textbf{TD}} are presented in both panels A and B.} \label{fig:emp_var_diff_criteria}
    \end{figure}
\end{center}

Since all estimators of nuisance parameters are $\sqrt{n}$-consistent, asymptotic variances  of the average treatment estimators are equal to the corresponding variance bounds (see Section~\ref{sec:semipar_est}). 
Furthermore, $\varphi_{td}$  from Equation \eqref{eq:eif_td} can be simplified because $Y \indep A|Z, C$ and $Z\indep C|A$ in the considered distribution.  The following expression was used to construct plug-in estimators of the ACE under Assumption \hyperref[ass:td]{\textbf{TD}}.
\begin{align*}
    &\varphi_{td}  =
    \left(Y-E(Y|Z, C)\right) \frac{p(Z|a^*)-p(Z|a)}{p(Z|A)} 
      +\left(E(Y|Z, C)-\sum_{z}E(Y|z,C)p(z|A)\right) \times \\
      & \times \frac{I(A=a^*) - I(A=a)}{p(A|C)} + \sum\limits_{z} E(Y|z,C)\big(p(z|a^*)- p(z|a)\big)-\theta. 
\end{align*}

Even though using additional restrictions might improve the estimation of nuisance parameters, an estimator based on this simplified influence function has the same influence function as the estimator based on $\varphi_{td}$ without the simplifications.

We consider samples of sizes $n=50, 100, 500, 1000, 5000, 10000, 20000, 30000,40000,$  $50000.$ For each sample size, $K=1000$ replicates were simulated. \autoref{fig:emp_var_diff_criteria} provides the lowest asymptotic variances for each set of assumptions considered, and empirical variances scaled by $n${:} $ns_{K-1}^2 = n/(K-1) \sum_{k=1}^K \big( \hat{\theta}_k - \bar{\hat{\theta}}\big)^2, $
where $\hat{\theta}_k$ is an estimate from replication $k$, and $\bar{\hat{\theta}} = \sum_{k=1}^K \hat{\theta}_k /K.$ Tables \ref{table: bias_sim_study_1} and \ref{table: scaled_empirical_variance_sim_study1} provide a summary of performance  of the semiparametric estimators considered together with a naive estimator,  $\hat{\theta}_{\text{naive}} = \frac{1}{\sum_{i=1}^n{A_i}} \sum_{i=1}^n A_i Y_i - \frac{1}{n-\sum_{i=1}^n{A_i}}\sum_{i=1}^n (1-A_i) Y_i.$

As expected theoretically,  the estimated bias of the naive estimator is much bigger than the bias of any of the semiparametric estimators considered (see \autoref{table: bias_sim_study_1}).
\autoref{fig:emp_var_diff_criteria} and \autoref{table: scaled_empirical_variance_sim_study1}  illustrate that the scaled empirical variances are asymptotically close to the respective bounds (the bounds are within 95\% confidence intervals for the variance of the estimators).
For the distribution in the simulation study, $var \varphi_{bd}$ and $var \varphi_{td}$ do not depend on the strength of the \mbox{$C$--$Y$} relationship (\ref{sec:appendix_expressions_simulation_study}).  Correspondingly, \autoref{fig:emp_var_diff_criteria} shows that the scaled empirical variances of $\hat{\theta}_{bd}$ and $\hat{\theta}_{td}$ vary around the same bound irrespective of $\gamma_2$ (panel A). On the other hand, $\varphi_{fd}$  depends on all nuisance parameters, including $\gamma_2. $
Note that the empirical variances of all estimators increase with a stronger \mbox{$Z$--$Y$} relationship (increasing $\gamma_1$, see  \autoref{table: scaled_empirical_variance_sim_study1}).

Panel A in \autoref{fig:emp_var_diff_criteria}  shows that, for a weaker \mbox{$A$--$Z$}  relationship ($\beta = 0.5$), the empirical variance of $\hat{\theta}_{td}$ is lower than the empirical variance of $\hat{\theta}_{bd}$. This is expected from the theoretical comparison of $var \varphi_{td}$ and $var \varphi_{bd}$  for the data generating mechanism in this simulation study (see \ref{sec:appendix_var_phi_td_sim_study}) and from the example in Appendix \ref{sec:example}.  The  estimator based on the \ref{ass:bd} assumption is more efficient than the two-door estimator for $\beta  = 1.5. $ Consistent with the findings in Sections \ref{sec:front_door_vs_two_door} and \ref{sec:front_door_vs_back_door}, the variance of the front-door estimators exceeds the variance of the two-door and the back-door estimators for bigger values of $\beta$ and $\gamma_2.$

\autoref{fig:emp_var_diff_criteria} and  \autoref{table: scaled_empirical_variance_sim_study1} confirm that the variance of the estimators constructed under a pair of assumptions (for example, the front-door and the two-door) tends to be lower than the variance of estimators constructed under only one set of assumptions of the pair (the front-door only or the two-door only). The efficient influence function $\varphi_{bd, fd, td}$ provides the most efficient estimators. 

In \ref{sec:appendix_sim_study_mv_cov}, we provide the comparison between the estimators' scaled empirical variances for a distribution with a more complicated covariate structure.

\subsection{Simulation Study 2}\label{sec: sim_study2}
These experiments aim to study the robustness of semiparametric estimators when some models are misspecified. According to Section~\ref{sec:semipar_est}, the estimators are still consistent even if some models are misspecified. Their asymptotic variances, however, might differ from the respective bounds. 

We consider the data generating mechanism from simulation study 1 with parameters $\beta = \gamma_1 = \gamma_2 = 1.5.$ Parameter values were selected so that the variance of $\hat{\theta}_{bd}$ was similar to the variance of other estimators. This is in contrast, for example, with the case where $\beta = \gamma_1 = 0.5$ and $var \varphi_{bd}$ is approximately four times larger than the other variances; see \autoref{table: scaled_empirical_variance_sim_study1}.  We chose $\beta = \gamma_1 = \gamma_2 = 1.5$ for symmetry in the coefficients.

We ran Monte Carlo simulations with $K=1000$ iterations and sample size $n=50000$ for the following four settings:
\begin{center}
    \begin{figure}[t]
        \centering
        \includegraphics{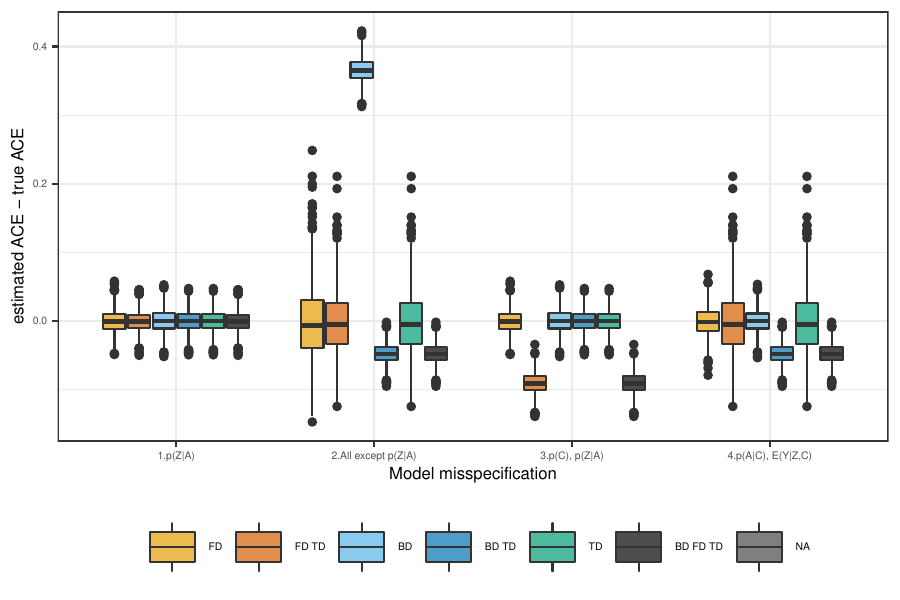}
        \caption{Boxplots of differences between estimates and the average causal effect over Monte Carlo iterations, by model misspecification setting and by estimation method.} \label{fig:est_misspecification}
    \end{figure}
\end{center}

\begin{enumerate}
    \item Only $p(Z|A)$ is misspecified ($A$ is omitted from the regression). In this setting all estimators are expected to be consistent according to Section~\ref{sec:semipar_est}. 
    \item All models are misspecified, except $p(Z|A).$ We use true $p(A=1) \approx 0.62$ but  $\hat{p}(A=1) = \frac{1}{4};$  $p(C=1) = \frac{1}{2},$ while $\hat{p}(C=1) = \frac{1}{4}; $   $C$  is omitted  from the linear regression of $Y$ on $A, C$ and from the logistic regression of $A$ on $C$;  $Z$ is omitted  from the linear regressions of $Y$ on $A, Z$ and $Y$ on $Z, C.$ 
    According to Section~\ref{sec:semipar_est},  $\hat{\theta}_{fd},$ $\hat{\theta}_{fd, td}, $ and $\hat{\theta}_{td}$ are  consistent in this setting, while other estimators are inconsistent. 
    \item $p(C)$ and $p(Z|A)$ are misspecified. We use biased $\hat{p}(C=1) = \frac{1}{4}.$ To misspecify $p(Z|A)$, $A$ is omitted  from the regression. This situation illustrates the case where both $\hat{\theta}_{fd}$ and $\hat{\theta}_{td}$ are consistent, but $\hat{\theta}_{fd,td}$ is not.
    \item $p(A|C)$  and  $E(Y|Z, C)$ are misspecified. $C$  is omitted  from the logistic regression of $A$ on $C$ and $Z$ is omitted  from the linear regressions of $Y$ on  $Z, C$.   Here, both $\hat{\theta}_{bd}$ and $\hat{\theta}_{td}$ are consistent, but $\hat{\theta}_{bd,td}$ is not. 
\end{enumerate}

\autoref{fig:est_misspecification} provides boxplots of $\hat{\theta}-\theta$ to illustrate the robustness properties outlined in Section~\ref{sec:semipar_est}.
Estimated biases, empirical standard errors, variances, and mean squared errors are provided in  \autoref{table: results_sim_study2}. \autoref{fig:est_misspecification} confirms the theoretical results from Section~\ref{sec:semipar_est}. 
When only $p(Z|A)$ is misspecified, all estimators have low bias. 
When only $p(Z|A)$ is correctly specified, $\hat{\theta}_{bd}, \hat{\theta}_{bd, td}, \hat{\theta}_{bd, fd, td}$ are clearly biased. The bias of $\hat{\theta}_{bd}$ is much bigger than the bias of the other estimators. 
When $p(C)$ and $p(Z|A)$ are misspecified, $\hat{\theta}_{fd, td}, \hat{\theta}_{bd, fd, td}$ are biased, while $\hat{\theta}_{fd}$ and $\hat{\theta}_{td}$ have low bias. 
As expected, when $p(A|C)$  and  $E(Y|Z, C) $ are incorrectly specified, $\hat{\theta}_{bd}$ and $\hat{\theta}_{td}$ are unbiased, but $\hat{\theta}_{bd, td}$ and $\hat{\theta}_{bd, fd, td}$ are biased. The asymptotic variance of $\hat{\theta}_{bd}$ is much lower than the asymptotic variance of $\hat{\theta}_{td}.$
\autoref{table: results_sim_study2} also shows that the most efficient estimators, for example, $\hat{\theta}_{bd, fd, td}$ under misspecification 4, do not always have the lowest mean squared error due to their high bias compared to the bias of other estimators.

\section{Discussion}\label{sec:discussion}

We have studied semiparametric inference about the average causal effect from observational data using identification models based on the back-, front- and two-door assumptions. In practice, several identification models may be plausible simultaneously. We have, therefore, derived the semiparametric efficiency bounds that hold under each of the different models separately, but also when several or all the models hold simultaneously (\autoref{sec:appendix_var_eif}). We have, for example, shown that none of the three identifying assumptions above, when considered separately, yields the lowest asymptotic efficiency bound, regardless of the observed data distribution (Propositions \ref{th:sufficient_conditions_td_vs_bd}, \ref{th:fd_vs_td}, \ref{th:fd_vs_bd}).  Simulation results in Section~\ref{sec:sim_study1} confirmed these theoretical results and demonstrated that the strength of the dependences between the observed variables determines which of the three identifying assumptions considered corresponds to the estimator with the lowest variance.  \citet{kuroki2004selection} came to similar conclusions in their comparison of the back- and front-door criteria in a parametric Gaussian causal model.  In contrast to  \citet{kuroki2004selection}  and our  semiparametric results, \citet{gupta2020estimating} and \citet{hayashi2014estimating} reported that using mediators and pre-treatment confounders (rather than estimators based on confounders, or mediators alone) improved the estimation accuracy in some specific parametric models. 
\begin{table}[h]
    \begin{center}
    \def\arraystretch{0.60}
        \begin{threeparttable}
        \footnotesize
        \setlength{\tabcolsep}{2pt} 
            \begin{tabular}{|c|c|cc|cc|cccc|}
                \hline
                Assumption/    & \multirow{2}{*}{$p(C)$} & \multirow{2}{*}{$p(A|C)$} & 
                       \multirow{2}{*}{$p(A)$} & \multirow{2}{*}{$p(Z|A)$} & 
                       \multirow{2}{*}{$p(Z|A,C)$} & \multirow{2}{*}{$E(Y|A, C)$} & 
                       \multirow{2}{*}{$E(Y|A, Z)$} & \multirow{2}{*}{$E(Y|Z, C)$} & \multirow{2}{*}{$E(Y|A, Z, C)$} \\
                Model &&&&&&&&&\\       
                \hline
                \multirow{2}{*}{\textbf{BD}} & & x& & & & & & & \\
                                    & & &  & & & x& & & \\
                \hline
                \multirow{2}{*}{\hyperref[ass:fd]{\textbf{FD}}} & & & & x&  &  &  & &   \\
                                    & &  & x &    &   & & x  &  & \\
                \hline
                \multirow{2}{*}{\hyperref[ass:td]{\textbf{TD}}} & &    &   &    & x    &  &  &  &\\
                                    & & x  &   &    & &  &  &  & x   \\
                \hline
                \multirow{3}{*}{\textbf{BD TD}} & & x  &   &    & &  &  & x&\\
                                       & &    &   &    & x    &  &  & x&\\
                                     & & x  &   &    & x    &  &  &  &\\
                \hline
                \multirow{2}{*}{\textbf{FD TD}}    & x & x  &   &    & &  &  &  & x   \\
                                          & &    &   & x  & &  &  &  &\\
                \hline 
                \multirow{3}{*}{\textbf{BD FD TD}} & x & x  &   &    & &  &  & x&\\
                                          & &    &   & x  & &  &  & x&\\
                                          & & x  &   & x  & &  &  &  &\\
                \hline   
            \end{tabular}
            \begin{tablenotes}
                \small {
                \item Note that $E(Y|Z, C) = E(Y|A, Z, C)$ under \textbf{BD TD} and \textbf{BD FD TD} assumptions  when $Y(a,z) \indep A|C$.} We also assume that $Z(a) \indep C|A$ under \textbf{FD TD} and \textbf{BD FD TD} assumptions. Note also that some conditions in Theorems \ref{th:efficiency_of_fd_and_td_est}--\ref{th:consistency_efficiency_of_bd_fd_td_est} include only $a$ and  $a^*.$ To save space,  $a$ and  $a^*$ are represented by $A$ in Table \ref{table: summary of consistency conditions} and Table \ref{table: summary of efficiency conditions}.
            \end{tablenotes}
            \caption{Summary of the robustness properties from} Section~\ref{sec:semipar_est}. For a specific identifying assumption in the first column, the estimation equation estimator of the ACE constructed using the corresponding efficient influence function   is consistent if the nuisance models marked with ``x'' within at least one row are consistently estimated.
            \label{table: summary of consistency conditions}
        \end{threeparttable}
    \end{center}
\end{table}
\begin{table}[h]
    \begin{center}    
        \def\arraystretch{0.60}
          \begin{threeparttable}
            \footnotesize
            \setlength{\tabcolsep}{2pt} 
                \begin{tabular}{|c|c|cc|cc|cccc|}
                    \hline
                    Assumption/    & \multirow{2}{*}{$p(C)$} & \multirow{2}{*}{$p(A|C)$} & 
                           \multirow{2}{*}{$p(A)$} & \multirow{2}{*}{$p(Z|A)$} & 
                           \multirow{2}{*}{$p(Z|A,C)$} & \multirow{2}{*}{$E(Y|A, C)$} & 
                           \multirow{2}{*}{$E(Y|A, Z)$} & \multirow{2}{*}{$E(Y|Z, C)$} & \multirow{2}{*}{$E(Y|A, Z, C)$} \\
                    Model &&&&&&&&&\\       
                    \hline
                    \textbf{BD} & & x& & & & x & & & \\
                    \hline
                    \multirow{2}{*}{\hyperref[ass:fd]{\textbf{FD}}} & & &x & x&  &  &  & &   \\
                                        & &  &  & x &   & & x  &  & \\
                    \hline
                    \multirow{2}{*}{\hyperref[ass:td]{\textbf{TD}}} & & x   &   &    & x    &  &  &  &\\
                                        & &  &   &    &x &  &  &  & x   \\
                    \hline
                    \multirow{3}{*}{\textbf{BD TD}} & & x  &   &    & &  &  & x&\\
                                           & &    &   &    & x    &  &  & x&\\
                                         & & x  &   &    & x    &  &  &  &\\
                    \hline
                    \multirow{4}{*}{\textbf{FD TD}}    &  &   &   & x   & &  &  &  & x   \\
                                              & &    & x   & x  & &  &  &  &\\
                                              & & x   &   & x  & &  &  &  &\\
                                              &x &    &   & x  & &  &  &  &\\
                    \hline 
                    \multirow{4}{*}{\textbf{BD FD TD}} &  & x  &   &    & &  &  & x&\\
                                              & &    &   & x  & &  &  & x&\\
                                              & & x  &   & x  & &  &  &  &\\
                                              & x &   &   & x  & &  &  &  &\\
                    \hline   
                \end{tabular}
                \caption{Summary of the efficiency properties of Section~\ref{sec:semipar_est}. Each row represents a product rate condition for the models marked with ``x'': the product of estimation errors in the marked nuisance models is $o_p(1/\sqrt{n}).$ For a specific identifying assumption, the corresponding estimator of the ACE among $\hat{\theta}_{bd}$, $\hat{\theta}_{fd},$ $\hat{\theta}_{fd},$ $\hat{\theta}_{bd,td},$ $\hat{\theta}_{fd,td}, $ and $\hat{\theta}_{bd, fd, td}$ is efficient if the product rate conditions in all corresponding rows are fulfilled.}
            \label{table: summary of efficiency conditions}
          \end{threeparttable}
    \end{center}
\end{table}

The choice of an identification model is difficult because it must rely on assumptions that cannot be tested empirically unless further information is available (see, for example, \citealp{deluna2006exogeneity, deluna2014testing}).  Another major challenge is the choice of an estimator given an identification model. Consider the case where we are ready to assume that several, or even all three, identification models studied hold. If all of the nuisance models were known, the ACE estimator based on the corresponding efficient influence function would reach the semiparametric efficiency bound. However, in practice, the nuisance models are unknown and must be fitted. We show how the estimation of nuisance models affects the robustness and efficiency properties of the resulting ACE estimator  (see Theorems \ref{th:efficiency_of_fd_and_td_est}--\ref{th:consistency_efficiency_of_bd_fd_td_est}, Tables \ref{table: summary of consistency conditions}--\ref{table: summary of efficiency conditions}, and simulations in Section~\ref{sec: sim_study2}). Even if some nuisance models are not estimated consistently, the ACE estimator may yet be consistent  (see a summary of the conditions in Table \ref{table: summary of consistency conditions}), but there is no efficiency guarantee. Conditions that do yield an efficient  estimation of the ACE are the $\sqrt{n}$-convergence rate of the product of the estimating errors of the nuisance models (Table \ref{table: summary of efficiency conditions}). This justifies the use of flexible estimation of nuisance models, by machine learning algorithms; see \citet{moosavi2021costs} for a review of post-machine learning valid causal inference.  

The results of this paper may assist in the choice of an estimation strategy among the estimators using pre-treatment confounders, mediators, and both mediators and pre-treatment confounders simultaneously.  The choice should recognize a bias-variance trade-off that depends on which nuisance models can be estimated consistently and at a high enough convergence rate.

\acks{This work was supported by the Marianne and Marcus Wallenberg Foundation (grant 2015.0060), FORTE (grant 2018-00852), the Swedish Research Council (grants 2018-02670 and 2016-00703) and Academy of Finland (grant number 311877).}

\appendix
\setcounter{table}{0}
\renewcommand{\thetable}{A\arabic{table}}
\renewcommand{\theassumption}{A\arabic{assumption}}

\section{Semiparametric Efficiency Bounds} \label{sec:appendix_semiparametric_efficiency}
In this section, we briefly outline the theory that leads to the semiparametric efficiency bounds; see \citet{newey1990semiparametric}, \citet{tsiatis2007semiparametric} for more rigorous descriptions.

The asymptotic variance of any regular semiparametric estimator of the parameter $\theta$ under semiparametric model $\mathcal{M}$ is no smaller than the supremum of the Cramer-Rao bounds for all regular parametric submodels of $\mathcal{M}$ \citep{newey1990semiparametric}. This supremum is called the semiparametric efficiency bound.  

To find the bound, the notion of pathwise differentiability of $\theta$ is important.  
The parameter $\theta(\alpha),$ where $\alpha$ parameterizes parametric submodels, is called pathwise differentiable if there exists a function $\varphi$ such that $E\varphi^2 < \infty$ and for all regular parametric submodels
\begin{equation}
    \frac{\partial \theta}{\partial \alpha}\bigg|_{\alpha=\alpha_0} = E(\varphi S_{\alpha}), \label{eq:pathwise_differentiability}
\end{equation}
where $\alpha_0$ corresponds to the true distribution and $S_{\alpha}$ denotes the score for a parametric submodel, evaluated at the truth. 
Let $\Lambda$ denote the tangent space under model $\mathcal{M},$ the mean square closure of all linear combinations of scores $S_{\alpha}$ for smooth parametric submodels of $\mathcal{M}$. 

For a pathwise differentiable parameter $\theta,$ the semiparametric efficiency bound is equal to the variance of the projection of $\varphi$ on the tangent space $\Lambda$ \citep{newey1990semiparametric}. The projection, $\varphi_{eff},$  is called the efficient influence function for $\theta.$ The asymptotic variance of any regular semiparametric estimator of $\theta$ under considered semiparametric model $\mathcal{M}$  is no smaller than the variance of the efficient influence function,  $var(\varphi_{eff}).$

Influence functions for a parameter may be used to construct asymptotically linear estimator as a solution $\hat{\theta}$ of the estimating equation $\sum_{i=1}^n \varphi_i(C_i, A_i, Z_i, Y_i, \eta, \theta) = 0,$ where $\eta$ represent nuisance parameters. If Equation \eqref{eq:pathwise_differentiability} holds, $\hat{\theta}$ has the influence function $\varphi,$ the asymptotic variance $var(\varphi)$ and is regular (see Theorem 2.2 in \citealp{newey1990semiparametric}). Correspondingly, the asymptotically linear estimators with influence function $\varphi_{eff}$ have the asymptotic variance $var(\varphi_{eff})$ and achieve the semiparametric efficiency bound under the model $\mathcal{M}.$

\section{ Expressions for the Semiparametric Efficiency Bounds}\label{sec:appendix_var_eif}
\ref{sec:appendix_proof_lemma_FD} shows that
    \begin{flalign}
          var\varphi_{fd} &=\sum_{z} (p(z|a^*)-p(z|a))^2  \sum\limits_{\bar{a}}\frac{p(\bar{a})}{p(z|\bar{a})}var(Y|\bar{a},z) \nonumber \\
         & + \sum\limits_{z}\left(\sum\limits_{\bar{a}} E(Y|\bar{a}, z) p(\bar{a})\right)^2 \left(\frac{p(z|a^*)}{p(a^*)} + \frac{p(z|a)}{p(a)}\right)\nonumber\\
         & -\frac{\big(\sum\limits_{\bar{a}, z} p(z|a^*) E(Y|\bar{a}, z) p(\bar{a})\big)^2}{p(a^*)}
         -\frac{\big(\sum\limits_{\bar{a}, z} p(z|a) E(Y|\bar{a},z) p(\bar{a})\big)^2}{p(a)}\nonumber\\
        &+\sum\limits_{\bar{a}}p(\bar{a})\left(\sum_{z}E(Y|\bar{a}, z) (p(z|a^*)-p(z|a))\right)^2-\theta^2. \label{eq:var_eif_fd} &
    \end{flalign} 

\ref{sec:appendix_proof_var_eif_td} derives
\begin{align} 
        & var \varphi_{td}=\sum_{z,c} (p(z|a^*,c)-p(z|a,c))^2  \sum\limits_{\bar{a}}\frac{p(\bar{a},c)}{p(z|\bar{a},c)}var(Y|\bar{a},z,c)\nonumber \\
        &+\sum\limits_{z,c}\frac{p(z|a^*,c)\left(\sum_{\bar{a}}E(Y|\bar{a}, z, c)p(\bar{a}|c)\right)^2p(c)}{p(a^*|c)}
        +\sum\limits_{z,c}\frac{p(z|a,c)\left(\sum_{\bar{a}}E(Y|\bar{a}, z, c)p(\bar{a}|c)\right)^2p(c)}{p(a|c)} \nonumber\\
        & -\sum\limits_{c}\left( \frac{\left(\sum_{\bar{a},z}E(Y|\bar{a}, z, c)p(z|a^*,c)p(\bar{a}|c)\right)^2}{p(a^*|c)}
        + \frac{\left(\sum_{\bar{a},z}E(Y|\bar{a}, z, c)p(z|a,c)p(\bar{a}|c)\right)^2}{p(a|c)}\right)p(c) \nonumber \\
        &+\sum\limits_{\bar{a},c}\left(\sum\limits_{z} E(Y|\bar{a}, z, c)\big(p(z|a^*, c)- p(z|a, c)\big)\right)^2p(\bar{a},c)-\theta^2.  \label{eq:var_eif_td}
    \end{align} 
Similarly to \ref{sec:appendix_proof_var_eif_td}, one can show that
\begin{flalign*}
        & var  \varphi_{bd, td}   = var \varphi_{td} \\
        & + \sum_{z,c} (p(z|a^*,c)-p(z|a,c))^2 p(c)var(Y|z,c) \bigg(\frac{1}{\sum_{\bar{a}} p(z|\bar{a},c)p(\bar{a}|c)} - \sum\limits_{\bar{a}}\frac{p(\bar{a}|c)}{p(z|\bar{a},c)} \bigg).&
    \end{flalign*} 
 \ref{sec:appendix_proof_th_fd_td_eif} derives    
\begin{flalign*}
       & var\varphi_{fd, td} = \sum_{z,c} (p(z|a^*)-p(z|a))^2  \sum\limits_{\bar{a},c}\frac{p(\bar{a},c)}{p(z|\bar{a})}var(Y|\bar{a},z,c)\\
        &+\sum\limits_{z} \frac{p(z|a^*)}{p(a^*)}\left(\sum\limits_{\bar{a},c}E(Y|\bar{a}, z, c)p(\bar{a}|c) p(c)\right)^2  - \frac{\left(\sum_{\bar{a},z,c}E(Y|\bar{a}, z, c)p(z|a^*)p(\bar{a}|c) p(c)\right)^2}{p(a^*)}\\
        &+\sum\limits_{z} \frac{p(z|a)}{p(a)}\left(\sum\limits_{\bar{a},c}E(Y|\bar{a}, z, c)p(\bar{a}|c) p(c)\right)^2  - \frac{\left(\sum_{\bar{a},z,c}E(Y|\bar{a}, z, c)p(z|a)p(\bar{a}|c) p(c)\right)^2}{p(a)} \\
    &+\sum\limits_{\bar{a},c}p(\bar{a},c)\left(\sum\limits_{z} E(Y|\bar{a}, z, c)\big(p(z|a^*)- p(z|a)\big)\right)^2-\theta^2.&
    \end{flalign*} 
     \ref{sec:appendix_proof_lemma_fd_td_bd} shows that
\begin{flalign*}
     & var \varphi_{bd, fd, td} = \sum_{z} (p(z|a^*)-p(z|a))^2 \sum_{c}\frac{p(c)var(Y|z,c)}{\sum_a p(a|c)p(z|a)}\\
            & + \sum_{z} \frac{p(z|a^*)}{p(a^*)} \left(\sum_{c}p(c)E(Y|z,c)\right)^2 - \frac{\left( \sum_{z,c} E(Y|z,c)p(z|a^*)p(c) \right)^2}{p(a^*)}\\
            & + \sum_{z} \frac{p(z|a)}{p(a)} \left(\sum_{c}p(c)E(Y|z,c)\right)^2 - \frac{\left( \sum_{z,c} E(Y|z,c)p(z|a)p(c) \right)^2}{p(a)}\\
            & +\sum_{c} p(c)\left(\sum_{z}E(Y|z,c) (p(z|a^*)- p(z|a))\right)^2  - \theta^2. &
    \end{flalign*}
\section{Proofs}

\subsection{Proof of Proposition \ref{th:sufficient_conditions_td_vs_bd}} \label{sec:appendix_proof_sufficient_conditions_td_less_than_bd}
\begin{proof}
    Under the assumptions of Section~\ref{sec:notation}, Assumptions \ref{ass:bd}, \hyperref[ass:td]{\textbf{TD}},  $Y (a, z)  \indep A | C$

    \begin{equation} 
    Y\indep A|Z,C  \label{eq:s1}
    \end{equation}
    since
    \begin{flalign*}
        p&(Y=y |A=a, Z=z, C=c)\stackrel{\text{consistency}}{=}p(Y(a,z)=y|A=a, Z(a)=z, C=c)\\
        &\stackrel{\ref{ass:td3}}{=}p(Y(a,z)=y| A=a, C=c)\stackrel{Y (a, z)  \indep A | C}{=}p(Y(a,z)=y|  C=c)\\
        &\stackrel{\ref{ass:td1}}{=}p(Y(a^*,z)=y| C=c)\stackrel{Y (a, z)  \indep A | C}{=}p(Y(a^*,z)=y| A=a^*, C=c)\\
        &\stackrel{\ref{ass:td3}}{=}p(Y(a^*,z)=y| A=a^*, Z=z, C=c)\stackrel{\text{consistency}}{=}p(Y=y| A=a^*, Z=z, C=c). &
    \end{flalign*}
   For a model compatible with a SWIG,  Assumption \ref{ass:bd} implies $Y (a, z)  \indep A | C$  as follows. Consider $ a=a'$ in Proposition 4 in \cite{malinsky2019potential} $a'' = \{a,z\}, Z= A$ and $Z(a) = Z(a,z) = A$ following the definition of one-step-ahead potential outcomes in   \cite{malinsky2019potential}). Similarly, let $X=C$ and $X(a) = X(a,z) = C$  since $C$ are pre-treatment and, therefore,  pre-mediator covariates.  Proposition 4 then reads: if 
$$(Y(a), a' \indep A|C)_{\mathcal{G}(a)}$$ then  
$$(Y(a,z), a' \indep A|C)_{\mathcal{G}(a,z)}. $$
 The implication $Y (a) \indep A | C \Rightarrow Y (a; z)  \indep A | C$  follows directly by considering marginal independency.
 
    Similarly to A2.1 in \citet{fulcher2020robust},
    \begin{flalign}
        E&[Y(a^*)|C]=\sum\limits_{z} E(Y(a^*)|Z(a^*)=z,C) p(Z(a^*)=z|C) \nonumber\\
        &\stackrel{\ref{ass:td2}}{=}\sum\limits_{z} E(Y(a^*,z)|Z(a^*)=z,C) p(Z(a^*)=z|A=a^*,C)\nonumber\\
        &\stackrel{\text{consistency}}{=}\sum\limits_{z}  E(Y(a^*,z)|Z(a^*)=z,C) p(Z=z|A=a^*,C)\nonumber\\
        &\stackrel{\ref{ass:td1}}{=}\sum\limits_{z} E(Y(z)|Z(a^*)=z,C) p(Z=z|A=a^*,C)\nonumber\\
        & =\sum\limits_{z} \bigg( \sum\limits_{a}E(Y(z)|A=a,Z(a^*)=z,C) p(a|Z(a^*)=z,C)\bigg)p(Z=z|A=a^*,C)\nonumber\\
        &\stackrel{\ref{ass:td3}, \ref{ass:td2}}{=}\sum\limits_{z} \bigg( \sum\limits_{a}E(Y(z)|A=a, C) p(a|C)\bigg)p(Z=z|A=a^*,C)\nonumber\\
        &=\sum\limits_{z}p(Z=z|A=a^*,C)\bigg( \sum\limits_{a}E(Y(z)|A=a,C) p(a|C)\big)\nonumber\\
        &\stackrel{\ref{ass:td3},\text{ consistency}}{=}\sum\limits_{z}p(Z=z|A=a^*,C)\bigg( \sum\limits_{a}E(Y(z)|A=a,Z=z,C) p(a|C)\bigg)\nonumber\\
        &\stackrel{\ref{ass:td1}}{=}\sum\limits_{z}p(Z=z|A=a^*,C)\bigg( \sum\limits_{a}E(Y|A=a,Z=z,C) p(a|C)\bigg)\nonumber\\
        & =\sum\limits_z p(z|a^*, C)\sum_aE(Y|a, z, C)p(a|C)\stackrel{\eqref{eq:s1}}{=}\sum\limits_z E(Y|z,C)p(z|a^*, C). \label{eq:s2} &
    \end{flalign}
    From  Equations \eqref{eq:eif_td}, \eqref{eq:s1}, \eqref{eq:s2}, , 
    \begin{flalign}
        \varphi&_{td}=\left(Y-E(Y|Z, C)\right) \frac{p(Z|a^*,C)-p(Z|a,C)}{p(Z|A,C)} \label{eq:two-door eif under td bd assumptions}\\
        & +\left(E(Y|Z, C)-\sum_{z}E(Y|z,C)p(z|A,C)\right)\frac{I(A=a^*)-I(A=a)}{p(A|C)}\nonumber\\
        &+ E(Y(a^*)|C)  -E(Y(a)| C)  - \theta  \nonumber &
    \end{flalign}
    and 
    \begin{flalign*}
         v&ar \varphi_{td}=\sum_{z,c} (p(z|a^*,c)-p(z|a,c))^2 p(c)var(Y|z,c) \sum\limits_{\bar{a}}\frac{p(\bar{a}|c)}{p(z|\bar{a},c)}\\
            &+  \sum\limits_{c}\bigg(\sum\limits_{z}E^2[Y|z,c]p(z|a^*,c)-E^2\big[Y(a^*)| c\big]\bigg)\frac{p(c)}{p(a^*| c)}\\
            &+ \sum\limits_{c}\bigg(\sum\limits_{z}E^2[Y|z,c]p(z|a,c)-E^2\big[Y(a)| c\big]\bigg)\frac{p(c)}{p(a| c)}+E\delta^2, &
    \end{flalign*}
    where $\delta = E\big[Y(a^*)|C\big]-E\big[Y(a)|C\big]-\theta.$ The following equalities are obtained by addition and subtraction of some terms and regrouping.
    \begin{flalign*}
    v&ar \varphi_{td}=
            \sum_{z,c} (p(z|a^*,c)-p(z|a,c))^2 p(c)var(Y|z,c) \sum\limits_{\bar{a}}\frac{p(\bar{a}| c)}{p(z|\bar{a},c)}\\
            &+\sum\limits_{z,c} \frac{p(c)}{p(a^*| c)}\bigg(E^2[Y|z,c]p(z|a^*,c)-E(Y^2|z,c) p(z|a^*,c)\bigg) \\
            &+\sum\limits_{z,c}\frac{p(c)}{p(a^*| c)}E(Y^2|z, c)p(z|a^*,c)-E\bigg(\frac{E^2\big[Y(a^*, Z(a^*))|C\big]}{p(a^*|C)}\bigg)\\
            &+ \sum\limits_{z,c} \frac{p(c)}{p(a| c)}\bigg(E^2[Y|z,c]p(z|a,c)-\sum\limits_{z,c}E(Y^2|z, c)p(z|a,c)\bigg)+ \\
            &+\sum\limits_{z,c}\frac{p(c)}{p(a| c)}E(Y^2|z, c)p(z|a,c)-E\bigg(\frac{E^2\big[Y(a, Z(a))|C\big]}{p(a|C)}\bigg)+E\delta^2\\
            &=\sum_{z,c}p(c)var(Y|z,c) \Bigg( (p(z|a^*,c)-p(z|a,c))^2  \sum\limits_{\bar{a}}\frac{p(\bar{a}| c)}{p(z|\bar{a},c)}\\
            &- \frac{p(z|a^*,c)}{p(a^*| c)}  - \frac{p(z|a,c)}{p(a| c)}  \Bigg) +E\bigg(\frac{var\big[Y(a^*)|C\big]}{p(a^*|C)}\bigg)+E\bigg(\frac{var\big[Y(a)|C\big]}{p(a|C)}\bigg)+E\delta^2.&
    \end{flalign*}
    \begin{align*}
    v&ar \varphi_{td}- var \varphi_{bd}\\
    &= \sum_{z,c}p(c)var(Y|z,c) \Bigg( (p(z|a^*,c)-p(z|a,c))^2  \sum\limits_{\bar{a}}\frac{p(\bar{a}| c)}{p(z|\bar{a},c)}- \frac{p(z|a^*,c)}{p(a^*| c)}  - \frac{p(z|a,c)}{p(a| c)}  \Bigg).
    \end{align*}
    
    $var \varphi_{td}- var \varphi_{bd}$  is a summation over all $z \in \mathcal{Z}$ and $c \in \mathcal{C}$. Each term is a product of a positive $p(c)var(Y|z,c)$ and a term of an unknown sign.
    The proof is completed by noting that if the term of an unknown sign is positive for all  $z \in \mathcal{Z}$ and $c \in \mathcal{C},$    $var \varphi_{td} - var \varphi_{bd}$ is also positive. If the term of an unknown sign is negative for all  $z \in \mathcal{Z}$ and $ c \in \mathcal{C},$   $var \varphi_{td} - var \varphi_{bd}$ is also negative. 
\end{proof}

\subsection{Proof of Proposition \ref{th:td_bd_eif} } \label{sec:proof of lemma EIF TD BD}
\begin{proof}
    The efficient influence function is a projection of an arbitrary influence function onto the tangent space for the model (see Equation 3.34 in Theorem 3.5 in \citealt{tsiatis2007semiparametric}). As an influence function, we use $\varphi_{td} $ that under Assumptions \hyperref[ass:td]{\textbf{TD}}  and  \ref{ass:bd}, may be simplified according to Equation \eqref{eq:two-door eif under td bd assumptions}.
    
    To find the projection, we first note that under the Assumptions of the Proposition, $Y \indep A|Z, C$ (see Appendix \ref{sec:appendix_proof_sufficient_conditions_td_less_than_bd}), and the density of the observed data is $p(c,a,z,y) = p(c) p(a| c) p(z|a,c)p(y|z,c).$  From Theorem 4.5 in \citet{tsiatis2007semiparametric}, the projection of any influence function $h$ on the tangent space is $\alpha(C) + \alpha(A, C) + \alpha(Z, A, C) + \alpha(Y, Z, C),$ where
    \begin{flalign*}
        & \alpha(C)  = E(h|C) - Eh\\
        & \alpha(A,C) = E(h|A, C) - E(h|C)\\
        & \alpha(Z, A, C) = E(h|Z, A,C) - E(h|A,C)\\
        & \alpha(Y, Z, C) = E(h|Y,Z,C) - E(h|Z,C).
    \end{flalign*}
    The projection of $\varphi_{td}$ onto the tangent space is then $\varphi_{bd, td} = E(\varphi_{td}|Y,Z,C) -E(\varphi_{td}|Z,C) + E(\varphi_{td}|Z, A,C).$ 
\end{proof}

\subsection{Consistency and Efficiency of the Estimators}
\label{sec:appendix_consistency_efficiency_estimators}
Let $X$ denote the set of relevant observed variables, for example, $X=(C,A,Z,Y)$ or $X=(C,A,Y).$ 
The considered estimators of the ACE take a form 
    $$\hat{\theta} = \frac{1}{n}\sum_{i=1}^n m(X_i, \hat{\eta}), $$
    where $m(X, \eta) = \varphi(X, \eta, \theta )  + \theta.$ Note that functions $\varphi$ (and $m$) vary depending on the model assumptions. For example, $\varphi$ and $m$ represent $\varphi_{fd}$  and $m_{fd}$ under Assumption \hyperref[ass:fd]{\textbf{FD}} and $\varphi_{td}$ and $m_{td}$ under Assumption \hyperref[ass:td]{\textbf{TD}}, etc.
    \subsubsection{Consistency} \label{sec:appendix_consistency_conditions}
    From Theorem 7.3 in \citet{boos2013essential}, $\hat{\theta} \rightarrow_p Em (X_1, \bar{\eta}) $  when  $\hat{\eta} \rightarrow_p \bar{\eta}$ and the following regularity \Cref{ass:regularity_consistency} is fulfilled:
    \begin{assumption}\label{ass:regularity_consistency} 
    \mbox{}
        \begin{itemize}
            \item[(a)] $X_1, \ldots, X_n $  are i.i.d
            \item[(b)]  $m(X, \eta)$ is differentiable with respect to $\eta, E| m{'} (X_1, \bar{\eta})|<\infty$
            \item[(c)]  there exist a function $M(x)$ such that $EM(X_1)<\infty$ and for all $\eta$ in a neighborhood of $\bar{\eta}$ and all components of $\eta$ $\eta_j,$ $$\vline \frac{\partial m(x, \eta)}{\partial \eta_j} \vline \leq M(x).$$
        \end{itemize} 
    \end{assumption}
    In order to prove the consistency of the estimators, we show that  $Em (X_1, \bar{\eta}) = \theta_0,$ that is, $E\varphi(X, \bar{\eta}, \theta_0) = 0,$ where $\theta_0$ is the true value of the ACE.  Then, from Theorem 7.3 in \citet{boos2013essential}, $\hat{\theta} \rightarrow_p \theta_0. $ 
    
\subsubsection{Efficiency} \label{sec:appendix_efficiency_conditions}
    To show the asymptotic normality (and the efficiency) of the estimators, we use the following decomposition:
    \begin{flalign*}
         \sqrt{n} (\hat{\theta} - \theta)  & = \frac{1}{\sqrt{n}}\sum_{i=1}^n \varphi(X_i, \hat{\eta}, \theta) =  \frac{1}{\sqrt{n}}\sum_{i=1}^n (m(X_i, \hat{\eta}) - \theta )= R_1 + R_2, \\
         R_1 & = \frac{1}{\sqrt{n}}  \sum_{i=1}^n \big( m(X_i, \hat{\eta}) - \mathbb{P} m(X, \hat{\eta}) \big), \\
         R_2 &  = \frac{1}{\sqrt{n}}  \sum_{i=1}^n \big( \mathbb{P} m(X, \hat{\eta}) - \theta \big) = \sqrt{n}\big( \mathbb{P} m(X, \hat{\eta}) - \theta \big). &
    \end{flalign*}
    Let  $\eta_0 $ represent the true value of $\eta.$  Assumption~\ref{ass:regularity_asympt_normality} ensures that $R_1 = \frac{1}{\sqrt{n}}  \sum_{i=1}^n \big( m(X_i, \eta_0) - \mathbb{P} m(X, \eta_0) \big) + o_p(1)$ from  Lemma 2 \citep{kennedy2020sharp} in case of sample splitting and from Lemma 19.24 \citep{vaart1998} when $m (X, \hat{\eta})$ belongs to a Donsker class. 
    For each model separately, we provide conditions for $R_2$ being $o_p(1).$
    
    Then $\hat{\theta}$  is asymptotically linear with influence function $\varphi, $
    $$\sqrt{n} (\hat{\theta} - \theta_0) =\frac{1}{\sqrt{n}}  \sum_{i=1}^n \big( m(X_i, \eta_0) - \mathbb{P} m(X, \eta_0) \big) + o_p(1) =  \frac{1}{\sqrt{n}}  \sum_{i=1}^n \varphi(X_i, \eta_0, \theta_0) + o_p(1).$$

\subsubsection{Boundedness assumptions in Theorems \ref{th:efficiency_of_fd_and_td_est}- \ref{th:consistency_efficiency_of_bd_fd_td_est}}
The following Assumption \ref{ass:an_td} is used in the proof of the efficiency of $\hat{\theta}_{td}$ in  \autoref{th:efficiency_of_fd_and_td_est}. 
\begin{assumption}\label{ass:an_td}~  
\begin{enumerate}
    \item $\exists \epsilon_1{:}$ $P(|\hat{E}(Y|A, Z, C) | < \epsilon_1) = 1 $  with probability $ 1 - \delta_{1n},$ where $\delta_{1n}$ are positive, and $\delta_{1n} \rightarrow 0,$ $n \rightarrow \infty$ (here and below in assumptions $P$ is over $Z, A, C$, but still depends on the sample used in $\hat{E}$ or $\hat{p}$).
    \item $\exists \epsilon_2>0{:}$  $P( p(Z|A, C) > \epsilon_2) = 1.$ 
    \item $\exists \epsilon_3>0{:}$ $P(\hat{p}(Z|A, C) > \epsilon_3)  = 1$  with probability $ 1 - \delta_{3n},$ where $\delta_{3n}$ are positive, and $ \delta_{3n} \rightarrow 0,$ $n \rightarrow \infty.$ 
    \item $\exists \epsilon_4{:}$ $P(|\hat{p}(Z|\breve{a}, C) | < \epsilon_4) = 1 $  with probability at least $ 1 - \delta_{4n},$ where $\delta_{4n}$ are positive, and $\delta_{4n} \rightarrow 0,$ $n \rightarrow \infty;$ $ \breve{a} \in \{a^*, a\}.$
    \item $\exists \epsilon_5>0{:}$  $P( p(A| C) > \epsilon_5) = 1.$  
     \item $\exists \epsilon_6>0{:}$ $P(\hat{p}(\breve{a}|C) > \epsilon_6)  = 1$  with probability  at least $ 1 - \delta_{6n},$ where $\delta_{6n}$ are positive, and $\delta_{6n} \rightarrow 0,$ $n \rightarrow \infty;$ $\breve{a} \in \{a^*, a\}.$
    \item $\exists \epsilon_7{:}$ $P(|\hat{p}(A| C) | < \epsilon_7) = 1 $ with probability $ 1 - \delta_{7n},$ where $\delta_{7n}$ are positive, and  $\delta_{7n} \rightarrow 0,$ $n \rightarrow \infty.$
    \item $\exists \epsilon_8{:}$ $P(|\hat{p}(\breve{a}| C) | < \epsilon_8) = 1 $  with probability at least $ 1 - \delta_{8n},$ where $\delta_{8n}$ are positive, and  $\delta_{8n} \rightarrow 0,$ $n \rightarrow \infty;$ $ \breve{a} \in \{a^*, a\}.$
\end{enumerate}
\end{assumption}
The following Assumption \ref{ass:an_fd} is used in the proof of the efficiency of $\hat{\theta}_{fd}$ in  \autoref{th:efficiency_of_fd_and_td_est}. 
\begin{assumption}\label{ass:an_fd}~ 
\begin{enumerate}
     \item $\exists \epsilon_1{:}$ $P(|\hat{E}(Y|A, Z) | < \epsilon_1) = 1 $  with probability $ 1 - \delta_{1n},$ where $\delta_{1n}$ are positive, and $\delta_{1n} \rightarrow 0,$ $n \rightarrow \infty$ (here and below in assumptions $P$ is over $Z, A, C$, but still depends on the sample used in $\hat{E}$ or $\hat{p}$).
    \item $\exists \epsilon_2>0{:}$  $P( p(Z|A) > \epsilon_2) = 1.$ 
    \item $\exists \epsilon_3>0{:}$ $P(\hat{p}(Z|A) > \epsilon_3)  = 1$  with probability $ 1 - \delta_{3n},$ where $\delta_{3n}$ are positive, and $ \delta_{3n} \rightarrow 0,$ $n \rightarrow \infty.$ 
    \item $\exists \epsilon_4{:}$ $P(|\hat{p}(Z|\breve{a}) | < \epsilon_4) = 1 $  with probability at least $ 1 - \delta_{4n},$  where $\delta_{4n}$ are positive, and $\delta_{4n} \rightarrow 0,$ $n \rightarrow \infty;$ $ \breve{a} \in \{a^*, a\}.$
    \item $\exists \epsilon_5>0{:}$  $P( p(A) > \epsilon_5) = 1.$  
     \item $\exists \epsilon_6>0{:}$ $\hat{p}(\breve{a}) > \epsilon_6 $   with probability at least $ 1 - \delta_{6n}$, where $\delta_{6n}$ are positive, and $\delta_{6n} \rightarrow 0,$ $n \rightarrow \infty; $ $\breve{a} \in \{a^*, a\}$. 
    \item $\exists \epsilon_7{:}$ $P(|\hat{p}(A) | < \epsilon_7) = 1 $ with probability $ 1 - \delta_{7n},$ where $\delta_{7n}$ are positive, and  $\delta_{7n} \rightarrow 0,$ $n \rightarrow \infty.$
    \item $\exists \epsilon_8{:}$ $|\hat{p}(\breve{a}) | < \epsilon_8  $  with probability at least $ 1 - \delta_{8n},$ where $\delta_{8n}$ are positive, and  $\delta_{8n} \rightarrow 0,$ $n \rightarrow \infty;$ $ \breve{a} \in \{a^*, a\}.$
\end{enumerate}
\end{assumption}
The following Assumption \ref{ass:an_bd_td} is used in the proof of the efficiency of $\hat{\theta}_{bd, td}$ in \autoref{th:consistency_efficiency_of_bd_td_est}. 
\begin{assumption}\label{ass:an_bd_td}~ 
\begin{enumerate}
    \item $\exists \epsilon_2>0{:}$  $P( p(Z|A, C) > \epsilon_2) = 1.$ 
    \item $\exists \epsilon_1>0{:}$ $P (\sum_{\bar{a}} \hat{p}(Z|\bar{a}, C) \hat{p}(\bar{a}|C)> 
     \epsilon_1) $ with probability at least $ 1 - \delta_{1n},$ where $\delta_{1n}$ are positive, and $\delta_{1n} \rightarrow 0,$ $n \rightarrow \infty;$ 
    \item $\exists \epsilon_3>0{:}$ $\sum_{\bar{a}} p(Z|\bar{a}, C) p(\bar{a}|C)> 
     \epsilon_3) = 1$
    \item $\exists \epsilon_4{:}$ $P(|\hat{p}(Z|\breve{a}, C) | < \epsilon_4) = 1 $  with probability at least $ 1 - \delta_{4n},$ where $\delta_{4n}$ are positive, and $\delta_{4n} \rightarrow 0,$ $n \rightarrow \infty;$ $ \breve{a} \in \{a^*, a\}.$
    \item $\exists \epsilon_5>0{:}$  $P( p(A| C) > \epsilon_5) = 1.$  
     \item $\exists \epsilon_6>0{:}$ $P(\hat{p}(\breve{a}|C) > \epsilon_6)  = 1$  with probability  at least $ 1 - \delta_{6n},$ where $\delta_{6n}$ are positive, and $\delta_{6n} \rightarrow 0,$ $n \rightarrow \infty;$ $\breve{a} \in \{a^*, a\}.$
\end{enumerate}
\end{assumption}
The following Assumption \ref{ass:an_fd_td} is used in the proof of the efficiency of $\hat{\theta}_{fd,td}$ in \autoref{th:consistency_efficiency_of_fd_td_est}.
\begin{assumption}\label{ass:an_fd_td}~  
\begin{enumerate}
    \item $\exists \epsilon_1{:}$ $P(|\hat{E}(Y|A, Z, C) | < \epsilon_1) = 1 $  with probability $ 1 - \delta_{1n},$ where $\delta_{1n}$ are positive, and $\delta_{1n} \rightarrow 0,$ $n \rightarrow \infty$ (here and below in assumptions $P$ is over $Z, A, C$, but still depends on the sample used in $\hat{E}$ or $\hat{p}$).
    \item $\exists \epsilon_2>0{:}$  $P( p(Z|A) > \epsilon_2) = 1.$ 
    \item $\exists \epsilon_3>0{:}$ $P(\hat{p}(Z|A) > \epsilon_3)  = 1$  with probability $ 1 - \delta_{3n},$ where $\delta_{3n}$ are positive, and $ \delta_{3n} \rightarrow 0,$ $n \rightarrow \infty.$ 
    \item $\exists \epsilon_4{:}$ $P(|\hat{p}(Z|\breve{a}) | < \epsilon_4) = 1 $  with probability at least $ 1 - \delta_{4n},$ where $\delta_{4n}$ are positive, and $\delta_{4n} \rightarrow 0,$ $n \rightarrow \infty;$ $ \breve{a} \in \{a^*, a\}.$
    \item $\exists \epsilon_5>0{:}$  $P( p(A| C) > \epsilon_5) = 1.$  
     \item $\exists \epsilon_6>0{:}$ $P(\hat{p}(\breve{a}) > \epsilon_6)  = 1$  with probability  at least $ 1 - \delta_{6n},$ where $\delta_{6n}$ are positive, and $\delta_{6n} \rightarrow 0,$ $n \rightarrow \infty;$ $\breve{a} \in \{a^*, a\}.$
    \item $\exists \epsilon_7{:}$ $P(|\hat{p}(A| C) | < \epsilon_7) = 1 $ with probability $ 1 - \delta_{7n},$ where $\delta_{7n}$ are positive, and  $\delta_{7n} \rightarrow 0,$ $n \rightarrow \infty.$
    \item $\exists \epsilon_8>0{:}$  $P( p( C) > \epsilon_5) = 1.$ 
    \item $\exists \epsilon_9{:}$ $P(|\hat{p}( C) | < \epsilon_7) = 1 $ with probability $ 1 - \delta_{9n},$ where $\delta_{9n}$ are positive, and  $\delta_{9n} \rightarrow 0,$ $n \rightarrow \infty.$
\end{enumerate}
\end{assumption}
The following Assumption \ref{ass:an_bd_fd_td} is used in the proof of the efficiency of $\hat{\theta}_{bd,fd,td}$ in \autoref{th:consistency_efficiency_of_bd_fd_td_est}.
\begin{assumption}\label{ass:an_bd_fd_td}~
    \begin{enumerate}
        \item $\exists \epsilon_1{:}$ $P(|\hat{E}(Y| Z, C) | < \epsilon_1) = 1 $  with probability $ 1 - \delta_{1n},$ where $\delta_{1n}$ are positive, and $\delta_{1n} \rightarrow 0,$ $n \rightarrow \infty$ (here and below in assumptions $P$ is over $Z, A, C$, but still depends on the sample used in $\hat{E}$ or $\hat{p}$).
        \item $\exists \epsilon_2>0{:}$  $P( p(Z|A) > \epsilon_2) = 1.$ 
        \item  $\exists \epsilon_3>0{:}$  $P( p(Z|\bar{a}) > \epsilon_3) = 1.$ 
        \item  $\exists \epsilon_4{:}$ $P(|\hat{p}(Z|\breve{a}) | < \epsilon_4) = 1 $  with probability at least $ 1 - \delta_{4n},$ where $\delta_{4n}$ are positive, and $\delta_{4n} \rightarrow 0,$ $n \rightarrow \infty;$ $ \breve{a} \in \{a^*, a\}.$
        \item  $\exists \epsilon_5>0{:}$ $P(\hat{p}(Z|\bar{a}) > \epsilon_5)  = 1$  with probability $ 1 - \delta_{5n},$  where $\delta_{5n}$ are positive and $ \delta_{5n} \rightarrow 0,$ $n \rightarrow \infty;$  $\bar{a} \in \mathcal{A}.$ 
        \item  $\exists \epsilon_6>0{:}$  $P( p(A| C) > \epsilon_6) = 1.$  
        \item $\exists \epsilon_7>0{:}$  $P( p(\bar{a}| C) > \epsilon_7) = 1; $ $\bar{a} \in \mathcal{A}.$ 
        \item $\exists \epsilon_8{:}$ $P(|\hat{p}(A| C) | < \epsilon_8) = 1 $ with probability $ 1 - \delta_{8n},$ where $\delta_{8n}$ are positive, and  $\delta_{8n} \rightarrow 0,$ $n \rightarrow \infty.$
        \item $\exists \epsilon_9>0{:}$ $P(\hat{p}(\bar{a}|C) > \epsilon_9)  = 1$  with probability  at least $ 1 - \delta_{9n},$ where $\delta_{9n}$ are positive, and $\delta_{9n} \rightarrow 0,$ $n \rightarrow \infty;$ $\bar{a} \in \mathcal{A}.$
        \item $\exists \epsilon_{10}>0{:}$ $P(\hat{p}(\breve{a}) > \epsilon_{10})  = 1$  with probability  at least $ 1 - \delta_{10n},$ where $\delta_{10n}$ are positive, and $\delta_{10n} \rightarrow 0,$ $n \rightarrow \infty;$ $\breve{a} \in \{a^*, a\}.$
        \item $\exists \epsilon_{11}>0{:}$  $P( p(C) > \epsilon_{11}) = 1.$ 
        \item $\exists \epsilon_{12}{:}$ $P(|\hat{p}( C) | < \epsilon_{12}) = 1 $ with probability $ 1 - \delta_{12n},$ where $\delta_{12n}$ are positive, and  $\delta_{12n} \rightarrow 0,$ $n \rightarrow \infty.$
    \end{enumerate}
\end{assumption}

Note that, if $A, Z, C$ are continuous,  they should have a bounded support to satisfy the boundedness assumptions in \Cref{ass:an_td}-\Cref{ass:an_bd_fd_td}.

\subsection{Proof of \autoref{th:consistency_efficiency_of_bd_td_est}} \label{sec:appendix_proof_lemma_consistency_efficiency_of_bd_td_est}

\begin{proof}
According to Appendix \ref{sec:appendix_consistency_conditions}, we prove the consistency of the estimator by showing that under each scenario $E\varphi_{bd, td}(X, \bar{\eta}, \theta_0) = 0,$ where $\theta_0$ is the true value of the ACE.
    
When $\bar{p}(A|C) = p(A|C),$  
    \begin{flalign*}
        & E\varphi_{bd, td}(X, \bar{\eta}, \theta) = E\bigg[\frac{\left(Y-\bar{E}(Y|Z, C)\right) (\bar{p}(Z|a^*,C)-\bar{p}(Z|a,C))}{\sum_a p(a|C)\bar{p}(Z|a, C)}\bigg.  \\
        & + \left(\bar{E}(Y|Z, C)-\sum_{z}\bar{E}(Y|z, C)\bar{p}(z|A,C)\right)\frac{I(A=a^*)-I(A=a)}{p(A|C)}\\
        & + \bigg. \sum\limits_{z } \bar{E}(Y|z, C)\bar{p}(z|a^*, C) - \sum\limits_{z} \bar{E}(Y|z, C)\bar{p}(z|a, C)-\theta \bigg] \\ 
        & = \sum_{c, z} \frac{\left(E(Y|z,c)-\bar{E}(Y|z, c)\right) (\bar{p}(z|a^*,c)-\bar{p}(z|a,c))}{\sum_a p(a|c)\bar{p}(z|a, c)} p(c)\sum_a p(a|c)p(z|a,c)\\
        & + \sum_{c, z} \bar{E}(Y|z, c) p(c)  
            \left( p(z|a^*, c) - \bar{p}(z|a^*, c) - p(z|a, c) + \bar{p}(z|a, c)\right)\\
        &  + \sum_{c, z} \bar{E}(Y|z, c) p(c)  
            \left( \bar{p}(z|a^*, c) - \bar{p}(z|a, c) \right) \\
        & -  \sum_{c, z} E(Y|z,c) p(c)\left( p(z|a^*, c) - p(z|a, c) \right)\\
        & = \sum_{c, z}\left(E(Y|z,c) - \bar{E}(Y|z, c)\right)p(c) \times\\
        & \times 
             \left( \frac{\sum_a p(a|c)p(z|a,c)}{\sum_a p(a|c)\bar{p}(z|a,c)}
             \left(\bar{p}(z|a^*,c) - \bar{p}(z|a,c)\right) -  p(z|a^*, c) + p(z|a, c) \right).&
    \end{flalign*}
    Therefore, when $\bar{p}(A|C) = p(A|C)$ and either $\bar{E}(Y|Z,  C) = E(Y|Z,  C)$  or $\bar{p}(Z|A, C) = p(Z|A, C),$ 
    $ E\varphi_{bd,td}(X, \bar{\eta}, \theta_0) = 0.$ When $\bar{p}(Z|A, C) = p(Z|A, C),$
    \begin{flalign*}
        &  E\varphi_{bd, td}(X, \bar{\eta}, \theta) =  E\bigg[\frac{\left(Y-\bar{E}(Y|Z, C)\right) (p(Z|a^*,C)- p(Z|a,C))}{\sum_a \bar{p}(a|C)p(Z|a, C)}\bigg.  \\
        & + \left(\bar{E}(Y|Z, C)-\sum_{z}\bar{E}(Y|z, C)p(z|A,C)\right)\frac{I(A=a^*)-I(A=a)}{\bar{p}(A|C)}\\
        & + \bigg. \sum\limits_{z } \bar{E}(Y|z, C)p(z|a^*, C) - \sum\limits_{z} \bar{E}(Y|z, C)p(z|a, C)-\theta \bigg] \\ 
        & = \sum_{c, z}\frac{\left(E(Y|z,c)-\bar{E}(Y|z, c)\right) (p(z|a^*,c)-p(z|a,c))}{\sum_a \bar{p}(a|c)p(z|a, c)} p(c)\sum_a p(z|a,c)p(a|c)\\
        & + \sum_{c, z}\left(\bar{E}(Y|z, c) - E(Y|z,c) \right)\left(p(z|a^*, c) - p(z|a, c) \right)p(c) \\
        & =  \sum_{c, z}\left(E(Y|z,c)-\bar{E}(Y|z, c)\right)
            \left(p(z|a^*, c) - p(z|a, c) \right) p(c)
            \left( \frac{\sum_a p(a|c)p(z|a,c)}{\sum_a \bar{p}(a|c)p(z|a, c)} - 1\right).&
    \end{flalign*} 
    The last expression is equal to 0 when either $\bar{E}(Y|Z,  C) = E(Y|Z,  C)$ or $\bar{p}(A|C) = p(A|C).$ When $\bar{E}(Y|Z,  C) = E(Y|Z,  C),$ 
    \begin{flalign*}
        & E\varphi_{bd, td}(X, \bar{\eta}, \theta) = E\bigg[\frac{\left(Y-E(Y|Z, C)\right) (\bar{p}(Z|a^*,C)-\bar{p}(Z|a,C))}{\sum_a \bar{p}(a|C)\bar{p}(Z|a, C)}\bigg.  \\
        & + \left(E(Y|Z, C)-\sum_{z}E(Y|z,C)\bar{p}(z|A,C)\right)
            \frac{I(A=a^*)-I(A=a)}{\bar{p}(A|C)}\\
        & + \bigg. \sum\limits_{z } E(Y|z,C)\bar{p}(z|a^*, C) 
            - \sum\limits_{z} E(Y|z,C)\bar{p}(z|a, C)-\theta \bigg] \\ 
        & = \sum_{c, z}E(Y|z,c)\left(p(z|a^*, c) - \bar{p}(z|a^*,c)\right)\left( \frac{ p(a^*|c)}{\bar{p}(a^*|c)} - 1 \right)p(c)\\
        & - \sum_{c, z}E(Y|z,c)\left(p(z|a, c) - \bar{p}(z|a,c)\right)\left( \frac{ p(a|c)}{\bar{p}(a|c)} - 1 \right) p(c) = 0&
    \end{flalign*}
   when, additionally to $\bar{E}(Y|Z,  C) = E(Y|Z,  C),$  either $\bar{p}(Z|A, C) = p(Z|A, C)$ or  $\bar{p}(A|C) = p(A|C).$ These conditions have already been mentioned above.
    
   To prove the efficiency of the estimator, according to Appendix \ref{sec:appendix_efficiency_conditions}, we now show that $\mathbb{P} m_{bd, td}(X, \hat{\eta}) - \theta = o_p(n^{-1/2}).$ 
    \begin{flalign*}
        & \mathbb{P}   m_{bd, td}(X, \hat{\eta}) - \theta \\
        & = \sum_{c, z}\left(E(Y|z,c) - \hat{E}(Y|z,c) \right)  
            \left(\hat{p}(z|a^*,c) - \hat{p}(z|a,c) \right)
            \frac{\sum_{\bar{a}} p(z|\bar{a},c) p(\bar{a}|c)
                 }{ \sum_{\bar{a}} \hat{p}(z|\bar{a}, c) \hat{p}(\bar{a}|c) 
                 }  p(c)\\
        & + \sum_{\bar{a}, c, z} \hat{E}(Y|z,c) ( p(z|\bar{a}, c) - \hat{p}(z|\bar{a},c)) \frac{I(\bar{a}=a^*)}{\hat{p}(\bar{a}|c)} p(\bar{a},c)\\
        & - \sum_{\bar{a}, c, z} \hat{E}(Y|z,c) ( p(z|\bar{a}, c) - \hat{p}(z|\bar{a},c)) \frac{I(\bar{a}=a)}{\hat{p}(\bar{a}|c)} p(\bar{a}, c)\\
        & + \sum_{c, z} \hat{E}(Y|z,c) ( \hat{p}(z|a^*, c) - \hat{p}(z|a,c))p(c)\\
        & - \sum_{c, z} E(Y|z,c) ( p(z|a^*, c) - p (z|a,c))p(c).& 
    \end{flalign*}
    Subtraction, addition of some terms, and rearranging allows us to re-write the expression as 
    \begin{flalign*}
        & \mathbb{P}   m_{bd, td}(X, \hat{\eta}) - \theta \\
        & = \sum_{c, z}\left(E(Y|z,c) - \hat{E}(Y|z,c) \right)  
            \left(\hat{p}(z|a^*,c) - \hat{p}(z|a,c) \right)
            \left( 
                \frac{
                    \sum_{\bar{a}} p(z|\bar{a},c) p(\bar{a}|c)
                    }{\sum_{\bar{a}} \hat{p}(z|\bar{a}, c) \hat{p}(\bar{a}|c)
                    } - 1 
            \right) p(c)\\
        & + \sum_{c, z}\left(E(Y|z,c) - \hat{E}(Y|z,c) \right)  
            \left(\hat{p}(z|a^*,c) - \hat{p}(z|a,c) \right) p(c)\\
        & + \sum_{\bar{a}, c, z} \hat{E}(Y|z,c) ( p(z|\bar{a}, c) - \hat{p}(z|\bar{a},c)) \frac{p(\bar{a}|c)}{\hat{p}(\bar{a}|c)} I(\bar{a}=a^*) p(c)\\
        & - \sum_{\bar{a}, c, z} \hat{E}(Y|z,c) ( p(z|\bar{a}, c) - \hat{p}(z|\bar{a},c))
        \frac{p(\bar{a}|c)}{\hat{p}(\bar{a}|c)}I(\bar{a}=a) p(c)\\
        & + \sum_{c, z} \hat{E}(Y|z,c) ( \hat{p}(z|a^*, c) - \hat{p}(z|a,c))p(c)\\
        & - \sum_{c, z} E(Y|z,c) ( p(z|a^*, c) - p (z|a,c))p(c).&
    \end{flalign*}
    The second last term and a part with $\hat{E}(z,c)$ of the second term are the same and cancel out. The other terms can be rearranged as follows.
    \begin{flalign*}
        & \mathbb{P}   m_{bd, td}(X, \hat{\eta}) - \theta \\
        & = \sum_{c, z}\left(E(Y|z,c) - \hat{E}(Y|z,c) \right)  
            \left(\hat{p}(z|a^*,c) - \hat{p}(z|a,c) \right)  \times\\
        & \times \frac{
                        \sum_{\bar{a}} p(z|\bar{a},c) p(\bar{a}|c) - \sum_{\bar{a}} \hat{p}(z|\bar{a}, c) \hat{p}(\bar{a}|c) \pm \sum_{\bar{a}} \hat{p}(z|\bar{a}, c)     p(\bar{a}|c)
                        }{ \sum_{\bar{a}} \hat{p}(z|\bar{a}, c) \hat{p}(\bar{a}|c)
                        } p(c)   \\
        & + \sum_{c, z} \left(\hat{E}(Y|z,c) \frac{p(a^*|c)}{\hat{p}(a^*|c)} - E(Y|z,c)\right)( p(z|a^*, c) - \hat{p}(z|a^*,c)) p(c)\\
        & - \sum_{c, z}  \left(\hat{E}(Y|z,c) \frac{p(a|c)}{\hat{p}(a|c)} - E(Y|z,c)\right)( p(z|a, c) - \hat{p}(z|a,c))p(c)\\
        & =  \sum_{c, z}\left(E(Y|z,c) - \hat{E}(Y|z,c) \right)  
            \left(\hat{p}(z|a^*,c) - \hat{p}(z|a,c) \right)  \times\\
        & \times \frac{
                        \sum_{\bar{a}}  (p(z|\bar{a}, c) - \hat{p}(z|\bar{a}, c)) p(\bar{a}|c) 
                         + \sum_{\bar{a}} \hat{p}(z|\bar{a}, c) (p(\bar{a}|c) - \hat{p}(\bar{a}|c))     
                        }{ \sum_{\bar{a}} \hat{p}(z|\bar{a}, c) \hat{p}(\bar{a}|c)
                        } p(c)  \\
        & + \sum_{c, z} \frac{ p(a^*|c) 
                        \left(\hat{E}(Y|z,c) - E(Y|z,c) 
                        \right) - E(Y|z,c) (\hat{p}(a^*|c) - p(a^*|c))
                    }{\hat{p}(a^*|c)
                    }  \times \\
        & \times    ( p(z|a^*, c) - \hat{p}(z|a^*,c)) p(c) \\
        & - \sum_{c, z} \frac{ p(a|c) 
                        \left(\hat{E}(Y|z,c) - E(Y|z,c) 
                        \right) - E(Y|z,c) (\hat{p}(a|c) - p(a|c))
                    }{\hat{p}(a|c)
                    } ( p(z|a, c) - \hat{p}(z|a,c)) p(c) \\
        & =  \mathbb{P} 
        \left[ 
            \left(\hat{E}(Y|Z,C) - E(Y|Z,C) \right)
            \left(\hat{p}(Z|A, C) - p(Z|A, C) \right)
            \frac{
                \left(\hat{p}(Z|a^*,C) - \hat{p}(Z|a,C) \right)
                }{
                {p(Z|A,C) \sum_{\bar{a}} \hat{p}(Z|\bar{a}, C) \hat{p}(\bar{a}|C)
                        } }
        \right] \\
        & + \mathbb{P} 
        \left[ 
            \left(\hat{E}(Y|Z,C) - E(Y|Z,C) \right)
            \left(\hat{p}(A| C) - p(A| C) \right)
            \frac{\hat{p}(Z|A,C)
                \left(\hat{p}(Z|a^*,C) - \hat{p}(Z|a,C) \right)
                }{
                {p(Z|A,C)p(A|C) \sum_{\bar{a}} \hat{p}(Z|\bar{a}, C) \hat{p}(\bar{a}|C)
                        } }
        \right] \\
        & + \mathbb{P} 
        \left[ 
            \left(\hat{E}(Y|Z,C) - E(Y|Z,C) \right)
            \left(p(Z|A, C) - \hat{p}(Z|A, C) \right)
            \frac{I(A = a^*)}{\hat{p}(a^*|C)
            p(Z|A, C)}
        \right]\\
        & - \mathbb{P} 
        \left[ 
            \left(p(Z|a^*, C) - \hat{p}(Z|a^*, C) \right)
            \left(\hat{p}(a^*|C) - p(a^*|C) \right)
            \frac{E(Y|Z,C)}{\hat{p}(a^*|C)
            \sum_{\bar{a}} p(Z|\bar{a}, C) p(\bar{a}|C)}
        \right]\\
        & - \mathbb{P} 
        \left[ 
            \left(\hat{E}(Y|Z,C) - E(Y|Z,C) \right)
            \left(p(Z|A, C) - \hat{p}(Z|A, C) \right)
            \frac{I(A = a)}{\hat{p}(a|C)
            p(Z|A, C)}
        \right]\\
        & + \mathbb{P} 
        \left[ 
            \left(p(Z|a, C) - \hat{p}(Z|a, C) \right)
            \left(\hat{p}(a|C) - p(a|C) \right)
            \frac{E(Y|Z,C)}{\hat{p}(a|C)
            \sum_{\bar{a}} p(Z|\bar{a}, C) p(\bar{a}|C)}
        \right].
    \end{flalign*}
 Let's consider the first term in the summation above.    
    \begin{align*}
    & \lim\limits_{n \rightarrow \infty} P\left( \left | \sqrt{n} \mathbb{P} 
                \left[ W_n V_n 
                \frac{
                \left(\hat{p}(Z|a^*,C) - \hat{p}(Z|a,C) \right)
                }{
                {p(Z|A,C) \sum_{\bar{a}} \hat{p}(Z|\bar{a}, C) \hat{p}(\bar{a}|C)
                        } } 
                \right] 
            \right |>\epsilon 
    \right) \\
    & \leq \lim\limits_{n \rightarrow \infty}  P\left( \left | \sqrt{n} \mathbb{P} 
                \left[ W_n V_n 
                Q 
                \right] 
            \right |>\epsilon
    \right)  \leq  \lim\limits_{n \rightarrow \infty}  P\left( \left | \sqrt{n} (\mathbb{P} 
                W_n^2)^{1/2} (\mathbb{P} V_n^2) ^2
            \right |>\epsilon/ Q
    \right)  = 0
\end{align*}
for some $Q,$ where the second last inequality follows from boundedness within   \Cref{ass:an_bd_td} and the last inequality follows from the Cauchy-Schwarz inequality.
$\mathbb{P} m_{bd, td}(X, \hat{\eta}) - \theta  = o_p(1/\sqrt{n})$ since  each term in the summation above is $o_p(1/\sqrt{n}). $
\end{proof}

\section{Example. Sufficient Condition for the Two-door Bound Being Lower than the Back-door Bound When All Observed Variables Are Binary}\label{sec:example}

When the treatment is binary, the expression in the sufficient condition from Proposition \ref{th:sufficient_conditions_td_vs_bd} can be simplified as follows.
    \begin{flalign*}
        &(p(z|a^*,c)-p(z|a,c))^2  \sum\limits_{\bar{a}}\frac{p(\bar{a}| c)}{p(z|\bar{a},c)}- \frac{p(z|a^*,c)}{p(a^*| c)}  - \frac{p(z|a,c)}{p(a| c)}  \\
        &=(p(z|a^*,c)-p(z|a,c))^2\Bigg[\frac{p(a^*| c)}{p(z|a^*,c)} + 
        \frac{p(a| c)}{p(z|a,c)}\Bigg]-\Bigg[\frac{p(z|a^*)}{p(a^*| c)}  +
        \frac{p(z|a,c)}{p(a| c)}\Bigg]\\
        &=\bigg(p(a| c)p(z|a^*,c)+p(a^*| c)p(z|a,c)\bigg) \bigg(\frac{(p(z|a^*,c)-p(z|a,c))^2}{p(z|a^*,c)p(z|a,c)}-\frac{1}{p(a| c)p(a^*| c)}\bigg).&
    \end{flalign*}
    The first term in the product is always positive under the positivity assumption of Section~\ref{sec:notation}.  The sufficient conditions correspond to the second term being nonpositive or nonnegative for all $z \in \mathcal{Z}$ and $c \in \mathcal{C}$.  The inequalities are solved as quadratic inequalities for $\frac{p(z|a^*,c)}{p(z|a,c)},$ and noting that $p(a| c) = 1-p(a^*| c)$ for binary treatment.

Consider pre-treatment covariate $C$ with no restrictions on its distribution and  $A|C \sim \Bernoulli (\expit(\alpha C)),$ where $\expit(x) = \exp(x)/\{1+ \exp(x)\}.$ Let $a^*=1,$ $a=0.$ The mediators $Z(a) \sim \Bernoulli (\expit \beta a), $ and the potential outcomes depend only on $Z$ and $C$. Such distribution satisfies Assumptions \ref{ass:bd} and \hyperref[ass:td]{\textbf{TD}}. 

Consider $\frac{(p(z|a^*,c)-p(z|a,c))^2}{p(z|a^*,c)p(z|a,c)}-\frac{1}{p(a| c)p(a^*| c)}>0$ for all $z \in \mathcal{Z}$ and $c \in \mathcal{C}$ as the sufficient condition  for $var \varphi_{td} > var \varphi_{bd}$. The condition corresponds to a system of inequalities:
\begin{equation*}
    \begin{cases}
        Z=1, C=1: \frac{(\expit(\beta)-0.5)^2}{0.5\expit(\beta)}- \frac{1}{\expit(\alpha) (1-\expit(\alpha))}>0\\
        Z=1, C=0: \frac{(\expit(\beta)-0.5)^2}{0.5\expit(\beta)}- \frac{1}{0.5^2}>0\\
        Z=0, C=0: \frac{(1-\expit(\beta)-0.5)^2}{0.5(1-\expit(\beta))}- \frac{1}{0.5^2}>0\\
        Z=0, C=1: \frac{(1-\expit(\beta)-0.5)^2}{0.5(1-\expit(\beta))}- \frac{1}{\expit(\alpha) (1-\expit(\alpha))}>0
    \end{cases}
\end{equation*}
This system has no solution since there is no joint solution for the second and the third inequalities. This means that we can only state sufficient conditions for $var \varphi_{td} \leq var \varphi_{bd}$ by solving the following system of equations:
\begin{equation*}
    \begin{cases}
        \frac{(\expit(\beta)-0.5)^2}{0.5\expit(\beta)}- \frac{1}{\expit(\alpha) (1-\expit(\alpha))} \leq 0\\
        \frac{(\expit(\beta)-0.5)^2}{0.5\expit(\beta)}- \frac{1}{0.5^2} \leq 0\\
        \frac{(1-\expit(\beta)-0.5)^2}{0.5(1-\expit(\beta))}- \frac{1}{0.5^2} \leq 0\\
        \frac{(1-\expit(\beta)-0.5)^2}{0.5(1-\expit(\beta))}- \frac{1}{\expit(\alpha) (1-\expit(\alpha))} \leq 0.
    \end{cases}
\end{equation*}
The solution is $\expit(\beta)  \in \left[\frac{3-2\sqrt{2}}{2};  \frac{2\sqrt{2}-1}{2}\right].$

\begin{figure}[t]
    \center
    \includegraphics[scale=0.8]{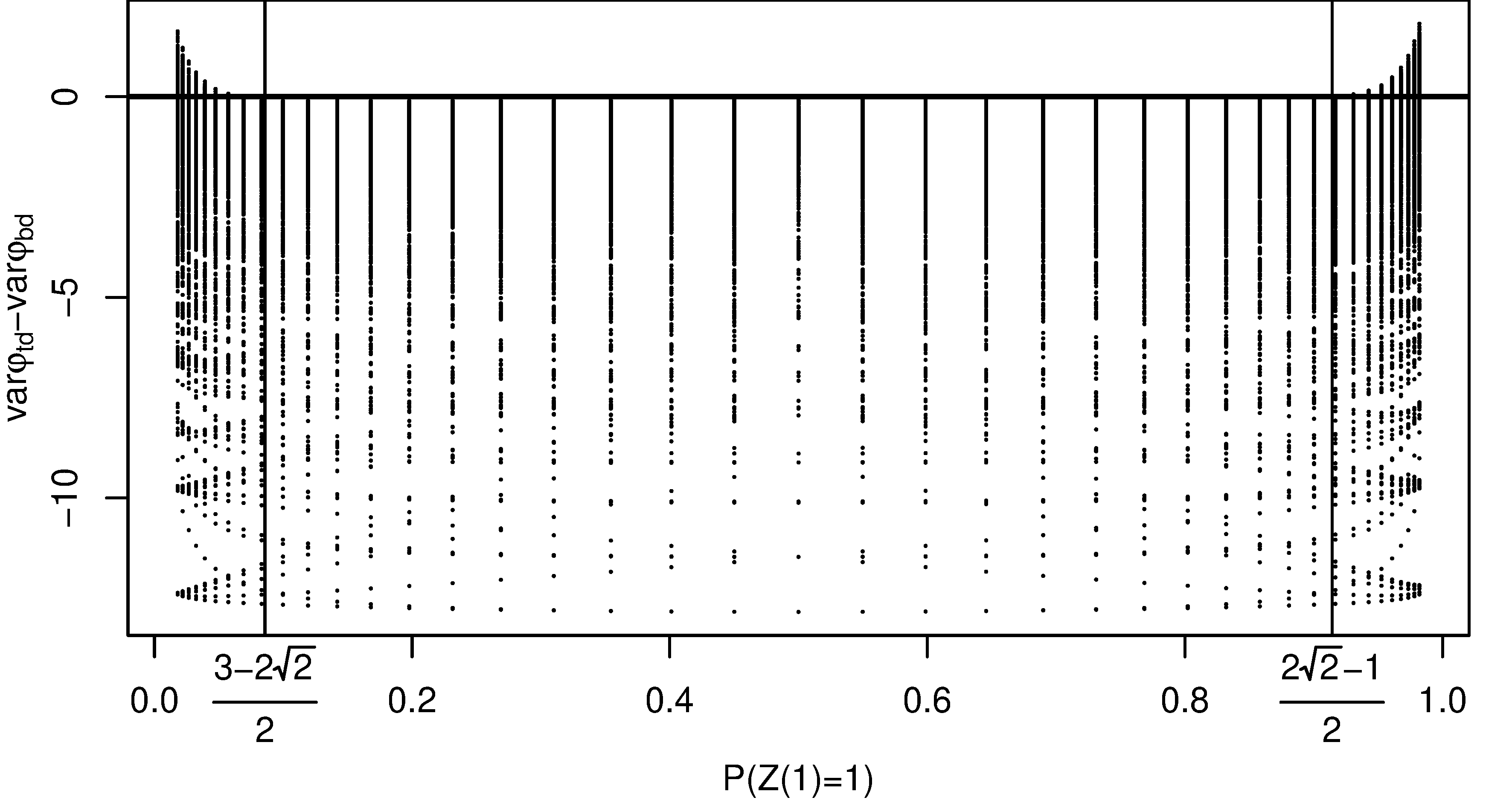}
    \caption{Difference between $var \varphi_{td}$  and $var \varphi_{bd}$ in the example.}\label{fig:td_vs_bd_allDAG}
 \end{figure}

For illustration, \autoref{fig:td_vs_bd_allDAG} shows $var \varphi_{td} - var \varphi_{bd}$ when $C \sim \Bernoulli (\expit(\beta_0))$ and\\  $Y(a)|Z, C \sim \Bernoulli (\expit(\gamma_1 Z + \gamma_2 C))$ for all the combinations of considered values of the parameters $\beta_0=0.1,0.3,0.6,0.9;$ $\alpha, \gamma_1, \gamma_2 \in \{ \m4, \m3,  \ldots, 3, 4\}$; $\beta \in \{ \m4, \m3.8,  \ldots, 3.8, 4\}$. A finer grid is chosen for $\beta$ since $\beta$ is a parameter in $Z(1)$ distribution.
\autoref{fig:td_vs_bd_allDAG} confirms that if  $p(Z(1)=1) = \expit(\beta)  \in \left[(3-2\sqrt{2})/2,  (2\sqrt{2}-1)/2\right],$ $var \varphi_{td}$ is at least as low as $var \varphi_{bd}$.

\section{Properties of the Semiparametric Estimators in the Simulation Studies}
\newpage\thispagestyle{empty}\label{sec:appendix_simulations_result}
\begin{table}[H]
    \tiny
    \setlength{\tabcolsep}{2pt}
    \centering
    \begin{threeparttable}
        \begin{tabular}{rrrrrrrrrrrrrrrrrr}
           \hline
                 $\beta$&  $\gamma1$ & $\gamma2$ & n & \multicolumn{7}{c}{Bias} \\
                 &  & & & Naive & BD & FD & TD & BD TD & FD TD & BD FD TD \\ 
            \hline
                0.5 & 0.5 & 0.5 & 50 &  0.119 (0.010) & -0.006 (0.011) & -0.008 (0.005) & -0.008 (0.006) & -0.007 (0.006) & -0.009 (0.005) & -0.008 (0.005) \\ 
                  0.5 & 0.5 & 0.5 & 100 &  0.126 (0.008) & -0.001 (0.008) & -0.001 (0.004) & -0.001 (0.004) & -0.001 (0.004) & -0.001 (0.004) & -0.001 (0.003) \\ 
                  0.5 & 0.5 & 0.5 & 500 &  0.124 (0.003) &  0.002 (0.003) &  0.002 (0.002) &  0.002 (0.002) &  0.002 (0.002) &  0.002 (0.002) &  0.002 (0.002) \\ 
                  0.5 & 0.5 & 0.5 & 1000 &  0.122 (0.002) &  0.000 (0.002) &  0.003 (0.001) &  0.002 (0.001) &  0.002 (0.001) &  0.003 (0.001) &  0.002 (0.001) \\ 
                  0.5 & 0.5 & 0.5 & 5000 &  0.122 (0.001) &  0.000 (0.001) & -0.001 (0.001) & -0.001 (0.001) &  0.000 (0.001) & -0.001 (0.001) &  0.000 (0.001) \\ 
                  0.5 & 0.5 & 0.5 & 10000 &  0.122 (0.001) &  0.001 (0.001) &  0.000 (0.000) &  0.000 (0.000) &  0.000 (0.000) &  0.000 (0.000) &  0.000 (0.000) \\ 
                  0.5 & 0.5 & 0.5 & 20000 &  0.122 (0.001) &  0.000 (0.001) &  0.000 (0.000) &  0.000 (0.000) &  0.000 (0.000) &  0.000 (0.000) &  0.000 (0.000) \\ 
                  0.5 & 0.5 & 0.5 & 30000 &  0.123 (0.000) &  0.001 (0.000) &  0.001 (0.000) &  0.000 (0.000) &  0.000 (0.000) &  0.001 (0.000) &  0.001 (0.000) \\ 
                  0.5 & 0.5 & 0.5 & 40000 &  0.122 (0.000) &  0.000 (0.000) &  0.000 (0.000) &  0.000 (0.000) &  0.000 (0.000) &  0.000 (0.000) &  0.000 (0.000) \\ 
                 0.5 & 0.5 & 0.5 & 50000 &  0.122 (0.000) &  0.000 (0.000) &  0.000 (0.000) &  0.000 (0.000) &  0.000 (0.000) &  0.000 (0.000) &  0.000 (0.000) \\ 
            \hline
                  0.5 & 0.5 & 1.5 & 50 &  0.368 (0.012) & -0.006 (0.011) &  0.002 (0.006) & -0.005 (0.006) & -0.005 (0.005) & -0.003 (0.005) & -0.003 (0.005) \\ 
                  0.5 & 0.5 & 1.5 & 100 &  0.355 (0.009) & -0.009 (0.008) &  0.001 (0.004) &  0.001 (0.004) &  0.001 (0.004) &  0.001 (0.004) &  0.001 (0.004) \\ 
                  0.5 & 0.5 & 1.5 & 500 &  0.363 (0.004) &  0.003 (0.003) &  0.000 (0.002) &  0.002 (0.002) &  0.002 (0.002) &  0.001 (0.002) &  0.001 (0.002) \\ 
                 0.5 & 0.5 & 1.5 & 1000 &  0.367 (0.003) &  0.001 (0.002) & -0.001 (0.001) &  0.000 (0.001) &  0.001 (0.001) &  0.000 (0.001) &  0.000 (0.001) \\ 
                  0.5 & 0.5 & 1.5 & 5000 &  0.366 (0.001) & -0.001 (0.001) &  0.000 (0.001) &  0.000 (0.001) &  0.000 (0.001) &  0.000 (0.001) &  0.000 (0.001) \\ 
                  0.5 & 0.5 & 1.5 & 10000 &  0.364 (0.001) & -0.002 (0.001) & -0.001 (0.000) & -0.001 (0.000) & -0.001 (0.000) & -0.001 (0.000) & -0.001 (0.000) \\ 
                  0.5 & 0.5 & 1.5 & 20000 &  0.367 (0.001) &  0.000 (0.001) &  0.000 (0.000) &  0.000 (0.000) &  0.000 (0.000) &  0.000 (0.000) &  0.000 (0.000) \\ 
                  0.5 & 0.5 & 1.5 & 30000 &  0.366 (0.000) &  0.000 (0.000) &  0.000 (0.000) &  0.000 (0.000) &  0.000 (0.000) &  0.000 (0.000) &  0.000 (0.000) \\ 
                  0.5 & 0.5 & 1.5 & 40000 &  0.366 (0.000) &  0.000 (0.000) &  0.000 (0.000) &  0.000 (0.000) &  0.000 (0.000) &  0.000 (0.000) &  0.000 (0.000) \\ 
                  0.5 & 0.5 & 1.5 & 50000 &  0.366 (0.000) &  0.000 (0.000) &  0.000 (0.000) &  0.000 (0.000) &  0.000 (0.000) &  0.000 (0.000) &  0.000 (0.000) \\ 
            \hline
                  0.5 & 1.5 & 0.5 & 50 &  0.131 (0.017) &  0.003 (0.017) &  0.005 (0.014) &  0.005 (0.015) &  0.008 (0.015) &  0.005 (0.014) &  0.009 (0.014) \\ 
                  0.5 & 1.5 & 0.5 & 100 &  0.109 (0.012) & -0.020 (0.012) & -0.008 (0.010) & -0.014 (0.011) & -0.015 (0.011) & -0.009 (0.010) & -0.010 (0.010) \\ 
                  0.5 & 1.5 & 0.5 & 500 &  0.118 (0.005) & -0.004 (0.006) & -0.007 (0.004) & -0.007 (0.005) & -0.006 (0.005) & -0.007 (0.004) & -0.007 (0.004) \\ 
                  0.5 & 1.5 & 0.5 & 1000 &  0.123 (0.004) &  0.000 (0.004) & -0.001 (0.003) & -0.002 (0.003) & -0.002 (0.003) & -0.001 (0.003) & -0.001 (0.003) \\ 
                  0.5 & 1.5 & 0.5 & 5000 &  0.124 (0.002) &  0.001 (0.002) &  0.001 (0.001) &  0.000 (0.001) &  0.000 (0.001) &  0.001 (0.001) &  0.001 (0.001) \\ 
                  0.5 & 1.5 & 0.5 & 10000 &  0.121 (0.001) &  0.000 (0.001) &  0.000 (0.001) &  0.000 (0.001) &  0.000 (0.001) &  0.000 (0.001) &  0.000 (0.001) \\ 
                  0.5 & 1.5 & 0.5 & 20000 &  0.121 (0.001) & -0.001 (0.001) & -0.001 (0.001) & -0.001 (0.001) & -0.001 (0.001) & -0.001 (0.001) & -0.001 (0.001) \\ 
                  0.5 & 1.5 & 0.5 & 30000 &  0.122 (0.001) &  0.000 (0.001) &  0.000 (0.001) &  0.000 (0.001) &  0.000 (0.001) &  0.000 (0.001) &  0.000 (0.001) \\ 
                  0.5 & 1.5 & 0.5 & 40000 &  0.122 (0.001) &  0.000 (0.001) &  0.000 (0.001) &  0.000 (0.001) &  0.000 (0.001) &  0.000 (0.001) &  0.000 (0.001) \\ 
                  0.5 & 1.5 & 0.5 & 50000 &  0.121 (0.001) &  0.000 (0.001) & -0.001 (0.000) &  0.000 (0.000) &  0.000 (0.000) & -0.001 (0.000) &  0.000 (0.000) \\ 
            \hline
                  0.5 & 1.5 & 1.5 & 50 &  0.387 (0.018) &  0.019 (0.018) &  0.020 (0.015) &  0.015 (0.015) &  0.013 (0.015) &  0.016 (0.015) &  0.014 (0.014) \\ 
                  0.5 & 1.5 & 1.5 & 100 &  0.366 (0.012) & -0.008 (0.012) & -0.008 (0.010) & -0.014 (0.010) & -0.014 (0.010) & -0.009 (0.010) & -0.009 (0.010) \\ 
                  0.5 & 1.5 & 1.5 & 500 &  0.370 (0.006) &  0.002 (0.006) &  0.005 (0.004) &  0.006 (0.005) &  0.006 (0.005) &  0.005 (0.004) &  0.005 (0.004) \\ 
                  0.5 & 1.5 & 1.5 & 1000 &  0.373 (0.004) &  0.007 (0.004) &  0.006 (0.003) &  0.009 (0.003) &  0.009 (0.003) &  0.007 (0.003) &  0.007 (0.003) \\ 
                  0.5 & 1.5 & 1.5 & 5000 &  0.365 (0.002) & -0.001 (0.002) & -0.001 (0.001) & -0.002 (0.001) & -0.002 (0.001) & -0.002 (0.001) & -0.002 (0.001) \\ 
                  0.5 & 1.5 & 1.5 & 10000 &  0.364 (0.001) & -0.002 (0.001) & -0.002 (0.001) & -0.002 (0.001) & -0.002 (0.001) & -0.002 (0.001) & -0.002 (0.001) \\ 
                  0.5 & 1.5 & 1.5 & 20000 &  0.365 (0.001) & -0.002 (0.001) & -0.001 (0.001) & -0.001 (0.001) & -0.001 (0.001) & -0.001 (0.001) & -0.001 (0.001) \\ 
                  0.5 & 1.5 & 1.5 & 30000 &  0.365 (0.001) & -0.001 (0.001) &  0.000 (0.001) &  0.000 (0.001) &  0.000 (0.001) &  0.000 (0.001) &  0.000 (0.001) \\ 
                  0.5 & 1.5 & 1.5 & 40000 &  0.365 (0.001) & -0.001 (0.001) &  0.000 (0.001) &  0.000 (0.001) &  0.000 (0.001) &  0.000 (0.000) &  0.000 (0.000) \\ 
                  0.5 & 1.5 & 1.5 & 50000 &  0.366 (0.001) &  0.000 (0.001) &  0.000 (0.000) &  0.000 (0.000) &  0.000 (0.000) &  0.000 (0.000) &  0.000 (0.000) \\ 
            \hline
                  1.5 & 0.5 & 0.5 & 50 &  0.136 (0.011) &  0.016 (0.011) &  0.015 (0.016) &  0.005 (0.015) &  0.006 (0.008) &  0.005 (0.015) &  0.005 (0.008) \\ 
                  1.5 & 0.5 & 0.5 & 100 &  0.117 (0.008) & -0.005 (0.008) & -0.001 (0.009) & -0.001 (0.009) & -0.007 (0.005) & -0.002 (0.009) & -0.007 (0.005) \\ 
                  1.5 & 0.5 & 0.5 & 500 &  0.123 (0.003) &  0.003 (0.003) &  0.009 (0.004) &  0.009 (0.004) &  0.001 (0.002) &  0.009 (0.004) &  0.001 (0.002) \\ 
                  1.5 & 0.5 & 0.5 & 1000 &  0.122 (0.002) & -0.001 (0.002) &  0.001 (0.003) &  0.001 (0.003) &  0.001 (0.002) &  0.001 (0.003) &  0.001 (0.002) \\ 
                  1.5 & 0.5 & 0.5 & 5000 &  0.123 (0.001) &  0.000 (0.001) & -0.001 (0.001) & -0.001 (0.001) &  0.000 (0.001) & -0.001 (0.001) &  0.000 (0.001) \\ 
                  1.5 & 0.5 & 0.5 & 10000 &  0.122 (0.001) &  0.000 (0.001) &  0.000 (0.001) &  0.000 (0.001) &  0.001 (0.001) &  0.000 (0.001) &  0.001 (0.001) \\ 
                  1.5 & 0.5 & 0.5 & 20000 &  0.122 (0.001) &  0.000 (0.001) &  0.000 (0.001) &  0.000 (0.001) &  0.000 (0.000) &  0.000 (0.001) &  0.000 (0.000) \\ 
                  1.5 & 0.5 & 0.5 & 30000 &  0.122 (0.000) &  0.000 (0.000) &  0.001 (0.001) &  0.000 (0.001) &  0.000 (0.000) &  0.000 (0.001) &  0.000 (0.000) \\ 
                  1.5 & 0.5 & 0.5 & 40000 &  0.122 (0.000) &  0.000 (0.000) &  0.000 (0.001) &  0.000 (0.000) &  0.000 (0.000) &  0.000 (0.000) &  0.000 (0.000) \\ 
                  1.5 & 0.5 & 0.5 & 50000 &  0.123 (0.000) &  0.001 (0.000) &  0.000 (0.000) &  0.000 (0.000) &  0.000 (0.000) &  0.000 (0.000) &  0.000 (0.000) \\ 
            \hline
                  1.5 & 0.5 & 1.5 & 50 &  0.355 (0.012) & -0.007 (0.011) & -0.023 (0.015) & -0.022 (0.013) & -0.014 (0.008) & -0.020 (0.013) & -0.012 (0.008) \\ 
                  1.5 & 0.5 & 1.5 & 100 &  0.371 (0.008) &  0.002 (0.008) &  0.002 (0.011) & -0.001 (0.009) & -0.003 (0.005) & -0.002 (0.009) & -0.004 (0.005) \\ 
                  1.5 & 0.5 & 1.5 & 500 &  0.356 (0.004) & -0.007 (0.003) & -0.003 (0.005) & -0.004 (0.004) & -0.004 (0.002) & -0.004 (0.004) & -0.004 (0.002) \\ 
                  1.5 & 0.5 & 1.5 & 1000 &  0.368 (0.003) &  0.001 (0.002) &  0.005 (0.004) &  0.003 (0.003) &  0.003 (0.002) &  0.003 (0.003) &  0.002 (0.002) \\ 
                  1.5 & 0.5 & 1.5 & 5000 &  0.368 (0.001) &  0.002 (0.001) & -0.003 (0.002) & -0.001 (0.001) &  0.001 (0.001) & -0.001 (0.001) &  0.001 (0.001) \\ 
                  1.5 & 0.5 & 1.5 & 10000 &  0.366 (0.001) &  0.000 (0.001) &  0.000 (0.001) &  0.000 (0.001) &  0.000 (0.001) &  0.000 (0.001) &  0.000 (0.000) \\ 
                  1.5 & 0.5 & 1.5 & 20000 &  0.367 (0.001) &  0.000 (0.001) &  0.000 (0.001) &  0.000 (0.001) &  0.000 (0.000) &  0.000 (0.001) &  0.000 (0.000) \\ 
                  1.5 & 0.5 & 1.5 & 30000 &  0.366 (0.001) &  0.000 (0.000) &  0.000 (0.001) &  0.000 (0.001) &  0.000 (0.000) &  0.000 (0.001) &  0.000 (0.000) \\ 
                  1.5 & 0.5 & 1.5 & 40000 &  0.366 (0.000) &  0.000 (0.000) & -0.001 (0.001) &  0.000 (0.000) &  0.000 (0.000) &  0.000 (0.000) &  0.000 (0.000) \\ 
                  1.5 & 0.5 & 1.5 & 50000 &  0.366 (0.000) &  0.000 (0.000) &  0.000 (0.001) &  0.000 (0.000) &  0.000 (0.000) &  0.000 (0.000) &  0.000 (0.000) \\ 
            \hline
                  1.5 & 1.5 & 0.5 &50 &  0.108 (0.016) & -0.020 (0.017) & -0.024 (0.018) & -0.029 (0.019) & -0.023 (0.016) & -0.021 (0.018) & -0.015 (0.015) \\ 
                  1.5 & 1.5 & 0.5 &100 &  0.124 (0.012) & -0.008 (0.013) &  0.012 (0.013) &  0.000 (0.013) & -0.009 (0.011) &  0.006 (0.012) & -0.003 (0.011) \\ 
                  1.5 & 1.5 & 0.5 &500 &  0.127 (0.005) &  0.008 (0.005) &  0.005 (0.006) &  0.004 (0.006) &  0.006 (0.005) &  0.003 (0.006) &  0.006 (0.005) \\ 
                  1.5 & 1.5 & 0.5 &1000 &  0.117 (0.004) & -0.005 (0.004) & -0.002 (0.004) & -0.001 (0.004) & -0.003 (0.003) & -0.002 (0.004) & -0.004 (0.003) \\ 
                  1.5 & 1.5 & 0.5& 5000 &  0.124 (0.002) &  0.002 (0.002) & -0.002 (0.002) & -0.001 (0.002) &  0.001 (0.002) & -0.002 (0.002) &  0.001 (0.001) \\ 
                  1.5 & 1.5 & 0.5 &10000 &  0.121 (0.001) &  0.000 (0.001) &  0.002 (0.001) &  0.002 (0.001) &  0.001 (0.001) &  0.002 (0.001) &  0.000 (0.001) \\ 
                  1.5 & 1.5 & 0.5& 20000 &  0.123 (0.001) &  0.000 (0.001) &  0.001 (0.001) &  0.001 (0.001) &  0.000 (0.001) &  0.001 (0.001) &  0.000 (0.001) \\ 
                  1.5 & 1.5 & 0.5 &30000 &  0.122 (0.001) &  0.000 (0.001) & -0.001 (0.001) & -0.001 (0.001) &  0.000 (0.001) & -0.001 (0.001) &  0.000 (0.001) \\ 
                  1.5 & 1.5 & 0.5& 40000 &  0.122 (0.001) &  0.000 (0.001) & -0.001 (0.001) & -0.001 (0.001) &  0.000 (0.001) & -0.001 (0.001) &  0.000 (0.001) \\ 
                  1.5 & 1.5 & 0.5& 50000 &  0.122 (0.001) &  0.000 (0.001) &  0.000 (0.001) &  0.000 (0.001) &  0.000 (0.000) &  0.000 (0.001) &  0.000 (0.000) \\ 
            \hline
                  1.5 & 1.5 & 1.5 &50 &  0.361 (0.018) &  0.009 (0.017) & -0.019 (0.018) & -0.002 (0.017) &  0.004 (0.015) & -0.003 (0.017) &  0.003 (0.015) \\ 
                  1.5 & 1.5 & 1.5 & 100 &  0.361 (0.013) &  0.003 (0.013) & -0.003 (0.015) &  0.004 (0.014) &  0.006 (0.011) &  0.001 (0.013) &  0.002 (0.011) \\ 
                  1.5 & 1.5 & 1.5 & 500 &  0.363 (0.006) & -0.001 (0.006) & -0.001 (0.007) &  0.000 (0.006) & -0.002 (0.005) &  0.000 (0.006) & -0.002 (0.005) \\ 
                  1.5 & 1.5 & 1.5& 1000 &  0.372 (0.004) &  0.007 (0.004) &  0.008 (0.005) &  0.010 (0.004) &  0.008 (0.003) &  0.010 (0.004) &  0.007 (0.003) \\ 
                  1.5 & 1.5 & 1.5 & 5000 &  0.366 (0.002) & -0.002 (0.002) & -0.001 (0.002) & -0.001 (0.002) & -0.001 (0.002) & -0.001 (0.002) & -0.001 (0.002) \\ 
                  1.5 & 1.5 & 1.5 & 10000 &  0.368 (0.001) &  0.001 (0.001) &  0.001 (0.002) &  0.000 (0.001) &  0.000 (0.001) &  0.000 (0.001) &  0.000 (0.001) \\ 
                  1.5 & 1.5 & 1.5 & 20000 &  0.366 (0.001) &  0.001 (0.001) & -0.002 (0.001) & -0.001 (0.001) &  0.000 (0.001) & -0.001 (0.001) &  0.000 (0.001) \\ 
                  1.5 & 1.5 & 1.5& 30000 &  0.366 (0.001) &  0.000 (0.001) & -0.001 (0.001) &  0.000 (0.001) &  0.000 (0.001) &  0.000 (0.001) &  0.000 (0.001) \\ 
                  1.5 & 1.5 & 1.5& 40000 &  0.366 (0.001) &  0.000 (0.001) &  0.000 (0.001) &  0.000 (0.001) &  0.000 (0.001) &  0.000 (0.001) &  0.000 (0.001) \\ 
                  1.5 & 1.5 & 1.5 & 50000 &  0.367 (0.001) &  0.001 (0.001) &  0.001 (0.001) &  0.001 (0.001) &  0.001 (0.000) &  0.001 (0.001) &  0.001 (0.000) \\ 
            \hline
        \end{tabular}
        \caption{Bias estimates (Monte Carlo SEs in parentheses) in simulation study 1}
        \label{table: bias_sim_study_1}
    \end{threeparttable}
    \thispagestyle{empty}
\end{table}

\newpage\thispagestyle{empty}
\begin{table}[H]
    \tiny
    \centering
    \setlength{\tabcolsep}{2pt}
    \begin{threeparttable}
        \begin{tabular}{rrrrrrrrrr}
          \hline
             $\beta$&  $\gamma1$ & $\gamma2$ & n & BD & FD & TD & BD TD & FD TD & BD FD TD \\ 
          \hline
               0.5 & 0.5 & 0.5 & 50 &  6.073 (0.272) &  1.451 (0.065) &  1.612 (0.072) &  1.540 (0.069) &  1.401 (0.063) &  1.339 (0.060) \\ 
               0.5 & 0.5 & 0.5 & 100 &  6.346 (0.284) &  1.253 (0.056) &  1.344 (0.060) &  1.276 (0.057) &  1.226 (0.055) &  1.153 (0.052) \\ 
               0.5 & 0.5 & 0.5 & 500 &  5.692 (0.255) &  1.457 (0.065) &  1.584 (0.071) &  1.509 (0.067) &  1.470 (0.066) &  1.399 (0.063) \\ 
               0.5 & 0.5 & 0.5 & 1000 &  5.924 (0.265) &  1.339 (0.060) &  1.397 (0.062) &  1.381 (0.062) &  1.311 (0.059) &  1.292 (0.058) \\ 
               0.5 & 0.5 & 0.5 & 5000 &  5.432 (0.243) &  1.355 (0.061) &  1.418 (0.063) &  1.353 (0.060) &  1.338 (0.060) &  1.277 (0.057) \\ 
               0.5 & 0.5 & 0.5 & 10000 &  5.611 (0.251) &  1.354 (0.061) &  1.401 (0.063) &  1.355 (0.061) &  1.340 (0.060) &  1.299 (0.058) \\ 
               0.5 & 0.5 & 0.5 & 20000 &  6.031 (0.270) &  1.371 (0.061) &  1.438 (0.064) &  1.402 (0.063) &  1.360 (0.061) &  1.326 (0.059) \\ 
               0.5 & 0.5 & 0.5 & 30000 &  5.651 (0.253) &  1.433 (0.064) &  1.469 (0.066) &  1.429 (0.064) &  1.406 (0.063) &  1.368 (0.061) \\ 
               0.5 & 0.5 & 0.5 & 40000 &  5.212 (0.233) &  1.322 (0.059) &  1.376 (0.062) &  1.350 (0.060) &  1.302 (0.058) &  1.276 (0.057) \\ 
               0.5 & 0.5 & 0.5 & 50000 &  5.266 (0.236) &  1.311 (0.059) &  1.392 (0.062) &  1.341 (0.060) &  1.295 (0.058) &  1.244 (0.056) \\ 
               0.5 & 0.5 & 0.5 & Bound& 5.68 & 1.36 & 1.42 & 1.38 & 1.34 & 1.30 \\ 
          \hline
              0.5 & 0.5 & 1.5 &50 &  6.426 (0.287) &  1.830 (0.082) &  1.555 (0.070) &  1.418 (0.063) &  1.451 (0.065) &  1.307 (0.058) \\ 
              0.5 & 0.5 & 1.5 &100 &  6.413 (0.287) &  1.633 (0.073) &  1.618 (0.072) &  1.535 (0.069) &  1.467 (0.066) &  1.390 (0.062) \\ 
              0.5 & 0.5 & 1.5 &500 &  5.933 (0.265) &  1.652 (0.074) &  1.472 (0.066) &  1.436 (0.064) &  1.440 (0.064) &  1.399 (0.063) \\ 
              0.5 & 0.5 & 1.5 &1000 &  5.276 (0.236) &  1.449 (0.065) &  1.335 (0.060) &  1.297 (0.058) &  1.266 (0.057) &  1.230 (0.055) \\ 
              0.5 & 0.5 & 1.5 &5000 &  5.890 (0.263) &  1.470 (0.066) &  1.471 (0.066) &  1.430 (0.064) &  1.356 (0.061) &  1.312 (0.059) \\ 
              0.5 & 0.5 & 1.5 &10000 &  5.681 (0.254) &  1.551 (0.069) &  1.417 (0.063) &  1.376 (0.062) &  1.357 (0.061) &  1.315 (0.059) \\ 
              0.5 & 0.5 & 1.5 &20000 &  5.830 (0.261) &  1.403 (0.063) &  1.370 (0.061) &  1.343 (0.060) &  1.285 (0.057) &  1.259 (0.056) \\ 
              0.5 & 0.5 & 1.5 &30000 &  5.114 (0.229) &  1.468 (0.066) &  1.376 (0.062) &  1.327 (0.059) &  1.307 (0.058) &  1.257 (0.056) \\ 
              0.5 & 0.5 & 1.5 & 40000 &  5.570 (0.249) &  1.417 (0.063) &  1.375 (0.062) &  1.362 (0.061) &  1.281 (0.057) &  1.265 (0.057) \\ 
              0.5 & 0.5 & 1.5 & 50000 &  5.436 (0.243) &  1.514 (0.068) &  1.367 (0.061) &  1.308 (0.058) &  1.312 (0.059) &  1.254 (0.056) \\ 
              0.5 & 0.5 & 1.5 & Bound & 5.68 & 1.49 & 1.42 & 1.38 & 1.34 & 1.30 \\ 
          \hline
              0.5 & 1.5 & 0.5 &50 & 15.312 (0.685) &  9.666 (0.432) & 10.713 (0.479) & 10.845 (0.485) &  9.608 (0.430) &  9.719 (0.435) \\ 
              0.5 & 1.5 & 0.5 &100 & 15.181 (0.679) & 10.131 (0.453) & 11.265 (0.504) & 11.109 (0.497) & 10.133 (0.453) &  9.973 (0.446) \\ 
              0.5 & 1.5 & 0.5 &500 & 15.750 (0.704) & 10.073 (0.450) & 10.728 (0.480) & 10.692 (0.478) & 10.071 (0.450) & 10.038 (0.449) \\ 
              0.5 & 1.5 & 0.5 &1000 & 14.864 (0.665) &  9.505 (0.425) & 10.283 (0.460) & 10.230 (0.457) &  9.514 (0.425) &  9.467 (0.423) \\ 
              0.5 & 1.5 & 0.5 &5000 & 14.993 (0.670) &  9.458 (0.423) & 10.103 (0.452) & 10.076 (0.451) &  9.435 (0.422) &  9.407 (0.421) \\ 
              0.5 & 1.5 & 0.5 &10000 & 15.968 (0.714) & 10.272 (0.459) & 11.032 (0.493) & 11.054 (0.494) & 10.263 (0.459) & 10.269 (0.459) \\ 
              0.5 & 1.5 & 0.5 &20000 & 14.384 (0.643) &  9.409 (0.421) & 10.162 (0.454) & 10.133 (0.453) &  9.409 (0.421) &  9.387 (0.420) \\ 
              0.5 & 1.5 & 0.5 &30000 & 14.059 (0.629) &  9.619 (0.430) & 10.251 (0.458) & 10.171 (0.455) &  9.594 (0.429) &  9.525 (0.426) \\ 
              0.5 & 1.5 & 0.5 &40000 & 14.855 (0.664) & 10.279 (0.460) & 10.742 (0.480) & 10.707 (0.479) & 10.225 (0.457) & 10.194 (0.456) \\ 
              0.5 & 1.5 & 0.5 &50000 & 15.777 (0.706) & 10.702 (0.479) & 11.724 (0.524) & 11.714 (0.524) & 10.718 (0.479) & 10.698 (0.478) \\ 
              0.5 & 1.5 & 0.5 &Bound & 14.77 & 9.81 & 10.51 & 10.46 & 9.79 & 9.75 \\ 
          \hline
              0.5 & 1.5 & 1.5 &50 & 16.013 (0.716) & 11.043 (0.494) & 11.679 (0.522) & 11.491 (0.514) & 10.618 (0.475) & 10.428 (0.466) \\ 
              0.5 & 1.5 & 1.5 &100 & 14.036 (0.628) &  9.950 (0.445) & 10.490 (0.469) & 10.422 (0.466) &  9.733 (0.435) &  9.658 (0.432) \\ 
              0.5 & 1.5 & 1.5 &500 & 15.254 (0.682) &  9.825 (0.439) & 10.498 (0.469) & 10.466 (0.468) &  9.681 (0.433) &  9.650 (0.432) \\ 
              0.5 & 1.5 & 1.5 &1000 & 15.080 (0.674) &  9.835 (0.440) & 10.576 (0.473) & 10.627 (0.475) &  9.744 (0.436) &  9.776 (0.437) \\ 
              0.5 & 1.5 & 1.5 &5000 & 15.244 (0.682) &  9.873 (0.442) & 10.517 (0.470) & 10.464 (0.468) &  9.688 (0.433) &  9.646 (0.431) \\ 
              0.5 & 1.5 & 1.5 &10000 & 14.444 (0.646) & 10.283 (0.460) & 10.677 (0.477) & 10.609 (0.474) & 10.016 (0.448) &  9.936 (0.444) \\ 
              0.5 & 1.5 & 1.5 &20000 & 14.214 (0.636) &  9.800 (0.438) & 10.313 (0.461) & 10.251 (0.458) &  9.651 (0.432) &  9.594 (0.429) \\ 
              0.5 & 1.5 & 1.5 &30000 & 14.794 (0.662) & 10.000 (0.447) & 10.864 (0.486) & 10.852 (0.485) &  9.916 (0.443) &  9.894 (0.442) \\ 
              0.5 & 1.5 & 1.5 &40000 & 14.125 (0.632) & 10.111 (0.452) & 10.530 (0.471) & 10.357 (0.463) &  9.927 (0.444) &  9.796 (0.438) \\ 
              0.5 & 1.5 & 1.5 &50000 & 15.302 (0.684) & 10.595 (0.474) & 11.506 (0.515) & 11.402 (0.510) & 10.469 (0.468) & 10.374 (0.464) \\
              0.5 & 1.5 & 1.5 &Bound & 14.77 & 9.94 & 10.51 & 10.46 & 9.79 & 9.75 \\ 
          \hline
              1.5 & 0.5 & 0.5 & 50 &  6.335 (0.283) & 12.389 (0.554) & 11.640 (0.521) &  3.205 (0.143) & 11.588 (0.518) &  3.109 (0.139) \\ 
              1.5 & 0.5 & 0.5 & 100 &  5.759 (0.258) &  8.319 (0.372) &  7.971 (0.356) &  2.814 (0.126) &  7.834 (0.350) &  2.735 (0.122) \\ 
              1.5 & 0.5 & 0.5 & 500 &  5.661 (0.253) &  9.692 (0.433) &  9.179 (0.410) &  2.863 (0.128) &  9.078 (0.406) &  2.813 (0.126) \\ 
              1.5 & 0.5 & 0.5 & 1000 &  5.601 (0.250) & 11.823 (0.529) & 11.022 (0.493) &  2.824 (0.126) & 10.956 (0.490) &  2.747 (0.123) \\ 
              1.5 & 0.5 & 0.5 & 5000 &  5.529 (0.247) &  9.894 (0.442) &  9.123 (0.408) &  2.612 (0.117) &  9.098 (0.407) &  2.553 (0.114) \\ 
              1.5 & 0.5 & 0.5 & 10000 &  5.726 (0.256) & 10.033 (0.449) &  9.706 (0.434) &  2.907 (0.130) &  9.614 (0.430) &  2.824 (0.126) \\ 
              1.5 & 0.5 & 0.5 & 20000 &  5.507 (0.246) &  9.660 (0.432) &  9.347 (0.418) &  2.704 (0.121) &  9.246 (0.413) &  2.634 (0.118) \\ 
              1.5 & 0.5 & 0.5 & 30000 &  6.050 (0.271) & 10.399 (0.465) & 10.123 (0.453) &  2.977 (0.133) & 10.021 (0.448) &  2.846 (0.127) \\ 
              1.5 & 0.5 & 0.5 & 40000 &  5.706 (0.255) & 10.420 (0.466) &  9.657 (0.432) &  2.696 (0.121) &  9.634 (0.431) &  2.645 (0.118) \\ 
              1.5 & 0.5 & 0.5 & 50000 &  5.963 (0.267) &  9.884 (0.442) &  9.420 (0.421) &  2.941 (0.132) &  9.368 (0.419) &  2.819 (0.126) \\ 
              1.5 & 0.5 & 0.5 & Bound & 5.68 & 10.04 & 9.62 & 2.78 & 9.54 & 2.70 \\ 
          \hline
              1.5 & 0.5 & 1.5 & 50 &  6.187 (0.277) & 10.940 (0.489) &  7.917 (0.354) &  2.886 (0.129) &  7.873 (0.352) &  2.824 (0.126) \\ 
              1.5 & 0.5 & 1.5 & 100 &  5.670 (0.254) & 11.330 (0.507) &  7.438 (0.333) &  2.866 (0.128) &  7.325 (0.328) &  2.731 (0.122) \\ 
              1.5 & 0.5 & 1.5 & 500 &  5.553 (0.248) & 12.964 (0.580) &  9.817 (0.439) &  2.815 (0.126) &  9.664 (0.432) &  2.706 (0.121) \\ 
              1.5 & 0.5 & 1.5 & 1000 &  5.638 (0.252) & 13.473 (0.603) &  9.432 (0.422) &  2.898 (0.130) &  9.228 (0.413) &  2.783 (0.124) \\ 
              1.5 & 0.5 & 1.5 & 5000 &  5.371 (0.240) & 13.357 (0.597) &  8.835 (0.395) &  2.763 (0.124) &  8.783 (0.393) &  2.717 (0.122) \\ 
              1.5 & 0.5 & 1.5 & 10000 &  5.561 (0.249) & 13.753 (0.615) &  9.749 (0.436) &  2.513 (0.112) &  9.669 (0.432) &  2.408 (0.108) \\ 
              1.5 & 0.5 & 1.5 & 20000 &  6.154 (0.275) & 12.880 (0.576) &  8.541 (0.382) &  2.816 (0.126) &  8.455 (0.378) &  2.746 (0.123) \\ 
              1.5 & 0.5 & 1.5 & 30000 &  5.905 (0.264) & 14.398 (0.644) &  9.406 (0.421) &  2.880 (0.129) &  9.357 (0.418) &  2.801 (0.125) \\ 
              1.5 & 0.5 & 1.5 & 40000 &  5.514 (0.247) & 13.076 (0.585) &  9.138 (0.409) &  2.776 (0.124) &  9.055 (0.405) &  2.695 (0.121) \\ 
              1.5 & 0.5 & 1.5 & 50000 &  5.405 (0.242) & 14.109 (0.631) &  9.279 (0.415) &  2.687 (0.120) &  9.237 (0.413) &  2.570 (0.115) \\ 
              1.5 & 0.5 & 1.5 & Bound&  5.68 & 14.05 & 9.62 & 2.78 & 9.54 & 2.70 \\ 
          \hline
              1.5 & 1.5 & 0.5 & 50 & 15.001 (0.671) & 15.897 (0.711) & 17.403 (0.778) & 12.843 (0.574) & 15.903 (0.711) & 11.325 (0.506) \\ 
              1.5 & 1.5 & 0.5 & 100 & 15.940 (0.713) & 17.196 (0.769) & 16.254 (0.727) & 12.756 (0.570) & 15.559 (0.696) & 11.954 (0.535) \\ 
              1.5 & 1.5 & 0.5 & 500 & 14.892 (0.666) & 19.569 (0.875) & 19.746 (0.883) & 12.032 (0.538) & 19.108 (0.855) & 11.215 (0.502) \\ 
              1.5 & 1.5 & 0.5 & 1000 & 14.416 (0.645) & 18.103 (0.810) & 18.076 (0.808) & 11.297 (0.505) & 17.437 (0.780) & 10.941 (0.489) \\ 
              1.5 & 1.5 & 0.5 & 5000 & 14.247 (0.637) & 18.685 (0.836) & 18.673 (0.835) & 11.922 (0.533) & 18.022 (0.806) & 11.229 (0.502) \\ 
              1.5 & 1.5 & 0.5 & 10000 & 15.814 (0.707) & 18.454 (0.825) & 18.679 (0.835) & 12.317 (0.551) & 18.012 (0.806) & 11.659 (0.521) \\ 
              1.5 & 1.5 & 0.5 & 20000 & 14.832 (0.663) & 16.081 (0.719) & 16.479 (0.737) & 11.058 (0.495) & 15.729 (0.703) & 10.459 (0.468) \\ 
              1.5 & 1.5 & 0.5 & 30000 & 15.315 (0.685) & 17.978 (0.804) & 18.019 (0.806) & 11.598 (0.519) & 17.492 (0.782) & 11.080 (0.496) \\ 
              1.5 & 1.5 & 0.5 & 40000 & 14.014 (0.627) & 20.114 (0.900) & 20.081 (0.898) & 11.747 (0.525) & 19.536 (0.874) & 11.193 (0.501) \\ 
              1.5 & 1.5 & 0.5 & 50000 & 14.008 (0.626) & 18.055 (0.807) & 18.619 (0.833) & 11.650 (0.521) & 17.604 (0.787) & 10.826 (0.484) \\
              1.5 & 1.5 & 0.5 &Bound & 14.77 & 18.50 & 18.71 & 11.86 & 18.00 & 11.15 \\ 
          \hline
              1.5 & 1.5 & 1.5 & 50 & 14.774 (0.661) & 16.934 (0.757) & 15.112 (0.676) & 11.568 (0.517) & 14.623 (0.654) & 10.807 (0.483) \\ 
              1.5 & 1.5 & 1.5 &100 & 15.778 (0.706) & 21.271 (0.951) & 18.720 (0.837) & 12.671 (0.567) & 17.205 (0.769) & 11.316 (0.506) \\ 
              1.5 & 1.5 & 1.5 & 500 & 15.286 (0.684) & 23.220 (1.038) & 18.696 (0.836) & 12.504 (0.559) & 18.217 (0.815) & 11.887 (0.532) \\ 
              1.5 & 1.5 & 1.5 &1000 & 15.036 (0.672) & 21.690 (0.970) & 19.233 (0.860) & 11.943 (0.534) & 18.030 (0.806) & 10.773 (0.482) \\ 
              1.5 & 1.5 & 1.5 &5000 & 15.595 (0.697) & 22.336 (0.999) & 19.361 (0.866) & 12.253 (0.548) & 18.594 (0.832) & 11.430 (0.511) \\ 
              1.5 & 1.5 & 1.5 &10000 & 14.362 (0.642) & 24.340 (1.089) & 19.288 (0.863) & 11.654 (0.521) & 18.727 (0.837) & 11.036 (0.494) \\ 
              1.5 & 1.5 & 1.5 &20000 & 15.077 (0.674) & 22.801 (1.020) & 18.733 (0.838) & 12.145 (0.543) & 17.919 (0.801) & 11.586 (0.518) \\ 
              1.5 & 1.5 & 1.5 &30000 & 15.012 (0.671) & 23.535 (1.053) & 19.487 (0.871) & 12.112 (0.542) & 18.880 (0.844) & 11.509 (0.515) \\ 
              1.5 & 1.5 & 1.5 &40000 & 13.836 (0.619) & 23.352 (1.044) & 18.700 (0.836) & 11.515 (0.515) & 18.055 (0.807) & 10.907 (0.488) \\ 
              1.5 & 1.5 & 1.5 &50000 & 14.747 (0.660) & 21.892 (0.979) & 18.650 (0.834) & 11.895 (0.532) & 17.583 (0.786) & 11.127 (0.498) \\ 
              1.5 & 1.5 & 1.5 & Bound & 14.77 & 22.50 & 18.71 & 11.86 & 18.00 & 11.15 \\ 
           \hline
        \end{tabular}
        \begin{tablenotes} 
            \item {\small *Monte Carlo SE of scaled empirical variance $ns_{K-1}^2$ is calculated as $\sqrt{2n^2s_{K-1}^4/K}$ (\citealp{boos2013essential}, p. 370)}
        \end{tablenotes}
        \caption{Scaled empirical variance (Monte Carlo SEs in parentheses) of estimators in simulation study 1.}
        \label{table: scaled_empirical_variance_sim_study1}
    \end{threeparttable}
\end{table}

\begin{table}[H]
    \setlength{\tabcolsep}{1pt}
    \tiny
    \centering
    \begin{threeparttable}
        \begin{tabular}{rlrrrrrr}
          \hline
            Measure* & Misspecification & BD & FD & TD & BD TD & FD TD & BD FD TD \\ 
          \hline
            Bias & 1.p(Z$|$A) & -0.0003(0.001) & -0.0005(0.001) & -0.0005(0.000) & -0.0005(0.000) & -0.0005(0.000) & -0.0005(0.000) \\ 
              Bias & 2.All except p(Z$|$A) &  0.3656(0.001) & -0.0027(0.002) & -0.0027(0.001) & -0.0479(0.000) & -0.0027(0.001) & -0.0479(0.000) \\ 
              Bias & 3.p(C), p(Z$|$A) & -0.0003(0.001) & -0.0005(0.001) & -0.0005(0.000) & -0.0005(0.000) & -0.0907(0.001) & -0.0907(0.001) \\ 
              Bias & 4.p(A$|$C), E(Y$|$Z,C) & -0.0004(0.001) & -0.0010(0.001) & -0.0027(0.001) & -0.0479(0.000) & -0.0027(0.001) & -0.0479(0.000) \\ 
              EmpSE & 1.p(Z$|$A) &  0.0171(0.000) &  0.0168(0.000) &  0.0154(0.000) &  0.0154(0.000) &  0.0152(0.000) &  0.0152(0.000) \\ 
              EmpSE & 2.All except p(Z$|$A) &  0.0182(0.000) &  0.0550(0.001) &  0.0461(0.001) &  0.0148(0.000) &  0.0461(0.001) &  0.0148(0.000) \\ 
              EmpSE & 3.p(C), p(Z$|$A) &  0.0171(0.000) &  0.0168(0.000) &  0.0154(0.000) &  0.0154(0.000) &  0.0160(0.000) &  0.0160(0.000) \\ 
              EmpSE & 4.p(A$|$C), E(Y$|$Z,C) &  0.0171(0.000) &  0.0216(0.000) &  0.0461(0.001) &  0.0148(0.000) &  0.0461(0.001) &  0.0148(0.000) \\ 
              MSE & 1.p(Z$|$A) &  0.0003(0.000) &  0.0003(0.000) &  0.0002(0.000) &  0.0002(0.000) &  0.0002(0.000) &  0.0002(0.000) \\ 
              MSE & 2.All except p(Z$|$A) &  0.1340(0.000) &  0.0030(0.000) &  0.0021(0.000) &  0.0025(0.000) &  0.0021(0.000) &  0.0025(0.000) \\ 
              MSE & 3.p(C), p(Z$|$A) &  0.0003(0.000) &  0.0003(0.000) &  0.0002(0.000) &  0.0002(0.000) &  0.0085(0.000) &  0.0085(0.000) \\ 
              MSE & 4.p(A$|$C), E(Y$|$Z,C) &  0.0003(0.000) &  0.0005(0.000) &  0.0021(0.000) &  0.0025(0.000) &  0.0021(0.000) &  0.0025(0.000) \\ 
              ScEmpVar & 1.p(Z$|$A) &  14.5695(0.652) &  14.0761(0.630) &  11.8535(0.530) &  11.8535(0.530) &  11.5470(0.516) &  11.5470(0.516) \\ 
              ScEmpVar & 2.All except p(Z$|$A) &  16.6099(0.743) & 151.0885(6.757) & 106.2572(4.752) &  11.0084(0.492) & 106.2572(4.752) &  11.0084(0.492) \\ 
              ScEmpVar & 3.p(C), p(Z$|$A) &  14.5695(0.652) &  14.0761(0.630) &  11.8535(0.530) &  11.8535(0.530) &  12.8098(0.573) &  12.8098(0.573) \\ 
             ScEmpVar & 4.p(A$|$C), E(Y$|$Z,C) &  14.5601(0.651) &  23.2271(1.039) & 106.2572(4.752) &  11.0084(0.492) & 106.2572(4.752) &  11.0084(0.492) \\ 
              Bound &  & 14.765 & 22.5 & 18.71 & 11.864 & 17.995 & 11.15 \\ 
           \hline
        \end{tabular}
        \begin{tablenotes} 
            \item {\small*Bias: estimated bias, EmpSE: empirical standard error, MSE: mean squared error. Scaled empirical variance, $\text{ScEmpVar}$ = $n\text{EmpSE}^2 = ns_{K-1}^2.$  Monte Carlo SE of scaled empirical variance $ns_{K-1}^2$ is calculated as $\sqrt{2n^2s_{K-1}^4/K}$ (\citealp{boos2013essential}, p. 370). } 
        \end{tablenotes}
        \caption{Estimates of performance for measures of interest (Monte Carlo SEs in parentheses) in simulation study 2.}
        \label{table: results_sim_study2}
    \end{threeparttable}
\end{table}

\section{Supplementary Material}
\renewcommand{\thesection}{Supplementary Material  S.\arabic{section}}

\subsection{Semiparametric efficiency bound under Assumption \hyperref[ass:fd]{\textbf{FD}}} \label{sec:appendix_proof_lemma_FD}
The semiparametric efficiency bound under  the assumptions of Section \ref{sec:notation} and Assumption \textbf{FD} is equal to $var\varphi_{fd}.$ Since $E\varphi_{fd} = 0, $ and $\{A=a^*\} \cap \{A=a\} = \emptyset,$
\begin{flalign*}
 &var\varphi_{fd}= E\varphi_{fd}^2\\
    & = E\alpha^2+E\beta^2+E\gamma^2+E\delta^2+2E(\alpha\beta)+2E(\alpha\gamma)+2E(\alpha\delta)+2E(\beta\delta)+2E(\gamma\delta), \text{ where}\\
    &\alpha  = \left(Y-E(Y|A,Z)\right) \frac{p(Z|a^*)-p(Z|a)}{p(Z|A)}, \nonumber \\
    &\beta  = \frac{I(A=a^*)}{p(A)}\left(\sum_{\bar{a}} E(Y|\bar{a}, Z)p(\bar{a})-EY(a^*)\right),  \nonumber \\
    &\gamma  = -\frac{I(A=a)}{p(A)}\left(\sum_{\bar{a}} E(Y|\bar{a}, Z)p(\bar{a})-EY(a)\right),\nonumber\\
    & \delta  = \sum_{z}E(Y|A, z)p(z|a^*)-\sum_{z}E(Y|A, z)p(z|a)-\theta.  &
\end{flalign*}
Let's consider each term  in $var\varphi_{fd}$ separately.  Note that the  terms that include $\gamma$ can be obtained from the respective terms with $\beta$ by substituting $a^*$ by $a.$
\begin{flalign*}
    E&\alpha^2=E\left[\left(Y^2-2YE(Y|A,Z)+E^2(Y|A,Z)\right) \frac{(p(Z|a^*)-p(Z|a))^2}{p^2(Z|A)}\right]\\
        & = \sum_z (p(z|a^*)-p(z|a))^2 \sum\limits_{\bar{a}} \frac{p(\bar{a})}{p(z|\bar{a})}\big( E(Y^2|\bar{a},z)- 2E^2(Y|\bar{a},z) + E^2(Y|\bar{a},z)\big)\\
        & = \sum_z (p(z|a^*)-p(z|a))^2  \sum\limits_{\bar{a}}\frac{p(\bar{a})}{p(z|\bar{a})}var(Y|\bar{a},z).\\
    E&\beta^2=\sum\limits_{z}\frac{\left(\sum\limits_{\bar{a}} E(Y|\bar{a}, z)p(\bar{a})-EY(a^*)\right)^2}{p^2(a^*)}p(a^*, z)\\
        &=\frac{\sum\limits_{z}p(z|a^*)\left(\sum\limits_{\bar{a}} E(Y|\bar{a}, z)p(\bar{a})\right)^2}{p(a^*)} -
        \frac{2EY(a^*)\sum\limits_{z}p(z|a^*)\sum\limits_{\bar{a}} E(Y|\bar{a}, z)p(\bar{a})}{p(a^*)} \\
        & + \frac{E^2[Y(a^*)]\sum\limits_{z}p(z|a^*)}{p(a^*)} =\frac{\sum\limits_{z}p(z|a^*)\left(\sum\limits_{\bar{a}} E(Y|\bar{a}, z)p(\bar{a})\right)^2}{p(a^*)} - \frac{E^2[Y(a^*)]}{p(a^*)}.&
\end{flalign*}
The last equality follows from the front-door identification of $EY(a^*).$
\begin{flalign*}
    E&\delta^2 =\sum\limits_{\bar{a}}p(\bar{a})\left(\sum_{z}E(Y|\bar{a}, z)(p(z|a^*)-p(z|a))-\theta\right)^2\\
        &=\sum\limits_{\bar{a}}p(\bar{a})\left(\sum_{z}E(Y|\bar{a}, z)(p(z|a^*)-p(z|a))\right)^2 \\
        & - 2\theta\sum\limits_{\bar{a}}p(\bar{a})\left(\sum_{z}E(Y|\bar{a}, z)(p(z|a^*)-p(z|a))\right)+\theta^2\\
        &=\sum\limits_{\bar{a}}p(\bar{a})\left(\sum_{z}E(Y|\bar{a}, z)(p(z|a^*)-p(z|a))\right)^2-\theta^2.&
\end{flalign*}
The last equality follows from the front-door identification of $EY(a^*).$
\begin{flalign*}
    E&(\alpha\beta) = \sum\limits_{\bar{z}} \sum_y\frac{p(a^*, \bar{z}, y)\left(y - E(Y|a^*, \bar{z})\right)p(\bar{z}|a^*)-p(\bar{z}|a)}{p(\bar{z}|a^*)}\frac{\sum\limits_{\bar{a}} E(Y|\bar{a}, \bar{z})p(\bar{a})-EY(a^*)}{p(a^*)}\\
        & =  0 &
\end{flalign*}
since $\sum_{y}(y-E(Y|a^*,\bar{z}))p(y|\bar{z},a^*)=0. $ Similarly,
\begin{flalign*}E&\alpha\delta=\sum\limits_{\bar{a},\bar{z}}p(\bar{a})p(\bar{z}|\bar{a})\sum\limits_{y}\left(y-E(Y|\bar{a}, \bar{z})\right)p(y|\bar{z},\bar{a}) \frac{p(\bar{z}|a^*)-p(\bar{z}|a)}{p(\bar{z}|\bar{a})} \times\\
        & \times \left(\sum_{z}E(Y|\bar{a}, z)p(z|a^*)-\sum_{z}E(Y|\bar{a}, z)p(z|a)-\theta\right) = 0.\\
    E&\beta\delta =\textstyle{\sum\limits_{\bar{z}}p(a^*) p(\bar{z}|a^*)\frac{\left(\sum_{\bar{a}} E(Y|\bar{a}, \bar{z})p(\bar{a})-EY(a^*)\right)\left(\sum_{z}E(Y|a^*,z) p(z|a^*)-\sum_{z}E(Y|a^*,z)p(z|a)-\theta\right)}{p(a^*)}}\\
        &= 0 \text{ since} \sum\limits_{\bar{z}} p(\bar{z}|a^*)\sum_{\bar{a}} E(Y|\bar{a}, \bar{z})p(\bar{a})-EY(a^*)=0.& 
\end{flalign*}
The expression for $var \varphi_{fd}$ is obtained by plugging in all the terms in $var \varphi_{fd}$ and is equal to
    \begin{flalign}
         & var\varphi_{fd}=\sum_{z} (p(z|a^*)-p(z|a))^2  \sum\limits_{\bar{a}}\frac{p(\bar{a})}{p(z|\bar{a})}var(Y|\bar{a},z) \nonumber \\
         & + \sum\limits_{z}\left(\sum\limits_{\bar{a}} E(Y|\bar{a}, z) p(\bar{a})\right)^2 \left(\frac{p(z|a^*)}{p(a^*)} + \frac{p(z|a)}{p(a)}\right)\nonumber\\
         & -\frac{\big(\sum\limits_{\bar{a}, z} p(z|a^*) E(Y|\bar{a}, z) p(\bar{a})\big)^2}{p(a^*)}
         -\frac{\big(\sum\limits_{\bar{a}, z} p(z|a) E(Y|\bar{a},z) p(\bar{a})\big)^2}{p(a)}\nonumber\\
        &+\sum\limits_{\bar{a}}p(\bar{a})\left(\sum_{z}E(Y|\bar{a}, z) (p(z|a^*)-p(z|a))\right)^2-\theta^2.&
    \end{flalign}

\newpage

\subsection{Semiparametric efficiency bounds under Assumption \hyperref[ass:td]{\textbf{TD}}}\label{sec:appendix_proof_var_eif_td}
Under Assumption \hyperref[ass:td]{\textbf{TD}}, Theorem 1 in \citet{fulcher2020robust} provides the efficient influence function for $EY(a^*)$.  The  expression for $\varphi_{td}$ in Equation \eqref{eq:eif_td} follows immediately due to the linearity of the differentiation operation.  The semiparametric efficiency bound for $\theta$ is equal to $var\varphi_{td}.$ 
\begin{flalign*}
     v& ar\varphi_{td}  = E\alpha^2+E\beta^2+E\gamma^2+E\delta^2+2E(\alpha\beta)+2E(\alpha\gamma)+2E(\alpha\delta)+2E(\beta\delta)+2E(\gamma\delta),
\end{flalign*}
where 
\begin{flalign*}
        &\alpha  = \left(Y-E(Y|A, Z, C)\right) \frac{p(Z|a^*,C)-p(Z|a,C)}{p(Z|A,C)}, \\
        &\beta = \frac{I(A=a^*)}{p(A|C)}\left(\sum_{\bar{a}}E(Y|\bar{a}, Z, C)p(\bar{a}|C)-\sum_{\bar{a},z}E(Y|\bar{a}, z, C)p(z|A,C)p(\bar{a}|C)\right),\\\
        &\gamma  =-\frac{I(A=a)}{p(A|C)}\left(\sum_{\bar{a}}E(Y|\bar{a}, Z, C)p(\bar{a}|C)-\sum_{\bar{a},z}E(Y|\bar{a}, z, C)p(z|A,C)p(\bar{a}|C)\right),\\\
        &\delta = \sum\limits_z E(Y|A, z, C)\big(p(z|a^*, C)- p(z|a, C)\big)-\theta, &
\end{flalign*}
and since $E(\beta\gamma)=0$ ($\{A=a^*\} \cap \{A=a\} = \emptyset$).
Let's consider each term in $var\varphi_{td}.$  Note that the  terms that include $\gamma$ can be obtained from the respective terms with $\beta$ by substituting $a^*$ by $a.$
\begin{flalign*}
    E&\alpha^2=E\left[\left(Y^2-2YE(Y|A,Z,C)+E^2(Y|A,Z,C)\right) \frac{(p(Z|a^*,C)-p(Z|a,C))^2}{p^2(Z|A,C)}\right]\\
        & = \sum_{z,c} (p(z|a^*,c)-p(z|a,c))^2 \sum\limits_{\bar{a}} \frac{p(\bar{a},c)}{p(z|\bar{a},c)}  
            \big [ E(Y^2|\bar{a},z,c)- E^2(Y|\bar{a},z,c) \big]\\
        & = \sum_{z,c} (p(z|a^*,c)-p(z|a,c))^2  \sum\limits_{\bar{a}}\frac{p(\bar{a},c)}{p(z|\bar{a},c)}var(Y|\bar{a},z,c).\\
    E&\beta^2=\sum\limits_{\bar{z},c}\frac{\left(\sum_{\bar{a}}E(Y|\bar{a}, \bar{z}, c)p(\bar{a}|c)-\sum_{\bar{a},z}E(Y|\bar{a}, z, c)p(z|a^*,c)p(\bar{a}|c)\right)^2}{p^2(a^*|c)}p(\bar{z}, a^*, c)\\
        &=\sum\limits_{\bar{z},c}\frac{\left(\sum_{\bar{a}}E(Y|\bar{a}, \bar{z}, c)p(\bar{a}|c)\right)^2p(\bar{z}|a^*,c)p(c)}{p(a^*|c)} \\
        &-2 \sum\limits_{\bar{z},c}\frac{\sum_{a'}E(Y|a', \bar{z}, c)p(a'|c)*\sum_{\bar{a},z}E(Y|\bar{a}, z, c)p(z|a^*,c)p(\bar{a}|c)}{p^2(a^*|c)}p(\bar{z}, a^*, c)\\
        & + \sum\limits_{c}\frac{\left(\sum_{\bar{a},z}E(Y|\bar{a}, z, c)p(z|a^*,c)p(\bar{a}|c)\right)^2p(c)}{p(a^*|c)} \\
        &=\sum\limits_{z,c}\frac{\left(\sum_{\bar{a}}E(Y|\bar{a}, z, c)p(\bar{a}|c)\right)^2p(z|a^*,c)p(c)}{p(a^*|c)} \\
        & -\sum\limits_{c}\frac{\left(\sum_{\bar{a},z}E(Y|\bar{a}, z, c)p(z|a^*,c)p(\bar{a}|c)\right)^2p(c)}{p(a^*|c)}.\\
    E&\delta^2 =\sum\limits_{\bar{a},c}\left(\sum\limits_z E(Y|\bar{a}, z, c)\big(p(z|a^*, c)- p(z|a, c)\big)-\theta\right)^2p(\bar{a},c)\\
        &=\sum\limits_{\bar{a},c}\left(\sum\limits_z E(Y|\bar{a}, z, c)\big(p(z|a^*, c)- p(z|a, c)\big)\right)^2p(\bar{a},c)\\
        & - 2\theta\sum\limits_{\bar{a},c}\left(\sum\limits_z E(Y|\bar{a}, z, c)\big(p(z|a^*, c)- p(z|a, c)\big)\right)p(\bar{a},c)+\theta^2\\
        &=\sum\limits_{\bar{a},c}\left(\sum\limits_z E(Y|\bar{a}, z, c)\big(p(z|a^*, c)- p(z|a, c)\big)\right)^2p(\bar{a},c)-\theta^2. &
    \end{flalign*}
    Here, the last equality follows from the two-door adjustment.
    \begin{flalign*}
    E&(\alpha\beta) = \sum\limits_{\bar{z},c, y} \frac{\left(y-E(Y|a^*, \bar{z} ,c)\right) \left(p(\bar{z}|a^*,c)-p(\bar{z}|a,c)\right)}{p(\bar{z}|a^*,c)}\\
        & \times\frac{\left(\sum_{\bar{a}}E(Y|\bar{a}, z, c)p(\bar{a}|c)-\sum_{\bar{a},z}E(Y|\bar{a}, z, c)p(z|a^*,c)p(\bar{a}|c)\right)}{p(a^*|c)}p(y, \bar{z}, a^*, c)\\
        & =  \sum\limits_{\bar{z},c}\left(E(Y|a^*, \bar{z} ,c)-E(Y|a^*, \bar{z} ,c)\right)  (p(\bar{z}|a^*,c)-p(\bar{z}|a,c)) \\
        & \times \left(\sum_{\bar{a}}E(Y|\bar{a}, z, c)p(\bar{a}|c)-\sum_{\bar{a},z}E(Y|\bar{a}, z, c)p(z|a^*,c)p(\bar{a}|c)\right)p(c) = 0 &
    \end{flalign*}
    since $\sum_{y}(y-E(Y|a^*, \bar{z}, c))p(y|\bar{z},a^*,c)=E(Y|a^*, \bar{z}, c)-E(Y|a^*, \bar{z}, c)=0.$ Similarly to $E(\alpha\beta),$
    \begin{flalign*}
    E&(\alpha\delta)= \sum\limits_{\bar{a},\bar{z},c, y}\left(y-E(Y|\bar{a}, \bar{z}, c)\right)p(y|\bar{z},\bar{a},c) 
        \frac{p(\bar{z}|a^*,c) - p(\bar{z}|a,c)}{p(\bar{z}|\bar{a},c)} \times\\
        & \times \left(\sum_{z}E(Y|\bar{a},  z,  c)(p(z|a^*,c)-p(z|a,c))-\theta\right)p(\bar{z}|\bar{a},c)p(\bar{a},c)=0. \\
    E&(\beta\delta) =\sum\limits_{\bar{z}, c} 
        \frac{\left(\sum_{\bar{a}} E(Y|\bar{a}, \bar{z}, c)p(\bar{a}|c)-\sum_{\bar{a},z}E(Y|\bar{a}, z, c) p(z|a^*,c) p(\bar{a}|c)\right)
        }{
        p(a^*|c)}\\
        &\times\left(\sum_{z}E(Y|a^*,  z',  c)(p(z'|a^*,c)-p(z'|a,c))-\theta\right) p(\bar{z}|a^*,c) p(a^*|c) p(c)= 0 &
\end{flalign*}
since $\sum\limits_{\bar{a},\bar{z}}E(Y|\bar{a}, \bar{z}, c) p(\bar{z}|a^*,c)p(\bar{a}|c)-\sum\limits_{\bar{a}, z} E(Y|\bar{a}, z, c) p(z|a^*,c)p(\bar{a}|c)\sum\limits_{\bar{z}}p(\bar{z}|a^*, c) = 0.$
The semiparametric efficiency bound  for $\theta$ is equal to 
    \begin{align} 
        & var \varphi_{td}=\sum_{z,c} (p(z|a^*,c)-p(z|a,c))^2  \sum\limits_{\bar{a}}\frac{p(\bar{a},c)}{p(z|\bar{a},c)}var(Y|\bar{a},z,c)\nonumber \\
        &+\sum\limits_{z,c}\frac{p(z|a^*,c)\left(\sum_{\bar{a}}E(Y|\bar{a}, z, c)p(\bar{a}|c)\right)^2p(c)}{p(a^*|c)}
        +\sum\limits_{z,c}\frac{p(z|a,c)\left(\sum_{\bar{a}}E(Y|\bar{a}, z, c)p(\bar{a}|c)\right)^2p(c)}{p(a|c)} \nonumber\\
        & -\sum\limits_{c}\left( \frac{\left(\sum_{\bar{a},z}E(Y|\bar{a}, z, c)p(z|a^*,c)p(\bar{a}|c)\right)^2}{p(a^*|c)}
        + \frac{\left(\sum_{\bar{a},z}E(Y|\bar{a}, z, c)p(z|a,c)p(\bar{a}|c)\right)^2}{p(a|c)}\right)p(c) \nonumber \\
        &+\sum\limits_{\bar{a},c}\left(\sum\limits_{z} E(Y|\bar{a}, z, c)\big(p(z|a^*, c)- p(z|a, c)\big)\right)^2p(\bar{a},c)-\theta^2.  
    \end{align}

\subsection{Proof of Proposition \ref{th:fd_vs_td}}\label{sec:appendix_proof_proposition_fd_vs_td}
\begin{proof}
    Under Assumptions \textbf{FD}, \textbf{TD}, and $Z(a)\indep C|A$,  $p(y|a,z)$ can be represented as $p(y|a,z) = \frac{\sum_c p(c)p(a|c)p(y|a,z,c)}{p(a)}.$ 
    Consider the expression for $var \varphi_{fd}$ in Equation \eqref{eq:var_eif_fd}. If $E(Y|A, Z, C) = \gamma_0 + \gamma_1Z + \gamma_2C+ \gamma_3A,$
    \begin{flalign*}
        & p(\bar{a})var(Y|\bar{a},z) = p(\bar{a})\sum_y y^2 p(y|\bar{a}, z) -  p(\bar{a})\left(\sum_y y p(y|\bar{a}, z) \right)^2.&
    \end{flalign*}
    Plugging in $p(y|a,z)$ and $E(Y|A, Z, C)$ provides 
    \begin{flalign*}
    & p(\bar{a})var(Y|\bar{a},z) =
        \sum_c p(\bar{a},c)E (Y^2|\bar{a},z,c) -\frac{ \left(\sum_c p(\bar{a},c)E(Y|\bar{a}, z, c))\right)^2} {p(\bar{a})}\\
        & = \sum_c p(\bar{a},c)\left( var(Y|\bar{a},z,c) +  (\gamma_0 + \gamma_1z+ \gamma_2c + \gamma_3\bar{a})^2 \right) \\
        & - \frac{\left(\sum_c p(\bar{a},c) (\gamma_0 + \gamma_1z+ \gamma_2c + \gamma_3\bar{a}))\right)^2}{p(\bar{a})} \\
        & = \sum_c p(\bar{a},c)\left( var(Y|\bar{a},z,c) +  (\gamma_0 + \gamma_1z+ \gamma_2c + \gamma_3\bar{a})^2 \right) \\
        & - p(\bar{a})\left(\gamma_0 + \gamma_1z + \gamma_2E(C|\bar{a}) + \gamma_3\bar{a})\right)^2 
            = \sum_c p(\bar{a},c)var(Y|\bar{a},z,c)  + \gamma_2^2 var(C|\bar{a})p(\bar{a}).&
    \end{flalign*}
    Similarly,
    \begin{flalign*}
        & \sum\limits_{z}\left(\sum\limits_{\bar{a}} E(Y|\bar{a}, z)p(\bar{a})\right)^2 \frac{p(z|a^*)}{p(a^*)} -\frac{\big(\sum\limits_{\bar{a}, z} p(z|a^*) E(Y|\bar{a}, z) p(\bar{a})\big)^2}{p(a^*)}\\
        & = \frac{\sum\limits_{z} \left(\sum\limits_{\bar{a}, c}  p(\bar{a},c)(\gamma_0 + \gamma_1z + \gamma_2c+ \gamma_3\bar{a})\right)^2 p(z|a^*)}{p(a^*)}
            - \frac{(\gamma_0 + \gamma_1E(Z|a^*) + \gamma_2 EC + \gamma_3EA)^2}{p(a^*)}\\
        & = \sum\limits_{z} \left(\gamma_0 + \gamma_1z + \gamma_2EC + \gamma_3EA\right)^2\frac{p(z|a^*)}{p(a^*)}
            - \frac{(\gamma_0 + \gamma_1E(Z|a^*) + \gamma_2 EC + \gamma_3EA)^2}{p(a^*)} \\
        &  = \frac{\gamma_1^2 var(Z|a^*)}{p(a^*)}.\\
        & \sum\limits_{\bar{a}}p(\bar{a})\left(\sum_{z}E(Y|\bar{a}, z)(p(z|a^*)-p(z|a))\right)^2- \theta^2 \\
        & = \sum_{\bar{a},c}p(\bar{a}, c)\gamma_1^2 (E(Z|a^*) - E(Z|a))^2 - \gamma_1^2(E(Z|a^*) - E(Z|a))^2 = 0. 
        \end{flalign*}
    Therefore,
    \begin{flalign*}
        & var\varphi_{fd}= \sum_{z} (p(z|a^*)-p(z|a))^2  \sum\limits_{\bar{a}, c}\frac{ p(\bar{a},c) }{p(z|\bar{a})}var(Y|\bar{a},z,c) \\
        &  + \sum_{z} (p(z|a^*)-p(z|a))^2 \sum_{\bar{a}}\frac{\gamma_2^2 var(C|\bar{a})p(\bar{a})}{p(z|\bar{a})} + \frac{\gamma_1^2 var(Z|a^*)}{p(a^*)} + \frac{\gamma_1^2 var(Z|a)}{p(a)}.&
    \end{flalign*}
    $var \varphi_{td}$ from Equation \eqref{eq:var_eif_td} can be simplified using the fact that
    \begin{flalign*}
        & \sum\limits_{z,c}\frac{p(z|a^*)\left(\sum_{\bar{a}}E(Y|\bar{a}, z, c)p(\bar{a}|c)\right)^2p(c)}{p(a^*|c)}          -\sum\limits_{c}\frac{\left(\sum_{\bar{a},z}E(Y|\bar{a}, z, c)p(z|a^*)p(\bar{a}|c)\right)^2p(c)}{p(a^*|c)}\\
        & = \sum\limits_{z,c}p(c)\frac{p(z|a^*)}{p(a^*|c)} (\gamma_0 + \gamma_1z + \gamma_2c+\gamma_3E(A|c))^2 \\
        &   - \sum\limits_{c}\frac{p(c)(\gamma_0 + \gamma_1 E(Z|a^*) + \gamma_2C + \gamma_3 E(A|c))^2}{p(a^*|c)}
            = \sum\limits_{c}\frac{p(c)}{p(a^*|c)} \gamma_1^2var(Z|a^*)&
     \end{flalign*}
    and
    \begin{flalign*}   
        & \sum\limits_{\bar{a},c}\left(\sum_{z}E(Y|\bar{a}, z, c)(p(z|a^*)-p(z|a))\right)^2p(\bar{a},c)- \theta^2 = 0.\\
        & var \varphi_{td} = \sum_{z} (p(z|a^*)-p(z|a))^2          \sum\limits_{\bar{a},c}\frac{p(\bar{a},c)}{p(z|\bar{a})}var(Y|\bar{a},z,c) \\
        & + \sum\limits_{c}\frac{p(c)}{p(a^*|c)} \gamma_1^2var(Z|a^*) + \sum\limits_{c}\frac{p(c)}{p(a|c)} \gamma_1^2var(Z|a). &
    \end{flalign*}
    Therefore, $var \varphi_{fd} - var \varphi_{td}$ is equal to
    \begin{flalign*}
        & var \varphi_{fd} - var \varphi_{td} =  \sum_{z} (p(z|a^*)-p(z|a))^2 \sum_{\bar{a}}\frac{\gamma_2^2 var(C|\bar{a})p(\bar{a})}{p(z|\bar{a})} \\ 
        & + \gamma_1^2 var(Z|a^*)\left( \frac{1}{p(a^*)} - \sum\limits_{c}\frac{p(c)}{p(a^*|c)}\right) + \gamma_1^2 var(Z|a)\left(\frac{1}{p(a)} - \sum\limits_{c}\frac{p(c)}{p(a|c)}\right). &
    \end{flalign*}
    Since the first term   $\sum_{z} (p(z|a^*)-p(z|a))^2 \sum_{\bar{a}}\frac{\gamma_2^2 var(C|\bar{a})p(\bar{a})}{p(z|\bar{a})}$ is always nonnegative, the following condition for $var \varphi_{fd} \geq var \varphi_{td}$ can be provided.
    When 
    \begin{align*}
        \frac{1}{p(a^*)} \geq \sum\limits_{c}\frac{p(c)}{p(a^*|c)} \text{ and }\frac{1}{p(a)} \geq\sum\limits_{c}\frac{p(c)}{p(a|c)}, 
    \end{align*}
    all the terms in $var \varphi_{fd} - var \varphi_{td}$ are nonnegative, and $var \varphi_{fd} \geq var \varphi_{td}.$
\end{proof}

\newpage

\subsection{Proof of Proposition \ref{th:fd_td_eif}} \label{sec:appendix_proof_th_fd_td_eif}
\begin{proof}
    Since  $ Z(a) \indep C |A,$ $p(z|a,c) = p(z|a)$ and
    the density of the observed data can be expressed as
    $p(c, a,z,y) = p(y|a,z,c)p(z|a)p(a|c)p(c).$ From Theorem 4.5  in \citet{tsiatis2007semiparametric}, the projection of any influence function $h$ on the tangent space is
    $E(h|Y,A,Z,C) - E(h|A,Z,C) + E(h|Z, A) - E(h|A) +   E(h|A, C) - E(h|C) + E(h|C) - Eh .$
    
    The projection of $\varphi_{td}$ onto the tangent space is then 
    \begin{flalign*}
    &\varphi_{bd, td} = E(\varphi_{td}|Y,A, Z,C) - E(\varphi_{td}|A,Z,C) 
        + E(\varphi_{td}|Z, A) - E(\varphi_{td}| A) +
        E(\varphi_{td}| A, C). \\
    & E(\varphi_{td}|Y,A, Z,C) - E(\varphi_{td}|A,Z,C) = \left(Y-E(Y|A,  Z, C)\right) \frac{p(Z|a^*)-p(Z|a)}{p(Z|A)} &
    \end{flalign*}
    since all other terms  in $\varphi_{td}$ are functions of $A, Z, C$ only. From the two-door adjustment of $\theta$ and since $\sum_y (y - E(Y|A,Z,C))p(y|A,Z) = 0$ and   $\sum_y (y - E(Y|A,Z,C))p(y|A) = 0,$
    \begin{flalign*}
    & E(\varphi_{td}| Z, A)  - E(\varphi_{td} | A)  = \sum_c\left(\sum\limits_{\bar{a}}E(Y|\bar{a}, Z, c)p(\bar{a}|c)
        - \sum\limits_{\bar{a}, z}E(Y|\bar{a}, z, c)p(z|A)p(\bar{a}|c)\right) \times \\
    & \times \frac{(I(A=a^*) - I(A=a))p(c)}{p(A)}.
     \end{flalign*}
    Similarly, $E(\varphi_{td}| A, C) = \sum\limits_z E(Y|A, z, C)\big(p(z|a^*)- p(z|a )\big)-\theta. $
    $var\varphi_{fd, td} = E\varphi_{fd, td}^2 = E\alpha^2 + E\beta^2 + E\gamma^2 + E\delta^2,$
    where 
    \begin{flalign*}
        &\alpha = \left(Y-E(Y|A,  Z, C)\right) \frac{p(Z|a^*)-p(Z|a)}{p(Z|A,C)}\\
        &\beta = \sum_c\left(\sum\limits_{\bar{a}}E(Y|\bar{a}, Z, c)p(\bar{a}|c)
        -\sum\limits_{\bar{a}, z}E(Y|\bar{a}, z, c)p(\bar{a}|c)p(z|A)\right) \frac{I(A=a^*) p(c)}{p(A)}\\
        &\gamma = \sum_c\left(\sum\limits_{\bar{a}}E(Y|\bar{a}, Z, c)p(\bar{a}|c)
        -\sum\limits_{\bar{a}, z}E(Y|\bar{a}, z, c)p(\bar{a}|c)p(z|A)\right) \frac{I(A=a) p(c)}{p(A)}\\
        &\delta = \sum\limits_z E(Y|A, z, C)\big(p(z|a^*)- p(z|a)\big)-\theta.&
    \end{flalign*}
    and since $E\alpha\beta =  E\alpha\gamma = E\alpha\delta =  E\beta\gamma =  E\beta\delta = E\gamma\delta = 0$ similar to the proof in \ref{sec:appendix_proof_var_eif_td}.  
    \begin{flalign*}
       & E\beta^2 = \sum\limits_z \left( \sum_c \left(\sum\limits_{\bar{a}}E(Y|\bar{a}, z, c)p(\bar{a}|c)-\sum_{\bar{a},z}E(Y|\bar{a}, z, c)p(z|a^*)p(\bar{a}|c)\right) \frac{p(c)}{p(a^*)}\right)^2 \times \\
       & \times p(z|a^*)p(a^*)\\
       & = \sum\limits_z \frac{p(z|a^*)}{p(a^*)}  \left(\sum\limits_{\bar{a},c}E(Y|\bar{a}, z, c)p(\bar{a}|c)p(c)-\sum_{\bar{a},z,c}E(Y|\bar{a}, z, c)p(z|a^*)p(\bar{a}|c)p(c)\right)^2\\
       & = \sum\limits_z \frac{p(z|a^*)}{p(a^*)}\left(\sum\limits_{\bar{a},c}E(Y|\bar{a}, z, c)p(\bar{a}|c)p(c)\right)^2  - \frac{\left(\sum_{\bar{a},z,c}E(Y|\bar{a}, z, c)p(z|a^*)p(\bar{a}|c)p(c)\right)^2}{p(a^*)}
    \end{flalign*}
    $E\gamma^2$ can be obtained from $E\beta^2$ by substituting $a^*$ for $a.$ The expression for $E\alpha^2$ and $E\delta^2$ can be found similar to the expressions for the corresponding terms in \ref{sec:appendix_proof_var_eif_td}.
\end{proof}

\subsection{Proof of Proposition \ref{th:fd_vs_bd}}\label{sec:appendix_proof_proposition_fd_vs_bd}
\begin{proof}
     The observed data distribution under the considered assumptions is  $p(c,a,z,y) = p(c)p(a|c)p(z|a)p(y|z,c)$ and $E(Y(a^*)|C)   = \sum_z p(z|a^*)E(Y|Z, C).$ 
    
    If additionally $E(Y|Z, C) = \gamma_0 + \gamma_1Z+ \gamma_2C,$  $var \varphi_{bd}$ can be simplified as follows:
    \begin{flalign*}
        & var\big[Y(a^*)|C]= \sum_z p(z|a^*)E(Y^2|z, C) - \left(\sum_z p(z|a^*)E(Y|z, C)\right)^2\\
        & = \sum_z p(z|a^*)var(Y|Z,C) +  \sum_z p(z|a^*))(\gamma_0 + \gamma_1Z+ \gamma_2C)^2 \\
        &   - \left(\sum_z p(z|a^*)(\gamma_0 + \gamma_1Z+ \gamma_2C)\right)^2 
            = \sum_z p(z|a^*)var(Y|z,C) + \gamma_1^2 var(Z|a^*).\\
        &var \varphi_{bd}  = E\left[\frac{var\big[Y(a^*)|C\big]}{p(a^*|C)}+\frac{var\big[Y(a)|C\big]}{p(a|C)}\right] + E\bigg(E\big[Y(a^*)|C\big]-E\big[Y(a)|C\big]-\theta\bigg)^2\\
        & =  \sum_ c \left( \frac{\sum_z p(z|a^*)var(Y|z,c) + \gamma_1^2 var(Z|a^*)}{p(a^*|c)}p(c)\right) \\
            & + \sum_c\left( \frac{\sum_z p(z|a)var(Y|z,c) + \gamma_1^2 var(Z|a)}{p(a|c)}p(c) \right) \\
            & + \gamma_0 + \gamma_1E(Z|a^*)+ \gamma_2C - \gamma_0 - \gamma_1E(Z|a)- \gamma_2C - \gamma_1(E(Z|a^*) - E(Z|a))\\
        & = \sum_ c \left( \frac{\sum_z p(z|a^*)var(Y|z,c) + \gamma_1^2 var(Z|a^*)}{p(a^*|c)}p(c)\right) \\
            & + \sum_c\left( \frac{\sum_z p(z|a)var(Y|z,c) + \gamma_1^2 var(Z|a)}{p(a|c)}p(c) \right). 
    \end{flalign*}
    Note that this expression holds even if the linearity assumption is not fulfilled, but \\
    $E(Y|Z, C) = f(Z) + g(C)$ for some $f$ and $g.$ Similar to \ref{sec:appendix_proof_proposition_fd_vs_td},
    \begin{flalign*}
        & var\varphi_{fd}= \sum_{z} (p(z|a^*)-p(z|a))^2  \sum\limits_{\bar{a}, c}\frac{ p(\bar{a},c) }{p(z|\bar{a})}var(Y|z,c)\\
        & + \sum_{z} (p(z|a^*) - p(z|a))^2 \sum_{\bar{a}}\frac{\gamma_2^2 var(C|\bar{a})p(\bar{a})}{p(z|\bar{a})} 
            + \frac{\gamma_1^2 var(Z|a^*)}{p(a^*)} + \frac{\gamma_1^2 var(Z|a)}{p(a)}.\\
        & var\varphi_{fd} - var\varphi_{bd} = \sum_{z,c}var(Y|z,c)p(c) \left(\sum_{\bar{a}}\frac{p(\bar{a}|c)(p(z|a^*)-p(z|a))^2}{p(z|\bar{a})}-\frac{p(z|a^*)}{p(a^*|c)}-\frac{p(z|a)}{p(a|c)}\right)\\
        & + \sum_{z} (p(z|a^*)-p(z|a))^2 \sum_{\bar{a}}\frac{\gamma_2^2var(C|\bar{a})p(\bar{a})}{p(z|\bar{a})}\\
        & +\gamma_1^2 var(Z|a^*)\left( \frac{1}{p(a^*)} - \sum\limits_{c}\frac{p(c)}{p(a^*|c)}\right) + \gamma_1^2 var(Z|a)\left(\frac{1}{p(a)} - \sum\limits_{c}\frac{p(c)}{p(a|c)}\right).
    \end{flalign*}
    The proof is completed by noting that  when all the terms in the summations are nonnegative, $var\varphi_{fd} - var\varphi_{bd} \geq 0.$ If all the terms are not positive, $var\varphi_{fd} - var\varphi_{bd} \leq 0.$
\end{proof}

\subsection{Proof of Proposition \ref{th:bd_fd_td_eif}} \label{sec:appendix_proof_lemma_fd_td_bd}
\begin{proof}
    Under the considered assumptions, the density of the observed data can be expressed as
    $p(c, a,z,y) = p(c)p(a|c)p(z|a)p(y|z,c).$
    Similar to Appendix \ref{sec:proof of lemma EIF TD BD} and from Theorem 4.5 in \citet{tsiatis2007semiparametric}, the projection of  influence function $\varphi_{td}$ onto the tangent space is then
    \begin{flalign*}
        & \varphi_{bd, fd, td} = E(\varphi_{td}|Y,Z,C) -E(\varphi_{td}|Z,C) + E(\varphi_{td}|Z, A) - E(\varphi_{td}|A)  + E(\varphi_{td}| A,C)\\
        & = E(\alpha(Y, Z, A, C)|Y, Z, C) - E(\alpha(Y, Z, A, C)| Z, C) \\
        & + E \big(\alpha( Z, A, C) | Z, A \big)  - E\big(\alpha( Z, A, C) | A \big)  + \alpha(C), &
    \end{flalign*}
where 
    \begin{flalign*}
         &\alpha(Y, Z, A, C)  = \big(Y-E(Y|Z, C)\big) \frac{p(Z|a^*,C)-p(Z|a,C)}{p(Z|A,C)}\\
         &\alpha(Z, A, C)  = \frac{I(A=a^*)-I(A=a)}{p(A|C)}\left(E(Y|Z, C)-\sum_{z}E(Y|z,C)p(z|A,C)\right) \\
         & \alpha(C) =  E(Y(a^*)|C)  -E(Y(a)| C)  - \theta \stackrel{\eqref{eq:s2}}{=} \sum\limits_{z} E(Y|z,C)(p(z|a^*, c) - p(z|a)) - \theta.& 
    \end{flalign*}
    Consider $E(\alpha(Y, Z, A, C)|Y, Z, C) - E(\alpha(Y, Z, A, C)| Z, C).$
    \begin{flalign*}
     & E(\alpha(Y, Z, A, C)|Y, Z, C) - E(\alpha(Y, Z, A, C)| Z, C)  \\
     &  = \left(Y - E(Y|Z, C)\right) (p(Z|a^*)-p(Z|a))\sum_a\frac{p(a|Y, Z, C)}{ p(Z|a)} \\
     &  = \left(Y - E(Y|Z, C)\right) (p(Z|a^*)-p(Z|a))\sum_a\frac{p(C)p(a|C)p(Z|a)p(Y|Z,C)}{ p(Z|a)p(Y|Z, C)p(Z, C)} \\
     &  = \frac{\left(Y - E(Y|Z, C)\right) (p(Z|a^*)-p(Z|a))}{\sum_a p(a|C)p(Z|a)} ,&
    \end{flalign*}
    Consider the term in $E \big(\alpha( Z, A, C) | Z, A \big)  - E\big(\alpha( Z, A, C) | A \big)$ that corresponds to $a^*:$
    \begin{flalign*} 
       & \sum_{c}\frac{I(A=a^*)}{p(A|c)}\left(E(Y|Z,c)-\sum_{z}E(Y|z,c)p(z|A,c)\right)p(c|A,Z) \\
       & - \sum_{z, c}\frac{I(A=a^*)}{p(A|c)}E(Y|z,c)p(z, c|A)  +\sum_{ c,z}\frac{I(A=a^*)}{p(A|c)}E(Y|z,c)p(z|A,c)p(c|A) \\
      & = \sum_{c}\frac{I(A=a^*)}{p(A|c)}E(Y|Z,c)p(c|A) -  \sum_{z, c}\frac{I(A=a^*)}{p(A|c)}E(Y|z,c)p(z, c|A)\\
      & = \frac{I(A=a^*)}{\sum_c{p(c)p(A|c)}} \sum_{c}p(c)\left(E(Y|Z,c) - \sum_z E(Y|z,c)p(z|A)\right),&
    \end{flalign*}
    where the second last equality follows from $Z(a) \indep C|A,$ and the last equality follows from $ p(c|A) = \frac{p(c)p(A|c)}{\sum_c{p(c)p(A|c)}}$ and $p(c,z|A) =  \frac{p(c)p(A|c)p(z|A)}{\sum_c{p(c)p(A|c)}}.$    Therefore, 
    \begin{flalign*}
        &E \big(\alpha( Z, A, C) | Z, A \big)  - E\big(\alpha( Z, A, C) | A \big) \\
        & = \sum_{c} \left(E(Y|Z,c) - \sum_z E(Y|z,c)p(z|A)\right)\frac{(I(A=a^*)-I(A=a))p(c)}{\sum_c{p(c)p(A|c)}}. &
    \end{flalign*}
    $\varphi_{fd, bd, td}$ is exactly the same as the efficient influence function provided by \citet{rotnizky2020efficient} under the data distribution compatible with the DAG  defined by paths $C \rightarrow A \rightarrow Z \rightarrow Y,$ $C \rightarrow Y.$
    \begin{flalign*}
        v&ar\varphi_{fd, bd, td} = E\varphi_{fd, bd, td}^2 = E\alpha^2 + E\beta^2 + E\gamma^2 + E\delta^2, \text{ where} \\
        & \alpha = \frac{\left(Y-E(Y|Z, C)\right) (p(Z|a^*)-p(Z|a))}{\sum_a p(a|C)p(Z|a)}\\
        & \beta = \frac{I(A=a^*)}{\sum_c{p(c)p(A|c)}} \left(\sum_{c}p(c)E(Y|Z,c) - \sum_{z,c} E(Y|z,c)p(z|A)p(c)\right)\\
        & \gamma = \frac{I(A=a)}{\sum_c{p(c)p(A|c)}} \left(\sum_{c}p(c)E(Y|Z,c) - \sum_{z,c} E(Y|z,c)p(z|A)p(c)\right)\\
        & \delta = E(Y(a^*)|C)  -E(Y(a)| C)  -\theta,&
    \end{flalign*}
    and since $E\alpha\beta =  E\alpha\gamma = E\alpha\delta =  E\beta\gamma =  E\beta\delta = E\gamma\delta = 0$ similar to $var \varphi_{td}$ in \ref{sec:appendix_proof_var_eif_td}.
    \begin{flalign*}
        & E\alpha^2 = \sum_z p(z|a^*)-p(z|a))^2 \sum_c\frac{p(c)\sum_a p(a|c)p(z|a)}{(\sum_a p(a|C)p(Z|a))^2}var(Y|z,c)\\
            & = \sum_z ( p(z|a^*)-p(z|a))^2 \sum_c\frac{p(c)var(Y|z,c)}{\sum_a p(a|c)p(Z|a)}\\
        & E\beta^2 = \sum_z \left(\frac{\left(\sum_{c}p(c)E(Y|z,c) - \sum_{z,c} E(Y|z,c)p(z|a^*)p(c)\right)}{\sum_c{p(c)p(a^*|c)}}    \right)^2p(a^*)p(z|a^*)\\
            & = \sum_z \frac{p(z|a^*)}{p(a^*)} \left(\sum_{c}p(c)E(Y|z,c) - \sum_{z,c} E(Y|z,c)p(z|a^*)p(c)   \right)^2 \\
            & = \sum_z \frac{p(z|a^*)}{p(a^*)} \left(\sum_{c}p(c)E(Y|z,c)\right)^2 - \frac{\left( \sum_{z,c} E(Y|z,c)p(z|a^*)p(c) \right)^2}{p(a^*)}\\
        & E\gamma^2 = \sum_z \frac{p(z|a)}{p(a)} \left(\sum_{c}p(c)E(Y|z,c)\right)^2 - \frac{\left( \sum_{z,c} E(Y|z,c)p(z|a)p(c) \right)^2}{p(a)}\\
        & E\delta^2 = \sum_c p(c)\left(\sum_{z}E(Y|z,c) (p(z|a^*)- p(z|a))\right)^2  - \theta^2. &
    \end{flalign*}
\end{proof}

\newpage
\subsection{Proof of \autoref{th:efficiency_of_fd_and_td_est}}\label{sec:appendix_proof_lemma_efficiency_of_td_and_fd}
\begin{proof}
According to Appendix \ref{sec:appendix_efficiency_conditions}, we first show that $\mathbb{P} m_{td}(X, \hat{\eta}) - \theta = o_p(1/\sqrt{n})$ and then show that $\mathbb{P} m_{fd}(X, \hat{\eta}) - \theta  = o_p(1/\sqrt{n}).$
\begin{flalign*}
    & \mathbb{P} m_{td}(X, \hat{\eta}) - \theta   \\
    & =  \sum_{c, \bar{a}, z} 
            \left( E(Y|\bar{a}, z, c) - \hat{E} (Y|\bar{a}, z, c)
            \right) 
            \frac{ \hat{p}(z|a^*, c) - \hat{p}(z|a, c)
             }{\hat{p}(z|\bar{a}, c)
             } p(z|\bar{a}, c) p(\bar{a}|c) p(c)\\
    & + \sum_{c, \bar{a}, z} \hat{E} (Y|\bar{a}, z, c) 
        \left( p(z|a^*, c) - \hat{p}(z|a^*, c) 
        \right) 
        \frac{ p(a^*|c) }{ \hat{p}(a^*|c) 
             } \hat{p} (\bar{a}|c) p(c) \\
    & - \sum_{c, \bar{a}, z} \hat{E} (Y|\bar{a}, z, c) 
        \left( p(z|a, c) - \hat{p}(z|a, c) 
        \right) 
        \frac{ p(a|c) }{ \hat{p}(a|c) 
             } \hat{p} (\bar{a}|c) p(c) \\
    & + \sum_{c, \bar{a}, z} \hat{E}(Y|\bar{a}, z, c) 
        \big(\hat{p}(z|a^*, c) - \hat{p}(z|a, c)
        \big)  p (\bar{a}|c) p(c) \\
    & -  \sum_{c, \bar{a}, z} E(Y|\bar{a}, z, c) 
        \big(p(z|a^*, c) - p(z|a, c) 
        \big) p (\bar{a}|c) p(c)\\
    & \pm \sum_{c, \bar{a}, z} 
            \left( E(Y|\bar{a}, z, c) - \hat{E} (Y|\bar{a}, z, c)
            \right) 
            \big(\hat{p}(z|a^*, c) - \hat{p}(z|a, c)
            \big)  p(\bar{a}|c) p(c)\\
    & \pm \sum_{c, \bar{a}, z} \hat{E} (Y|\bar{a}, z, c) 
        \left( p(z|a^*, c) - \hat{p}(z|a^*, c) 
        \right) 
        \hat{p} (\bar{a}|c) p(c)\\
    & \pm \sum_{c, \bar{a}, z} \hat{E} (Y|\bar{a}, z, c) 
        \left( p(z|a, c) - \hat{p}(z|a, c) 
        \right) 
        \hat{p} (\bar{a}|c) p(c) &
\end{flalign*}
Here, the third last line will be combined with the first line,  the second last line will be combined with the second line, the last line will be combined with the third line.
\begin{flalign*}
    &  \mathbb{P} m_{td}(X, \hat{\eta}) - \theta\\
    & =   \sum_{c, \bar{a}, z} 
            \left( E(Y|\bar{a}, z, c) - \hat{E} (Y|\bar{a}, z, c)
            \right) 
            \big(\hat{p}(z|a^*, c) - \hat{p}(z|a, c)
            \big)
            \frac{p(z|\bar{a}, c) - \hat{p}(z|\bar{a}, c) 
             }{\hat{p}(z|\bar{a}, c)
             }  p(\bar{a}|c) p(c)\\
    & +  \sum_{c, \bar{a}, z} \hat{E} (Y|\bar{a}, z, c) 
        \left( p(z|a^*, c) - \hat{p}(z|a^*, c) 
        \right) 
        \frac{ p(a^*|c) - \hat{p}(a^*|c) }{ \hat{p}(a^*|c) 
             } \hat{p} (\bar{a}|c) p(c) \\
    & -  \sum_{c, \bar{a}, z} \hat{E} (Y|\bar{a}, z, c) 
        \left( p(z|a, c) - \hat{p}(z|a, c) 
        \right) 
        \frac{ p(a|c) - \hat{p}(a|c) }{ \hat{p}(a|c) 
             } \hat{p} (\bar{a}|c) p(c) \\
    & +  \sum_{c, \bar{a}, z} \hat{E}(Y|\bar{a}, z, c) 
        \big(\hat{p}(z|a^*, c) - \hat{p}(z|a, c)
        \big) p(\bar{a}|c) p(c) \\
    & - \sum_{c, \bar{a}, z} E(Y|\bar{a}, z, c) 
        \big(p(z|a^*, c) - p(z|a, c) 
        \big) p(\bar{a}|c) p(c) \\
    & +  \sum_{c, \bar{a}, z} 
            \left( E(Y|\bar{a}, z, c) - \hat{E} (Y|\bar{a}, z, c)
            \right) 
            \big(\hat{p}(z|a^*, c) - \hat{p}(z|a, c)
            \big)  p(\bar{a}|c) p(c) \\
    & +  \sum_{c, \bar{a}, z} \hat{E} (Y|\bar{a}, z, c) 
        \left( p(z|a^*, c) - \hat{p}(z|a^*, c) 
        \right) 
        \hat{p} (\bar{a}|c) p(c) \\
     & -  \sum_{c, \bar{a}, z} \hat{E} (Y|\bar{a}, z, c) 
        \left( p(z|a, c) - \hat{p}(z|a, c) 
        \right) 
        \hat{p} (\bar{a}|c) p(c) \\
    & \pm \sum_{c, \bar{a}, z}\hat{E}(Y|\bar{a}, z, c) (p(z|a^*, c) - p(z|a,c) p(\bar{a}|c)) \\
    & = \mathbb{P} 
     \left[ \left( E(Y|A, Z, C) - \hat{E} (Y|A, Z, C)
            \right) 
            \big(p(Z|A, C) - \hat{p}(Z|A, C) 
            \big)
            \frac{\hat{p}(Z|a^*, C) - \hat{p}(Z|a, C) 
             }{\hat{p}(Z|A, C)p(Z|A, C)
             } 
     \right] \\
    & + \mathbb{P} 
     \left[  \hat{E} (Y|A, Z, C)
            \big(p(Z|a^*, C) - \hat{p}(Z|a^*, C) 
            \big)
            \frac{ p(a^*|C) - \hat{p}(a^*|C) }{ \hat{p}(a^*|C) p(Z|A, C)p(A|C)}
     \right] \\
    & - \mathbb{P} 
     \left[  \hat{E} (Y|A, Z, C)
            \big(p(Z|a, C) - \hat{p}(Z|a, C) 
            \big)
            \frac{ p(a|C) - \hat{p}(a|C) }{ \hat{p}(a|C) p(Z|A, C)p(A|C)}
     \right] \\
    & + \mathbb{P} 
     \left[  \hat{E} (Y|A, Z, C)
            \big(p(Z|a^*, C) - \hat{p}(Z|a^*, C) 
            \big)
            \frac{ \hat{p}( A|C) - p(A|C) }{ p(Z|A,C)p(A|C) }
     \right] \\
    & - \mathbb{P} 
     \left[  \hat{E} (Y|A, Z, C)
            \big(p(Z|a, C) - \hat{p}(Z|a, C) 
            \big)
            \frac{ \hat{p}( A|C) - p(A|C) }{ p(Z|A,C)p(A|C) }
     \right] \\
     & + \mathbb{P} 
     \left[
        \big( \hat{E} (Y|A, Z, C) -  E (Y|A, Z, C)  
        \big)
        \frac{p (Z|a^*, C) - \hat{p}(Z|a^*, C)
        }{p(Z|A,C)}
    \right]\\ 
    & - \mathbb{P} 
     \left[
        \big( \hat{E} (Y|A, Z, C) -  E (Y|A, Z, C)  
        \big)
        \frac{p (Z|a, C) - \hat{p}(Z|a, C)
        }{p(Z|A,C)}
    \right]. &
\end{flalign*}
$\mathbb{P} m_{td}(X, \hat{\eta}) - \theta  = o_p(1/\sqrt{n})$  from the Cauchy-Schwartz inequality, \Cref{ass:an_td}, and the assumptions of the theorem since then each term in the summation above is $o_p(1/\sqrt{n}). $ Similarly,
\begin{flalign*}
    & \mathbb{P} m_{fd}(X, \hat{\eta}) - \theta  \\
    & = \sum_{ \bar{a}, z} 
            \left( E(Y|\bar{a}, z) - \hat{E} (Y|\bar{a}, z)
            \right) 
            \frac{ \hat{p}(z|a^*) - \hat{p}(z|a)
             }{\hat{p}(z|\bar{a})
             } p(z|\bar{a}) p(\bar{a})\\
    & + \sum_{\bar{a}, z} \hat{E} (Y|\bar{a}, z) 
        \left(
            \left( p(z|a^*) - \hat{p}(z|a^*) 
            \right) 
            \frac{ p(a^*) }{ \hat{p}(a^*) 
                } 
             - 
            \left( p(z|a) - \hat{p}(z|a) 
        \right) 
        \frac{ p(a) }{ \hat{p}(a) 
             }
        \right) \hat{p} (\bar{a})  \\
    & + \sum_{\bar{a}, z} \hat{E}(Y|\bar{a}, z) 
        \big(\hat{p}(z|a^*) - \hat{p}(z|a)
        \big) p(\bar{a})
        - \sum_{\bar{a}, z} E(Y|\bar{a}, z) 
        \big(p(z|a^*) - p(z|a) 
        \big) p(\bar{a})\\
    & = \sum_{\bar{a}, z} 
            \left( E(Y|\bar{a}, z) - \hat{E} (Y|\bar{a}, z)
            \right) 
            \big(\hat{p}(z|a^*) - \hat{p}(z|a)
            \big)
            \frac{p(z|\bar{a}) - \hat{p}(z|\bar{a}) 
             }{\hat{p}(z|\bar{a})
             }  p(\bar{a})\\
    & + \sum_{\bar{a}, z} 
            \left( E(Y|\bar{a}, z) - \hat{E} (Y|\bar{a}, z)
            \right) 
            \big(\hat{p}(z|a^*) - \hat{p}(z|a)
            \big)  p(\bar{a})\\
    & + \sum_{\bar{a}, z} \hat{E} (Y|\bar{a}, z)
        \left( 
            \left( p(z|a^*) - \hat{p}(z|a^*) 
            \right) 
            \frac{ p(a^*) - \hat{p}(a^*) }{ \hat{p}(a^*) 
             }  
             - 
             \left( p(z|a) - \hat{p}(z|a) 
            \right) 
            \frac{ p(a) - \hat{p}(a) }{ \hat{p}(a) 
             } 
        \right) \hat{p} (\bar{a})  \\
    & + \sum_{\bar{a}, z} \hat{E} (Y|\bar{a}, z) 
        \left( p(z|a^*) - \hat{p}(z|a^*) - p(z|a) + \hat{p}(z|a)
        \right) 
        \hat{p} (\bar{a})  \\
    & + \sum_{\bar{a}, z} \hat{E}(Y|\bar{a}, z) 
        \big(\hat{p}(z|a^*) - \hat{p}(z|a)
        \big) p(\bar{a}) 
      - \sum_{\bar{a}, z} E(Y|\bar{a}, z) 
        \big(p(z|a^*) - p(z|a) 
        \big) p(\bar{a})  \\
    & = \sum_{\bar{a}, z} 
            \left( E(Y|\bar{a}, z) - \hat{E} (Y|\bar{a}, z)
            \right) 
            \big(\hat{p}(z|a^*) - \hat{p}(z|a)
            \big)
            \frac{p(z|\bar{a}) - \hat{p}(z|\bar{a}) 
             }{\hat{p}(z|\bar{a})
             }  p(\bar{a})\\
    & + \sum_{\bar{a}, z} \hat{E} (Y|\bar{a}, z)
        \left( 
            \left( p(z|a^*) - \hat{p}(z|a^*) 
            \right) 
            \frac{ p(a^*) - \hat{p}(a^*) }{ \hat{p}(a^*) 
             }  
             - 
             \left( p(z|a) - \hat{p}(z|a) 
            \right) 
            \frac{ p(a) - \hat{p}(a) }{ \hat{p}(a) 
             } 
        \right) \hat{p} (\bar{a})  \\
    & + \sum_{\bar{a}, z} \hat{E} (Y|\bar{a}, z) 
        \left( p(z|a^*) - \hat{p}(z|a^*) - p(z|a) + \hat{p}(z|a)
        \right) 
        \hat{p} (\bar{a})  \\
    & + \sum_{\bar{a}, z} 
             E(Y|\bar{a}, z) 
            \big(\hat{p}(z|a^*) - \hat{p}(z|a)
      - p(z|a^*) + p(z|a) 
        \big) p(\bar{a})  \\
    & \pm \sum_{\bar{a}, z} \hat{E} (Y|\bar{a}, z) 
        \left( p(z|a^*) - \hat{p}(z|a^*) - p(z|a) + \hat{p}(z|a)
        \right) 
        p (\bar{a})  \\
    & = \mathbb{P} 
    \left[ 
        \left( E(Y|A, Z) - \hat{E} (Y|A, Z)
        \right) 
        \big(\hat{p}(Z|a^*) - \hat{p}(Z|a)
        \big)
            \frac{p(Z|A) - \hat{p}(Z|A) 
             }{\hat{p}(Z|A)p(Z|A)
             }  
    \right]\\
    & + \mathbb{P} 
    \left[ 
    \hat{E} (Y|A, Z)
            \left( p(Z|a^*) - \hat{p}(Z|a^*) 
            \right) 
            \frac{ p(a^*) - \hat{p}(a^*) }{ \hat{p}(a^*) 
             } \frac{\hat{p} (A) }{p(Z|A)p(A)}
    \right] \\
   & - \mathbb{P} 
    \left[ 
    \hat{E} (Y|A, Z)
             \left( p(Z|a) - \hat{p}(Z|a) 
            \right) 
            \frac{ p(a) - \hat{p}(a) }{ \hat{p}(a) 
             } 
         \frac{\hat{p} (A) }{p(Z|A)p(A)}
    \right] \\
    & +  \mathbb{P}
    \left[\hat{E} (Y|A, Z)
            \big(p (Z|a^*) - \hat{p}(Z|a^*)
            \big)
            \big(\hat{p} (A) - p(A)
            \big) \frac{1}{p(Z|A)p(A)}
    \right]    \\
    & -  \mathbb{P}
    \left[\hat{E} (Y|A, Z)
            \big(p (Z|a) - \hat{p}(Z|a)
            \big)
            \big(\hat{p} (A) - p(A)
            \big)\frac{1}{p(Z|A)p(A)} 
    \right]    \\
    & + \mathbb{P}
    \left[
        \big( \hat{E} (Y|A, Z) -  E (Y|A, Z) 
        \big)
        \big(p (Z|a^*) - \hat{p}(Z|a^*) - p (Z|a) + \hat{p}(Z|a)
        \big) \frac{1}{p(Z|A)}
    \right].
\end{flalign*}
$\mathbb{P} m_{fd}(X, \hat{\eta}) - \theta  = o_p(1/\sqrt{n})$  from the Cauchy-Schwartz inequality, \Cref{ass:an_fd}, and the assumptions of the theorem since each term in the summation above is then  $o_p(1/\sqrt{n}). $
\end{proof}

\subsection{Proof of \autoref{th:consistency_efficiency_of_fd_td_est}}\label{sec:appendix_proof_lemma_consistency_of_fd_td_est}
\begin{proof}

According to Appendix \ref{sec:appendix_consistency_conditions}, we prove the consistency of the estimator by showing that under each scenario $E\varphi_{fd, td}(X, \bar{\eta}, \theta_0) = 0,$ where $\theta_0$ is the true value of the average causal effect.

When $\bar{E}(Y|A, Z, C) = E(Y|A, Z, C),$
    \begin{flalign*}
        & E\varphi_{fd, td}(X, \bar{\eta}, \theta_0) = \\
        & E\bigg[\bigg. \left(Y-E(Y|A,  Z, C)\right) \frac{\bar{p}(Z|a^*)-\bar{p}(Z|a)}{\bar{p}(Z|A)}   \\
            &  + \sum_{c}\left(\sum\limits_{\bar{a}}E(Y|\bar{a}, Z, c)\bar{p}(\bar{a}|c)
                - \sum_{\bar{a},z}E(Y|\bar{a}, z, c)\bar{p}(z|A)\bar{p}(\bar{a}|c)\right) \frac{(I(A=a^*)-I(A=a)) \bar{p}(c)}{\bar{p}(A)}\\
            & + \sum\limits_{z} E(Y|A, z, C)\big(\bar{p}(z|a^*)- \bar{p}(z|a)\big)-\theta_0 \bigg. \bigg]\\
            & = 0  + \sum_{c, z} \left(\sum\limits_{\bar{a}}E(Y|\bar{a}, z, c)\bar{p}(\bar{a}|c)
                -\sum_{\bar{a},z}E(Y|\bar{a}, z, c)\bar{p}(z|a^*)\bar{p}(\bar{a}|c)\right) \frac{p(z|a^*)p(a^*)\bar{p}(c)}{\bar{p}(a^*)} \\
            & - \sum_{c, z} \left(\sum\limits_{\bar{a}}E(Y|\bar{a}, z, c)\bar{p}(\bar{a}|c)
                - \sum_{\bar{a},z}E(Y|\bar{a}, z, c)\bar{p}(z|a)\bar{p}(\bar{a}|c)\right) \frac{p(z|a)p(a)\bar{p}(c)}{\bar{p}(a)}\\
            & + \sum\limits_{ c, a, z} E(Y|a, z, c)\bar{p}(z|a^*)p(a|c)p(c)
            - \sum\limits_{c, a, z} E(Y|a, z, c)\bar{p}(z|a)p(a|c)p(c)- \theta_0\\
            & = \sum\limits_{ c, \bar{a}, z}E(Y|\bar{a}, z, c)\bar{p}(\bar{a}|c)\bar{p}(c)\left(\frac{p(a^*)}{\bar{p}(a^*)}\left(p(z|a^*)-\bar{p}(z|a^*) \right)
                - \frac{p(a)}{\bar{p}(a)}\left(p(z|a)-\bar{p}(z|a) \right)
                 \right)\\
            &  + \sum\limits_{ c, \bar{a}, z}E(Y|\bar{a}, z, c)  p(\bar{a}|c)p(c)
                \left(\bar{p}(z|a^*)  - \bar{p}(z|a) - p(z|a^*) + p(z|a) \right).
    \end{flalign*}    
    The expression above is equal to 0 when either $\bar{p}(A|C) = p(A|C)$ and $\bar{p}(C) = p(C)$  or  $\bar{p}(Z|A)=p(Z|A).$ The conditions of the theorem cover these cases. 
    
    When $\bar{p}(Z|A) = p(Z|A),$
    \begin{flalign*}
        & E\varphi_{fd, td}(X, \bar{\eta}, \theta_0) = \\
        & E\bigg[\bigg. \left(Y-\bar{E}(Y|A, Z, C)\right) \frac{p(Z|a^*)-p(Z|a)}{p(Z|A)}   \\
            &  + \sum_{c}\left(\sum\limits_{\bar{a}}\bar{E}(Y|\bar{a}, Z, c)\bar{p}(\bar{a}|c)
                - \sum_{\bar{a},z}\bar{E}(Y|\bar{a}, z, c)p(z|A)\bar{p}(\bar{a}|c)\right) \frac{(I(A=a^*)-I(A=a)) \bar{p}(c)}{\bar{p}(A)}\\
            & + \sum\limits_{z} \bar{E}(Y|A, z, C)\big(p(z|a^*)- p(z|a)\big)-\theta_0 \bigg. \bigg]\\
            & = \theta_0- \sum_{c, \bar{a}, z}\bar{E}(Y|\bar{a}, z, c)(p(z|a^*)-p(z|a))p(\bar{a}|c)p(c)\\
                & +\sum\limits_{ c, \bar{a}, z}\bar{E}(Y|\bar{a}, z, c)\bar{p}(\bar{a}|c)\bar{p}(c)\left(\frac{p(a^*)}{\bar{p}(a^*)}\left(p(z|a^*)-p(z|a^*) \right)
                - \frac{p(a)}{\bar{p}(a)}\left(p(z|a)-p(z|a) \right)
                 \right)\\
            &  + \sum\limits_{ c, \bar{a}, z}\bar{E}(Y|\bar{a}, z, c)  
                \left( p(z|a^*) - p(z|a) \right)p(\bar{a}|c)p(c) - \theta_0 = 0
    \end{flalign*}
     
   We show below that other conditions for the consistency are covered by the conditions of the theorem.
  
   When $\bar{p}(A|C) = p(A|C),$ 
    \begin{flalign*}
        & E\varphi_{fd, td}(X, \bar{\eta}, \theta_0) = \\
        & E\bigg[\bigg. \left(Y-\bar{E}(Y|A, Z, C)\right) \frac{\bar{p}(Z|a^*) - \bar{p}(Z|a)}{\bar{p}(Z|A)}   \\
            &  + \sum_{c}\left(\sum\limits_{\bar{a}}\bar{E}(Y|\bar{a}, Z, c)p(\bar{a}|c)
                - \sum_{\bar{a},z} \bar{E}(Y|\bar{a}, z, c)\bar{p}(z|A)p(\bar{a}|c)\right) \frac{(I(A=a^*)-I(A=a)) \bar{p}(c)}{\sum_{c} p(A|c) \bar{p}(c)}\\
            & + \sum\limits_{z} \bar{E}(Y|A, z, C)\big(\bar{p}(z|a^*)- \bar{p}(z|a)\big)-\theta_0 \bigg. \bigg]\\
            & = \sum_{c, z, \bar{a}}\left(E(Y|\bar{a}, z, c)-\bar{E}(Y|\bar{a}, z, c)\right)\frac{(\bar{p}(z|a^*) - \bar{p}(z|a))p(z|\bar{a})p(\bar{a}|c)p(c)}{\bar{p}(z|\bar{a})} \\
            & + \sum_{c,\bar{a} z} \bar{E}(Y|\bar{a}, z, c)p(\bar{a}|c)
            \left(p(z|a^*)  - \bar{p}(z|a^*) \right) 
            \frac{\bar{p}(c)p(a^*)}{\sum_{c} p(a^*|c)\bar{p}(c)}\\
            & - \sum_{c,\bar{a} z} \bar{E}(Y|\bar{a}, z, c)p(\bar{a}|c)
            \left(p(z|a)  - \bar{p}(z|a) \right) 
            \frac{\bar{p}(c)p(a)}{\sum_{c} p(a|c)\bar{p}(c)}\\
            & + \sum_{c,\bar{a}, z}  \left( \bar{E}(Y|\bar{a}, z, c)\big(\bar{p}(z|a^*)- \bar{p}(z|a)\big) - E(Y|\bar{a}, z,  c)\big(p(z|a^*)- p(z|a)\big)\right) p(\bar{a}|c)p(c).
    \end{flalign*}
   
    The last expression is equal to 0 when either $\bar{p}(C)  = p(C)$ and  \\
    $\bar{E}(Y|A, Z, C) = E(Y|A, Z, C)$  or 
   $\bar{p}(Z|A) = p(Z|A).$ These cases are covered by the conditions of the theorem.
    
    When $\bar{p}(C) = p(C),$
    \begin{flalign*}
        &E\bar{\varphi}_{fd, td}(X, \bar{\eta}, \theta_0) =  E\bigg[ \bigg.\left(Y-\bar{E}(Y|A, Z, C)\right) \frac{\bar{p}(Z|a^*) - \bar{p}(Z|a)}{\bar{p}(Z|A)}   \\
            &  + \sum_{c}\left(\sum\limits_{\bar{a}}\bar{E}(Y|\bar{a}, Z, c)\bar{p}(\bar{a}|c)
            - \sum_{\bar{a},z}\bar{E}(Y|\bar{a}, z, c)\bar{p}(z|A)\bar{p}(\bar{a}|c)\right) \frac{(I(A=a^*)-I(A=a)) p(c)}{\bar{p}(A)}\\
            & + \sum\limits_{z} \bar{E}(Y|A, z, C)\big(\bar{p}(z|a^*)- \bar{p}(z|a)\big)-\theta_0\bigg. \bigg]\\
            & = \sum_{c, \bar{a}, z} \left(E(Y|\bar{a}, z, c)-\bar{E}(Y|\bar{a}, z,c)\right)\frac{\bar{p}(z|a^*) - \bar{p}(z|a)}{\bar{p}(z|\bar{a})}p(z|\bar{a})p(\bar{a}|c)p(c)\\
            & + \sum_{c, z} \left(\sum\limits_{\bar{a}}\bar{E}(Y|\bar{a}, z, c)\bar{p}(\bar{a}|c)
                -\sum_{\bar{a},z}\bar{E}(Y|\bar{a}, z, c)\bar{p}(z|a^*)\bar{p}(\bar{a}|c)\right) \frac{p(z|a^*)p(a^*)p(c)}{\bar{p}(a^*)} \\
            & - \sum_{c, z} \left(\sum\limits_{\bar{a}}\bar{E}(Y|\bar{a}, z, c)\bar{p}(\bar{a}|c)
                - \sum_{\bar{a},z}\bar{E}(Y|\bar{a}, z, c)\bar{p}(z|a)\bar{p}(\bar{a}|c)\right) \frac{p(z|a)p(a)p(c)}{\bar{p}(a)}\\
            & + \sum\limits_{ c, \bar{a}, z} \bar{E}(Y|\bar{a}, z, c)\bar{p}(z|a^*)p(\bar{a}|c)p(c)
            - \sum\limits_{c, \bar{a}, z} \bar{E}(Y|\bar{a}, z, c)\bar{p}(z|a)p(\bar{a}|c)p(c)- \theta_0.
    \end{flalign*}
    The last expression is equal to 0 when $\bar{p}(Z|A) = p(Z|A)$ or  $\bar{E}(Y|A, Z, C) = E(Y|A, Z, C)$  and $\bar{p}(A|C) = p(A|C). $  The proof of the consistency is completed by noting that this situation is also covered by the conditions of the theorem. 
    
    According to Appendix \ref{sec:appendix_efficiency_conditions}, to show the efficiency of the estimator,  we provide conditions for  $\mathbb{P}  m_{fd,td}(X, \hat{\eta}) - \theta = o_p(n^{-1/2}).$
    \begin{flalign*}
        \mathbb{P}  m_{fd, td}(X_i, \hat{\eta}) - \theta = f(a^*) - f(a),&
    \end{flalign*}
    where 
    \begin{flalign*}
        f & (\breve{a}) = 
            \sum_{c, \bar{a}, z}
                \left( E(Y|\bar{a},z,c) - \hat{E}(Y|\bar{a},z,c) 
                \right) 
                \frac{ \hat{p}(z|\breve{a})
                    }{ \hat{p}(z|\bar{a})
                } 
                p(z| \bar{a}) p( \bar{a}| c) p(c)  \\
            & + \sum_{c, \bar{a}, z} \hat{E}(Y|\bar{a}, z, c) 
                \left( p(z|\breve{a}) - \hat{p}(z|\breve{a}) 
                \right) 
                \hat{p}(\bar{a}|c) \hat{p}(c) 
                \frac{ p(\breve{a})
                    }{ \hat{p}(\breve{a})
                }  \\
            & + \sum_{c, \bar{a}, z} 
                \left( \hat{E}(Y|\bar{a}, z, c) \hat{p}(z|\breve{a}) 
                    - E(Y|\bar{a}, z, c) p(z|\breve{a})
                \right) 
                p(\bar{a}|c) p(c) \\
            & = \sum_{c, \bar{a}, z} 
                \left( E(Y|\bar{a},z,c) - \hat{E}(Y|\bar{a},z,c) 
                \right) 
                \left( 
                    \frac{p(z| \bar{a})
                        }{\hat{p}(z|\bar{a})
                    }  - 1
                \right)
                \hat{p}(z|\breve{a}) p( \bar{a}| c) p(c) \\
            & + \sum_{c, \bar{a}, z} 
                \left( E(Y|\bar{a},z,c) - \hat{E}(Y|\bar{a},z,c) 
                \right) 
                \hat{p}(z|\breve{a}) p( \bar{a}| c) p(c) \\
            & + \sum_{c, \bar{a}, z} \hat{E}(Y|\bar{a}, z, c)
                \left( p(z|\breve{a}) - \hat{p}(z|\breve{a}) 
                \right) 
                \hat{p}(\bar{a}|c) \hat{p}(c) 
                \left( 
                    \frac{ p(\breve{a})
                        }{ \hat{p}(\breve{a})
                    } - 1
                \right) \\
            & + \sum_{c, \bar{a}, z} \hat{E}(Y|\bar{a}, z, c) 
                \left( p(z|\breve{a}) - \hat{p}(z|\breve{a}) 
                \right) 
                \hat{p}(\bar{a}|c) \hat{p}(c) \\
            & + \sum_{c, \bar{a}, z} 
                \left( \hat{E}(Y|\bar{a}, z, c) \hat{p}(z|\breve{a}) 
                    - E(Y|\bar{a}, z, c) p(z|\breve{a})
                \right) 
                p(\bar{a}|c) p(c).&
    \end{flalign*}
   The equality above is obtained by addition and subtraction of\\  $\sum_{c, \bar{a}, z} 
                \left( E(Y|\bar{a},z,c) - \hat{E}(Y|\bar{a},z,c) 
                \right) 
                \hat{p}(z|\breve{a}) p( \bar{a}| c) p(c)$ and \\$\sum_{c, \bar{a}, z} \hat{E}(Y|\bar{a}, z, c) 
                \left( p(z|\breve{a}) - \hat{p}(z|\breve{a}) 
                \right) 
                \hat{p}(\bar{a}|c) \hat{p}(c).$ Rearranging allows us to write
    \begin{flalign*}
        & f(\breve{a}) = \\
            & = \sum_{c, \bar{a}, z} 
                \left( E(Y|\bar{a},z,c) - \hat{E}(Y|\bar{a},z,c) 
                \right) 
                \left( 
                    \frac{p(z| \bar{a}) - \hat{p}(z|\bar{a}) 
                        }{\hat{p}(z|\bar{a})
                        } 
                \right)
                \hat{p}(z|\breve{a}) p( \bar{a}| c) p(c) \\  
            & + \sum_{c, \bar{a}, z} \hat{E}(Y|\bar{a}, z, c)
                \left( p(z|\breve{a}) - \hat{p}(z|\breve{a}) 
                \right) 
                \hat{p}(\bar{a}|c) \hat{p}(c)  
                \frac{ 
                    p(\breve{a}) - \hat{p}(\breve{a}) 
                    }{ \hat{p}(\breve{a})
                }                  \\
            &  + \sum_{c, \bar{a}, z} 
                \left( E(Y|\bar{a},z,c) - \hat{E}(Y|\bar{a},z,c) 
                \right) 
                \hat{p}(z|\breve{a}) p( \bar{a}| c) p(c) \\
            & + \sum_{c, \bar{a}, z} \hat{E}(Y|\bar{a}, z, c) 
                \left( p(z|\breve{a}) - \hat{p}(z|\breve{a}) 
                \right) 
                \hat{p}(\bar{a}|c) \hat{p}(c) \\
            & + \sum_{c, \bar{a}, z} 
                \left( \hat{E}(Y|\bar{a}, z, c) \hat{p}(z|\breve{a}) 
                    - E(Y|\bar{a}, z, c) p(z|\breve{a})
                \right) 
                p(\bar{a}|c) p(c)\\
            & = \sum_{c, \bar{a}, z} 
                \left( E(Y|\bar{a},z,c) - \hat{E}(Y|\bar{a},z,c) 
                \right) 
                \left( 
                    \frac{p(z| \bar{a}) - \hat{p}(z|\bar{a}) 
                        }{\hat{p}(z|\bar{a})
                        } 
                \right)
                \hat{p}(z|\breve{a}) p( \bar{a}| c) p(c) \\  
            & + \sum_{c, \bar{a}, z} \hat{E}(Y|\bar{a}, z, c)
                \left( p(z|\breve{a}) - \hat{p}(z|\breve{a}) 
                \right) 
                \hat{p}(\bar{a}|c) \hat{p}(c) \frac{ 
                    p(\breve{a}) - \hat{p}(\breve{a}) 
                    }{ \hat{p}(\breve{a})
                }  \\
            & + \sum_{c, \bar{a}, z} E(Y|\bar{a},z,c)
                \left( \hat{p}(z|\breve{a}) - p(z|\breve{a}) 
                \right) 
                \big ( p(\bar{a}|c) ( p(c) - \hat{p}(c)) + \hat{p}(c) (p(\bar{a}|c) - \hat{p}(\bar{a}|c)) 
                \big)\\
            & +  \sum_{c, \bar{a}, z} 
                \left( E(Y|\bar{a},z,c) - \hat{E}(Y|\bar{a},z,c) 
                \right) 
                \big ( \hat{p}(z|\breve{a}) - p (z|\breve{a})
                \big) \hat{p}( \bar{a}| c) \hat{p}(c) \\
            & = \mathbb{P} 
            \left[ 
                \left( E(Y|A,Z,C) - \hat{E}(Y|A,Z,C) 
                \right) 
                \left( p(Z| A) - \hat{p}(Z|A) 
                \right)
                \frac{\hat{p}(Z|\breve{a})
                }{\hat{p}(Z|A)p(Z|A)
                        }
            \right] \\ 
            & + \mathbb{P} 
            \left[
            \hat{E}(Y|A, Z, C)
                \left( p(Z|\breve{a}) - \hat{p}(Z|\breve{a}) 
                \right) 
                \left( p(\breve{a}) - \hat{p}(\breve{a})
                \right)
                \frac{\hat{p}(A|C) \hat{p}(C)
                }{p(Z|A)p(A|C)\hat{p}(\breve{a}) }
            \right]    \\
            & + \mathbb{P} 
            \left[
            E(Y|A, Z, C)
                \left( \hat{p}(Z|\breve{a}) - p(Z|\breve{a}) 
                \right) 
                \left( p(C) - \hat{p}(C)
                \right)
                \frac{1
                }{p(Z|A)p(C)}
            \right]\\
            & - \mathbb{P} 
            \left[
            E(Y|A, Z, C)
                \left( \hat{p}(Z|\breve{a}) - p(Z|\breve{a}) 
                \right) 
                \left( p(A|C) - \hat{p}(A|C)
                \right)
                \frac{\hat{p}(C)
                }{p(Z|A)p(A|C)p(C)}
            \right]\\
            & + \mathbb{P} 
            \left[ 
                \left( E(Y|A,Z,C) - \hat{E}(Y|A,Z,C) 
                \right) 
                \left( \hat{p}(Z|\breve{a} ) - p(Z|\breve{a} )
                \right)
                \frac{\hat{p}(A|C) \hat{p}(C)
                }{p(Z|A)p(A|C)p(C)
                        }
            \right]. &
    \end{flalign*}
    $\mathbb{P} m_{fd,td}(X, \hat{\eta}) - \theta  = o_p(1/\sqrt{n})$  from the Cauchy-Schwartz inequality, \Cref{ass:an_fd_td}, and the assumptions of the theorem since each term in the summation above is then  $o_p(1/\sqrt{n}). $
\end{proof}

\newpage

\subsection{Proof of  \autoref{th:consistency_efficiency_of_bd_fd_td_est}}\label{sec:appendix_proof_lemma_consistency_efficiency_of_bd_fd_td_est}
\begin{proof}
According to Appendix \ref{sec:appendix_consistency_conditions}, we have to show that under each scenario, \\ $E\varphi_{bd, fd, td}(X, \bar{\eta}, \theta_0) = 0,$ where $\theta_0$ is the true value of the average causal effect. 

When $\bar{E}(Y|Z, C) = E(Y|Z, C),$
    \begin{flalign*}
        & E \bar{\varphi}_{bd, fd, td}(X, \bar{\eta}, \theta_0)   = E \bigg[ \bigg. \frac{\left(Y-E(Y|Z, C)\right) (\bar{p}(Z|a^*)-\bar{p}(Z|a))}{\sum_a \bar{p}(Z|a)\bar{p}(a|C)}  \\
        & +  \left(\sum_{c}E(Y|Z,c) - \sum_{z,c} E(Y|z,c)\bar{p}(z|A)\right)
        \frac{I(A=a^*) - I(A=a)}{\sum_{c}{\bar{p}(A|c)\bar{p}(c)}}\bar{p}(c) \\
        & + \sum\limits_{z} E(Y|z, C)\big(\bar{p}(z|a^*) - \bar{p}(z|a)\big) -\theta_0\bigg.\bigg] \\
        &  = 0 + \sum_{z}\left(\sum_{c}E(Y|z,c) - \sum_{z,c} E(Y|z,c)\bar{p}(z|a^*)\right)
        \frac{p(z|a^*)p(a^*)}{\sum_{c}{\bar{p}(a^*|c)\bar{p}(c)}}\bar{p}(c) \\
        & - \sum_{z}\left(\sum_{c}E(Y|z,c) - \sum_{z,c} E(Y|z,c)\bar{p}(z|a)\right)
        \frac{p(z|a)p(a)}{\sum_{c}{\bar{p}(a|c)\bar{p}(c)}}\bar{p}(c)\\
        & + \sum\limits_{z, c} E(Y|z,c)\bar{p}(z|a^*)p(c) - \sum\limits_{ c,z} E(Y|z,c)\bar{p}(z|a)p(c)- \theta_0\\
        & = \sum_{c, z} E(Y|z,c)  
        \frac{ p(a^*)\bar{p}(c)}{\sum_{c} \bar{p}(a^*|c)\bar{p}(c)} \left(p(z|a^*) - \bar{p}(z|a^*) \right)\\
        & - \sum_{c, z} E(Y|z,c)  
        \frac{p(a)\bar{p}(c) }{\sum_{c} \bar{p}(a|c) \bar{p}(c)} \left(p(z|a) - \bar{p}(z|a) \right)\\
        & + \sum\limits_{ c,z} E(Y|z,c)p(c) \left(\bar{p}(z|a^*)-\bar{p}(z|a) - p(z|a^*)+p(z|a)\right)&
    \end{flalign*}
    The last expression is 0 when $\bar{p}(Z|A) = p(Z|A).$ Additionally, the expression is 0 when $\bar{p}(C) = p(C)$ and $\bar{p}(A|C) = p(A|C).$ These conditions correspond to the conditions of the theorem.
    
    When $\bar{p}(Z|A) = p(Z|A)$,
    \begin{flalign*}
        & E \bar{\varphi}_{bd, fd, td}(X, \bar{\eta}, \theta_0)  = E \bigg[ \bigg. \frac{\left(Y - \bar{E}(Y|Z, C)\right) (p(Z|a^*)-p(Z|a))}{\sum_a p(Z|a)\bar{p}(a|C)}  \\
        & +  \left(\sum_{c}\bar{E}(Y|Z, c) 
                 - \sum_{z,c} \bar{E}(Y|z, c)p(z|A)\right)
            \frac{I(A=a^*) - I(A=a)}{\sum_{c}{\bar{p}(c)\bar{p}(A|c)}}\bar{p}(c) \\
        & + \sum\limits_{z} \bar{E}(Y|z, C)\big(p(z|a^*)- p(z|a)\big) -\theta_0\bigg.\bigg] \\
        & = \sum_{c, z} \frac{\left(E(Y|z,c) - \bar{E}(Y|z, c)\right) 
            (p(z|a^*)-p(z|a))}{\sum_a p(z|a)\bar{p}(a|c)}p(c)\sum_a p(z|a)p(a|c)\\
        & + \sum_{c, z} \bar{E}(Y|z, c)  
            \frac{ p(a^*) \bar{p}(c)}{ \sum_{c}\bar{p}(a^*|c)\bar{p}(c)} 
            \left(p(z|a^*) - p(z|a^*) \right)\\
        & - \sum_{c, z} \bar{E}(Y|z, c)  
            \frac{ p(a) \bar{p}(c)}{ \sum_{c}\bar{p}(a|c)\bar{p}(c)} 
            \left(p(z|a) - p(z|a) \right)\\
        & + \sum_{c, z}(\bar{E}(Y|z, c) - E(Y|z,c))
           \big(p(z|a^*)- p(z|a)\big) p(c) \\
        & = \sum_{c, z}(E(Y|z,c) - \bar{E}(Y|z, c))( p(z|a^*)- p(z|a)) p(c)
        \left(\frac{\sum_a p(z|a)p(a|c)}{\sum_a p(z|a)\bar{p}(a|c)}-1\right) =0 & 
    \end{flalign*}
    when additionally either  $\bar{E}(Y|Z, C) = E(Y|Z, C)$ (already mentioned above) or $\bar{p}(A|C) = p(A|C).$
    
    Below we show that the conditions for $E \bar{\varphi}_{bd, fd, td}(X, \bar{\eta}, \theta_0) = 0$ under $\bar{p}(A|C) = p(A|C) $ or $\bar{p}(C) = p(C)$  are covered by already mentioned conditions of the theorem. 
    
    When $\bar{p}(A|C) = p(A|C),$
    \begin{flalign*}
        & E \bar{\varphi}_{bd, fd, td}(X, \bar{\eta}, \theta_0)    = E \bigg[ \bigg. \frac{\left(Y - \bar{E}(Y|Z, C)\right) 
            (\bar{p}(Z|a^*) - \bar{p}(Z|a))}{\sum_a \bar{p}(Z|a) p(a|C)}  \\
        & +  \left(\sum_{c}\bar{E}(Y|Z, c) 
                 - \sum_{z,c} \bar{E}(Y|z, c)\bar{p}(z|A)\right)
            \frac{I(A=a^*) - I(A=a)}{\sum_{c} p(A|c)\bar{p}(c)}\bar{p}(c) \\
        & + \sum\limits_{z} \bar{E}(Y|z, C)\big(\bar{p}(z|a^*) - \bar{p}(z|a)\big) -\theta_0\bigg.\bigg] \\
        & = \sum_{c, z} \left(E(Y|z,c) - \bar{E}(Y|z, c)\right) 
            \left(\bar{p}(z|a^*) - \bar{p}(z|a)\right)
            \frac{\sum_a p(z|a) p(a|c) }{\sum_a \bar{p}(z|a) p(a|c) }p(c)\\
        & + \sum_{c, z} \bar{E}(Y|z, c) 
                 \left(p(z|a^*)- \bar{p}(z|a^*)\right) 
                 \frac{\sum_{c}p(a^*|c)p(c)}{\sum_{c}p(a^*|c)\bar{p}(c)}\bar{p}(c)\\
        & - \sum_{c, z} \bar{E}(Y|z, c) 
                 \left(p(z|a)- \bar{p}(z|a)\right) 
                 \frac{\sum_{c}p(a|c)p(c)}{\sum_{c}p(a|c)\bar{p}(c)}\bar{p}(c)\\
        & + \sum_{c, z} \bar{E}(Y|z, c) 
                 \left(\bar{p}(z|a^*)- \bar{p}(z|a)\right) p(c) 
             - \sum_{c, z} E(Y|z,c) 
                \left(p(z|a^*)- p(z|a)\right) p(c) &
    \end{flalign*}
    the last expression is equal to 0 under the conditions of the theorem.
  
    When $p(C)$ is correctly specified,
    \begin{flalign*}
        & E \bar{\varphi}_{bd, fd, td}(X, \bar{\eta}, \theta_0)  = E \bigg[ \bigg. \frac{\left(Y - \bar{E}(Y|Z, C)\right) 
            (\bar{p}(Z|a^*) - \bar{p}(Z|a))}{\sum_a \bar{p}(Z|a) \bar{p}(a|C)}  \\
        & +  \left(\sum_{c}\bar{E}(Y|Z, c) 
                 - \sum_{z,c} \bar{E}(Y|z, c)\bar{p}(z|A)\right)
            \frac{I(A=a^*) - I(A=a)}{\sum_{c} \bar{p}(A|c)p(c)}p(c) \\
        & + \sum\limits_{z} \bar{E}(Y|z, C)\big(\bar{p}(z|a^*) - \bar{p}(z|a)\big) -\theta_0\bigg.\bigg] \\
        & = \sum_{c, z} \left(E(Y|z,c) - \bar{E}(Y|z, c)\right) 
            \left(\bar{p}(z|a^*) - \bar{p}(z|a)\right)
            \frac{\sum_a p(z|a) p(a|c) }{\sum_a \bar{p}(z|a) \bar{p}(a|c) }p(c)\\
        & + \sum_{c, z} \bar{E}(Y|z, c) 
                 \left(p(z|a^*)- \bar{p}(z|a^*)\right) 
                 \frac{\sum_{c}p(a^*|c)p(c)}{\sum_{c}\bar{p}(a^*|c)p(c)}p(c)\\
        & - \sum_{c, z} \bar{E}(Y|z, c) 
                 \left(p(z|a)- \bar{p}(z|a)\right) 
                 \frac{\sum_{c}p(a|c)p(c)}{\sum_{c}\bar{p}(a|c)p(c)}p(c)\\
        & + \sum_{c, z} \bar{E}(Y|z, c) 
                 \left(\bar{p}(z|a^*)- \bar{p}(z|a)\right) p(c) 
             - \sum_{c, z} E(Y|z,c) 
                \left(p(z|a^*)- p(z|a)\right) p(c). &
    \end{flalign*}
    The last expression equals 0 under already mentioned conditions of the theorem.
    
    As for the efficiency of the estimators, according to Appendix \ref{sec:appendix_efficiency_conditions}, we provide conditions for  $\mathbb{P}  m_{bd, fd, td}(X, \hat{\eta}) - \theta = o_p(n^{-1/2}).$
    \begin{flalign*}
       \mathbb{P} m_{bd, fd, td}(X, \hat{\eta}) - \theta = f(a^*) - f(a),&
    \end{flalign*}
    where 
    \begin{flalign*}
        & f(\breve{a}) =  \sum_{c, z} \left( E(Y|z,c) - \hat{E}(Y|z,c)
                \right) \hat{p}(z|\breve{a})
            \frac{\sum_{\bar{a}} p(z|\bar{a}) p(\bar{a}|c)
                }{ 
                \sum_{\bar{a }}  \hat{p}(z|\bar{a}) \hat{p}(\bar{a}|c)
                }  p(c)  \\
        & + \sum_{c, z} \hat{E}(Y|z,c) 
            \big( p(z|\breve{a}) - \hat{p}(z|\breve{a}) 
            \big) 
            \frac{ p(\breve{a})
            }{ \hat{p}(\breve{a}) } \hat{p}(c) \\
        & + \sum_{c, z} 
            \left(
                \hat{E}(Y|z,c) \hat{p}(z|\breve{a}) - E(Y|z,c) p(z|\breve{a})
            \right) p(c) &
    \end{flalign*}
    Addition and subtraction of several terms and rearranging implies the following.
    \begin{flalign*}
       & f(\breve{a})  = \sum_{c, z} \left( E(Y|z,c) - \hat{E}(Y|z,c)
                \right) \hat{p}(z|\breve{a})
            \frac{\sum_{\bar{a}} p(z|\bar{a}) p(\bar{a}|c) 
                 - \sum_a  \hat{p}(z|\bar{a}) \hat{p}(\bar{a}|c)
                }{ 
                 \sum_{\bar{a }}  \hat{p}(z|\bar{a}) \hat{p}(\bar{a}|c)
                }  p(c)  \\
        & + \sum_{c, z} \left( E(Y|z,c) - \hat{E}(Y|z,c)
                \right) \hat{p}(z|\breve{a}) p(c) \\
        & + \sum_{c, z} \hat{E}(Y|z,c) 
            \big( p(z|\breve{a}) - \hat{p}(z|\breve{a}) 
            \big) 
            \frac{p(\breve{a}) - \hat{p}(\breve{a})}{ \hat{p}(\breve{a})}
             \hat{p}(c)\\
        & + \sum_{c, z} \hat{E}(Y|z,c) 
            \big( p(z|\breve{a}) - \hat{p}(z|\breve{a}) 
            \big) \hat{p}(c)\\
        & + \sum_{c, z} 
            \left(
                \hat{E}(Y|z,c) \hat{p}(z|\breve{a}) - E(Y|z,c) p(z|\breve{a})
            \right) p(c)\\
        & = \sum_{c, z} 
                \left( E(Y|z,c) - \hat{E}(Y|z,c)
                \right) \hat{p}(z|\breve{a}) \times\\
        & \times \frac{\sum_{\bar{a}} p(z|\bar{a}) 
                        \big( p(\bar{a}|c) - \hat{p} (\bar{a}|c)
                        \big)
                     - \sum_a  \hat{p}(\bar{a}|c) 
                        \big( \hat{p}(z|\bar{a}) - p (z|\bar{a})
                        \big) 
                }{ 
                 \sum_{\bar{a }} \hat{p}(z|\bar{a}) \hat{p}(\bar{a}|c)
                }  p(c)  \\
        & + \sum_{c, z} \hat{E}(Y|z,c) 
            \big( p(z|\breve{a}) - \hat{p}(z|\breve{a}) 
            \big) 
            \frac{p(\breve{a}) - \hat{p}(\breve{a})}{ \hat{p}(\breve{a})}
             \hat{p}(c)\\
        & + \sum_{c, z} E(Y|z,c)\hat{p}(z|\breve{a}) p(c) 
         -  \sum_{c, z}\hat{E}(Y|z,c)
                \hat{p}(z|\breve{a}) p(c) \\
        & + \sum_{c, z} \hat{E}(Y|z,c) p(z|\breve{a})\hat{p}(c) 
        - \sum_{c, z} \hat{E}(Y|z,c)\hat{p}(z|\breve{a}) \hat{p}(c)\\
        & + \sum_{c, z} \hat{E}(Y|z,c) \hat{p}(z|\breve{a})p(c) 
        - \sum_{c, z} E(Y|z,c) p(z|\breve{a})p(c) \\
        & = \sum_{c, z} 
                \left( E(Y|z,c) - \hat{E}(Y|z,c)
                \right) \hat{p}(z|\breve{a}) \times\\
        & \times \frac{\sum_{\bar{a}} p(z|\bar{a}) 
                        \big( p(\bar{a}|c) - \hat{p} (\bar{a}|c)
                        \big)
                     - \sum_a  \hat{p}(\bar{a}|c) 
                        \big( \hat{p}(z|\bar{a}) - p (z|\bar{a})
                        \big) 
                }{ 
                 \sum_{\bar{a }} \hat{p}(z|\bar{a}) \hat{p}(\bar{a}|c)
                }  p(c)  \\
        & + \sum_{c, z} \hat{E}(Y|z,c) 
            \big( p(z|\breve{a}) - \hat{p}(z|\breve{a}) 
            \big) 
            \frac{p(\breve{a}) - \hat{p}(\breve{a})}{ \hat{p}(\breve{a})}
             \hat{p}(c)\\     
        &  +  \sum_{c, z} 
                \big( \hat{p} (z|\breve{a}) - p (z|\breve{a})
                \big)
                \left( 
                    \big(E(Y|z,c) - \hat{E}(Y|z,c)
                    \big)\hat{p}(c) +
                    E(Y|z,c) 
                    \big( p(c) - \hat{p}(c)
                    \big)
                \right)   \\
        & =  \mathbb{P} 
        \left[
            \left( E(Y|Z,C) - \hat{E}(Y|Z,C)  \right) 
            \big( p(A|C) - \hat{p} (A|C) \big)
        \frac{\hat{p}(Z|\breve{a})
            }{p(A|C)\sum_{\bar{a}}  \hat{p}(Z|\bar{a}) \hat{p}(\bar{a}|C)}
        \right] \\
        & - \mathbb{P} 
        \left[
            \left( E(Y|Z,C) - \hat{E}(Y|Z,C)  \right) 
            \big( \hat{p}(Z|A) - p (Z|A) \big)
        \frac{\hat{p}(Z|\breve{a})\hat{p}(A|C)
        }{ p(A|C) p(Z|A)\sum_a  \hat{p}(Z|\bar{a}) \hat{p}(\bar{a}|C)}
        \right]\\
        & + \mathbb{P} 
        \left[ \hat{E}(Y|Z,C)
        \big( p(Z|\breve{a}) - \hat{p}(Z|\breve{a})
        \frac{p(\breve{a}) - \hat{p}(\breve{a})
        }{ \hat{p}(\breve{a}) p(C)\sum_{\bar{a}} p(\bar{a}|C)p(Z|\bar{a})}
        \right]\\
        & + \mathbb{P} 
        \left[
           \left( E(Y|Z,C) - \hat{E}(Y|Z,C)  \right) 
            \big( \hat{p}(Z|A) - p (Z|A) \big) \frac{\hat{p}(C) I(A=\breve{a})}{p(Z, A,C)}
        \right]\\
        & + \mathbb{P} 
        \left[
            E(Y|Z,C)
            \big( \hat{p}(Z|\breve{a}) - p (Z|\breve{a}) \big)
            \left(  p(C) - \hat{p}(C)\right) 
            \frac{1}{p(C)\sum_{\bar{a}} p(\bar{a}|C)p(Z|\bar{a})}
        \right].&
    \end{flalign*}
$\mathbb{P} m_{bd, fd,td}(X, \hat{\eta}) - \theta  = o_p(1/\sqrt{n})$  from the Cauchy-Schwartz inequality, \Cref{ass:an_bd_fd_td}, and the assumptions of the theorem since each term in the summation above is then  $o_p(1/\sqrt{n}). $
\end{proof}

\newpage

\subsection{Expressions related to the example and simulation study 1} \label{sec:appendix_expressions_simulation_study}
\subsubsection{ \texorpdfstring{$var \varphi_{bd}$}{var varphi bd}}
From the law of total variance,
\begin{flalign*}
    & var[Y(a)\big)|C]=\sigma_y^2+var(\gamma_0+\gamma_1Z+\gamma_2C|a, C)= \sigma_y^2 + \gamma_1^2\sigma_z^2.&
\end{flalign*}
Since the back-door identification holds,
\begin{flalign*}
    & E(Y(a)| C)   = E(Y|a, C) = \gamma_0+\gamma_1(\beta a)+\gamma_2C.\\
    & E(Y(a^*)|C)  -E(Y(a)| C)   - \theta=\gamma_1\beta(a^*-a) - \theta=0.\\
    & var \varphi_{bd}  = (\sigma_y^2 + \gamma_1^2\sigma_z^2)\sum_c\Big(\frac{p(c)}{p(a^*|c)}+\frac{p(c)}{p(a|c)}\Big).&
\end{flalign*}
For the distribution in the simulation study, 
$$var \varphi_{bd} = (1+\gamma_1^2)\left(2+\frac{0.5}{\expit(\alpha)} + \frac{0.5}{1-\expit(\alpha)}\right).$$

\subsubsection{\texorpdfstring{$var \varphi_{fd}$}{var varphi fd}} \label{sec:appendix_var_phi_fd_in_the_simulation_study}
\begin{flalign*}
    & p(y|a,z) = \frac{\sum_c p(a, c)p(y|z, c)}{p(a)}.\\
    & E(Y|A,Z) = E\big[ E(Y|Z,A,C)| A,Z \big]= \gamma_0+\gamma_1Z+\gamma_2E(C|A).\\
    & E(C|A)  = p(C=1|A)=\frac{p(A|C=1)p(C=1)}{p(A)}=              A\frac{\expit(\alpha)p_c}{ \expit(\alpha)p_c +     (1-p_c)\expit(0)} \\
        & + (1-A)\frac{(1-\expit(\alpha))p_c}{     (1-\expit(\alpha))p_c + (1-p_c)(1-\expit(0))} \\
    & var(C|A) = p(C=1|A) (1-p(C=1|A))&
\end{flalign*} 
where $\alpha = 1 $ and $p_c =0.5$ for the simulation study.
\begin{flalign*}  
    & var(Y|\bar{a},z)   = \frac{\sum_c p(\bar{a},c)E(Y^2|z,c)}{p(\bar{a})} -E^2(Y|z,\bar{a}) \\
        & = \sum_c p(c|\bar{a})(\sigma_y^2 + (\gamma_0 + \gamma_1z+     \gamma_2c)^2)  
        - (\gamma_0+\gamma_1z+\gamma_2E(C|\bar{a}))^2\\
        & = \sigma_y^2 + \gamma_2^2(E(C^2|\bar{a}) - E^2(C|\bar{a})).\\
    & \frac{\sum_{z} (p(z|a^*)-p(z|a))^2} {p(z|a)}
        = \frac{\sum_{z} (p(z|a^*)-p(z|a))^2}{p(z|a^*)} 
        = e^{\beta^2/\sigma_z^2}-1.\\
    & \sum\limits_{z}\left(\sum\limits_{\bar{a}} E(Y|\bar{a}, z)p(\bar{a})\right)^2\frac{p(z|a)}{p(a)}  - \frac{E^2Y(a)}{p(a)}\\
        &  = \sum\limits_{z}p(z|a)\frac{\left(\gamma_0 + \gamma_1z+ \gamma_2 P(C=1)\right)^2}{p(a)}
        - \frac{(\gamma_0+\gamma_1E(Z|a)+\gamma_2EC)^2}{p(a)}
        = \frac{\gamma_1^2 \sigma_z^2}{p(a)}. \\
    & \sum\limits_{\bar{a}}p(\bar{a})\left(\sum_{z}E(Y|\bar{a}, z)(p(z|a^*)-p(z|a))\right)^2 -\theta^2\\
        & = \sum\limits_{\bar{a}}p(\bar{a})\left(\gamma_1E(Z|a^*)-\gamma_1E(Z|a)\right)^2 -\theta^2 =  \gamma_1^2\beta^2(a^* - a)^2 -\theta^2= 0.\\
   & var \varphi_{fd} =  \left(e^{\beta^2/\sigma_z^2} - 1 \right)
    \left(\sigma_y^2 + \gamma_2^2 p(c=1) - \gamma_2^2 \frac{p^2(c=1)p^2(a|c=1)}{p(a)} \right.\\
    &- \left. \gamma_2^2\frac{p^2(c=1)p^2(a^*|c=1)}{p(a^*)}\right)  + \gamma_1^2 \sigma_z^2 \left(\frac{1}{p(a^*)} +  \frac{1}{p(a)}\right).&
\end{flalign*}
In the simulation study,
\begin{flalign*}
& var \varphi_{fd} = \left(e^{\beta^2} - 1 \right) \left(1+ \frac{\gamma_2^2}{2} 
    - \frac{\gamma_2^2(1 - \expit(\alpha))^2}{1 + 2(1-\expit(\alpha))} 
    - \frac{\gamma_2^2\expit^2(\alpha)}{1 + 2(\expit(\alpha))}\right) \\
    & + \gamma_1^2 \left( 
        \frac{1}{0.5^2 + 0.5\expit(\alpha)} + \frac{1}{0.5^2 + 0.5(1-\expit(\alpha))}
        \right) &
\end{flalign*}

\subsubsection{\texorpdfstring{$var \varphi_{td}$}{var varphi td}} \label{sec:appendix_var_phi_td_sim_study}
\begin{flalign*}
    & var \varphi_{td} = var \varphi_{bd} + \sum_{z,c}p(c)\sigma_y^2 \Bigg( (p(z|a^*) - p(z|a))^2  
        \sum\limits_{\bar{a}}\frac{p(\bar{a}|c)}{p(z|\bar{a})}- \frac{p(z|a^*)}{p(a^*|c)}  - \frac{p(z|a)}{p(a|c)}  \Bigg)  \\
        & = var \varphi_{bd} 
            + \sigma_y^2\sum_z (p(z|a^*) - p(z|a))^2 
                \left( \frac{p(a)}{p(z|a)} 
                + \frac{p(a^*)}{p(z|a^*)} 
                \right) - \sigma_y^2\sum_c\Big(\frac{p(c)}{p(a^*|c)}+\frac{p(c)}{p(a|c)}\Big) \\
        & = var \varphi_{bd}  
            + \sigma_y^2\left(e^{\beta^2/\sigma_z^2} - 1 \right)
            -  \sigma_y^2\sum_c\Big(\frac{p(c)}{p(a^*|c)}+\frac{p(c)}{p(a|c)}\Big).&
\end{flalign*} 

For the distribution in the simulation study, 
$$var \varphi_{td} =  var \varphi_{bd} + \left(e^{\beta^2} - 1 \right) - \left(2+\frac{1}{2\expit(\alpha)} + \frac{1}{2(1-\expit(\alpha))}\right).$$

$var \varphi_{td} < var \varphi_{bd} $ when $e^{\beta^2}  < 3 + \frac{1}{2\expit(\alpha)} + \frac{1}{2(1-\expit(\alpha))}.$ 
When $\alpha = 1,$ $var \varphi_{td} < var \varphi_{bd}$ when $|\beta|< \sqrt{\log( 3 + \frac{1}{2\expit(\alpha)} + \frac{1}{2(1-\expit(\alpha))})} \approx 1.3.$

\newpage

\subsection{Simulation study  with multivariate covariates}
\label{sec:appendix_sim_study_mv_cov}

This simulation study follows simulation study 1 in the paper but considers multivariate covariates.  The observed data is generated as follows:
\begin{align*}
    \bm{C} & = (C_1, C_2, \bm{C}_3)^T\\
    C_1 &\sim \text{Bernoulli}(0.5) \\
    C_2 & \sim N(0,1)\\
    \bm{C}_3 & \sim N\left (\mathbf{0}, \begin{pmatrix}
                                        1 & 0 & 0\\
                                        0 & 1 & 0\\
                                        0 & 0 & 1
                                     \end{pmatrix}\right)\\
    A |\bm{C} & \sim \text{Bernoulli}( \expit (\bm{\alpha} \mathbf{C}))\\
    Z | A, \bm{C}  & \sim N(\beta A,1), \\ 
    Y| Z, A, \bm{C} & \sim N(\gamma_1Z+\bm{\gamma_2}\bm{C},1).
\end{align*}

We consider the following values of the parameters:
$\bm{\alpha} = (\alpha,\alpha,\alpha,\alpha,\alpha), $ where $\alpha = 0$ represents no effect of the covariates on treatment $A$ (when $var \varphi_{fd}$ is bigger than $var \varphi_{td}$ from comparisons in Section \ref{sec:front_door_vs_back_door}) and $\alpha = 1$ to represent stronger effect of the covariates on treatment $A$. Similarly,  $\bm{\gamma_2} = (\gamma_2,\gamma_2,\gamma_2,\gamma_2,\gamma_2),$ where $\gamma_2 = 0.1$  and $\gamma_2 = 1 $ to represent, correspondingly, weaker and stronger effect of the covariates on outcome $Y. $ The parameter $\beta$ takes values $0$ to study the situation when the sufficient condition for $var \varphi_{td} \leq var \varphi_{bd}$ in Proposition 1 is fulfilled and $\beta = 1.5$ as in the simulation study 1. Parameter $\gamma_1$ takes values $0.5$ and $1.5$ as in simulation study 1. 

\begin{center}
    \begin{figure}[t!]
        \centering
        \includegraphics{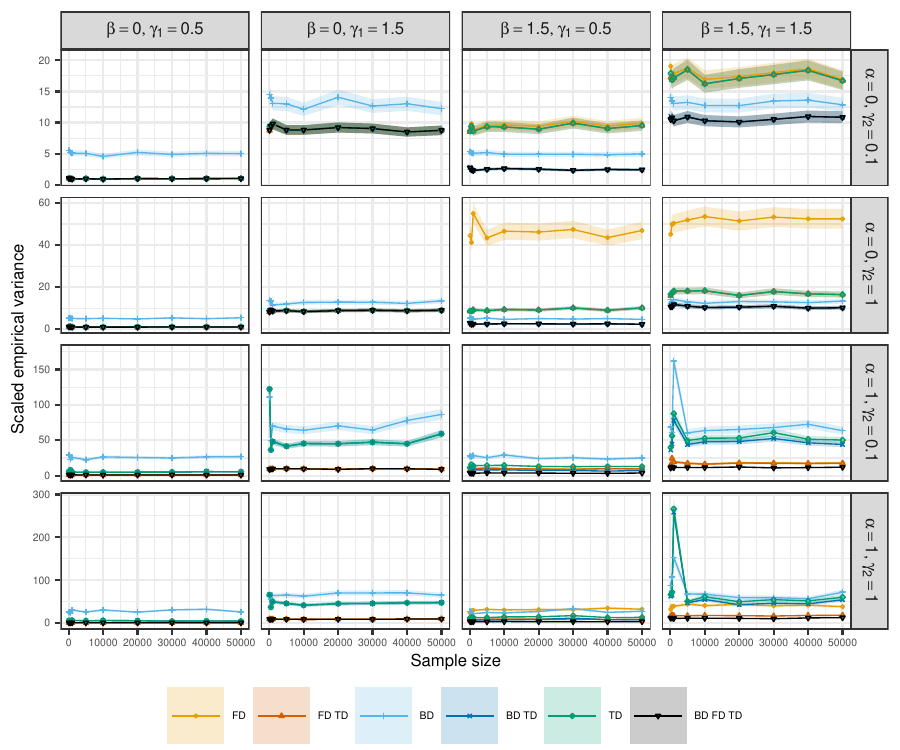}
        \caption{Scaled empirical variance of semiparametric estimators based on influence functions under Assumption \ref{ass:bd};  \hyperref[ass:fd]{\textbf{FD}}; \hyperref[ass:td]{\textbf{TD}}; \ref{ass:bd} and \hyperref[ass:td]{\textbf{TD}};  \ref{ass:bd},  \hyperref[ass:fd]{\textbf{FD}} and  \hyperref[ass:td]{\textbf{TD}}.  Shaded areas represent 95\% confidence intervals for the estimated scaled variance, based on the normal approximation.  Note the difference in scales between the panels} \label{fig:emp_var_sim_study_mv_cov}
    \end{figure}
\end{center}

We consider samples of sizes $n= 100, 500, 1000, 5000, 10000, 20000, 30000, 40000,$  $50000.$ For each sample size, $K=1000$ replicates were simulated. \autoref{fig:emp_var_sim_study_mv_cov} provides empirical variances scaled by $n${:} $ns_{K-1}^2 = n/(K-1) \sum_{k=1}^K \big( \hat{\theta}_k - \bar{\hat{\theta}}\big)^2, $
where $\hat{\theta}_k$ is an estimate from replication $k$, and $\bar{\hat{\theta}} = \sum_{k=1}^K \hat{\theta}_k /K.$ 

The simulation results show that the scaled empirical variance of the estimator based on Assumption \textbf{TD} is lower than the one for the estimator based on Assumption \textbf{BD} when the treatment does not affect the mediator ($\beta = 0$)  according to the sufficient condition of Proposition \ref{th:sufficient_conditions_td_vs_bd}.

As expected from the comparison of $var \varphi_{fd}$ and $var \varphi_{td},$ when the covariates are not related to the treatment ($\alpha = 0$), the sufficient condition in Proposition \ref{th:fd_vs_td} is fulfilled and the estimates based on Assumption \textbf{FD}   have close to or at least as large empirical variance as the estimates based on the \textbf{TD} assumption. The differences are more pronounced for a stronger relationship between the covariates and the outcome ($\gamma_2 = 1$).

Also,  \textbf{FD} outperforms \textbf{BD} when pre-treatment covariates do not affect the treatment ($\bm{\alpha} = 0$)  and the treatment does not affect the mediator  ($\beta = 0$) according to a sufficient condition of Proposition 5.  

\newpage

\begin{table}[ht]
\centering
\tiny
    \setlength{\tabcolsep}{2pt}
\begin{tabular}{rrrrrrrrrrrr}
  \hline
DGP* &  $\alpha$ & $\beta$ & $\gamma_1$ & $\gamma_2$&  n & BD & FD & TD & BD TD & FD TD & BD FD TD \\ 
  \hline
1 & 0.0 & 0.0 & 0.5 & 0.1 & 100 &   5.526 ( 0.247) &   1.035 ( 0.046) &   1.076 ( 0.048) &   1.086 ( 0.049) &   1.033 ( 0.046) &   1.041 ( 0.047) \\ 
  1 &  0.0 & 0.0 & 0.5 & 0.1 &500 &   5.184 ( 0.232) &   1.027 ( 0.046) &   1.047 ( 0.047) &   1.045 ( 0.047) &   1.031 ( 0.046) &   1.029 ( 0.046) \\ 
  1 &  0.0 & 0.0 & 0.5 & 0.1 &1000 &   5.102 ( 0.228) &   0.938 ( 0.042) &   0.932 ( 0.042) &   0.933 ( 0.042) &   0.935 ( 0.042) &   0.936 ( 0.042) \\ 
  1 & 0.0 & 0.0 & 0.5 & 0.1 &5000 &   5.066 ( 0.227) &   0.984 ( 0.044) &   0.986 ( 0.044) &   0.986 ( 0.044) &   0.984 ( 0.044) &   0.984 ( 0.044) \\ 
  1 & 0.0 & 0.0 & 0.5 & 0.1 &10000 &   4.571 ( 0.204) &   0.933 ( 0.042) &   0.934 ( 0.042) &   0.934 ( 0.042) &   0.933 ( 0.042) &   0.933 ( 0.042) \\ 
  1 &  0.0 & 0.0 & 0.5 & 0.1 &20000 &   5.226 ( 0.234) &   1.009 ( 0.045) &   1.009 ( 0.045) &   1.009 ( 0.045) &   1.009 ( 0.045) &   1.009 ( 0.045) \\ 
  1 &  0.0 & 0.0 & 0.5 & 0.1 &30000 &   4.886 ( 0.219) &   0.990 ( 0.044) &   0.990 ( 0.044) &   0.990 ( 0.044) &   0.990 ( 0.044) &   0.990 ( 0.044) \\ 
  1 &  0.0 & 0.0 & 0.5 & 0.1 &40000 &   5.074 ( 0.227) &   1.005 ( 0.045) &   1.006 ( 0.045) &   1.006 ( 0.045) &   1.005 ( 0.045) &   1.005 ( 0.045) \\ 
  1 & 0.0 & 0.0 & 0.5 & 0.1 & 50000 &   5.019 ( 0.224) &   1.022 ( 0.046) &   1.022 ( 0.046) &   1.022 ( 0.046) &   1.022 ( 0.046) &   1.022 ( 0.046) \\ 
  2 & 1.0 & 0.0 & 0.5 & 0.1 & 100 &  29.059 ( 1.300) &   1.139 ( 0.051) &   5.713 ( 0.255) &   5.718 ( 0.256) &   1.135 ( 0.051) &   1.137 ( 0.051) \\ 
  2 & 1.0 & 0.0 & 0.5 & 0.1 & 500 &  23.142 ( 1.035) &   0.971 ( 0.043) &   8.002 ( 0.358) &   7.999 ( 0.358) &   0.971 ( 0.043) &   0.970 ( 0.043) \\ 
  2 & 1.0 & 0.0 & 0.5 & 0.1 & 1000 &  25.807 ( 1.154) &   1.087 ( 0.049) &   5.001 ( 0.224) &   5.001 ( 0.224) &   1.086 ( 0.049) &   1.085 ( 0.049) \\ 
  2 & 1.0 & 0.0 & 0.5 & 0.1 & 5000 &  22.428 ( 1.003) &   1.016 ( 0.045) &   4.791 ( 0.214) &   4.791 ( 0.214) &   1.015 ( 0.045) &   1.016 ( 0.045) \\ 
  2 & 1.0 & 0.0 & 0.5 & 0.1 & 10000 &  26.541 ( 1.187) &   0.923 ( 0.041) &   4.822 ( 0.216) &   4.822 ( 0.216) &   0.923 ( 0.041) &   0.923 ( 0.041) \\ 
  2 & 1.0 & 0.0 & 0.5 & 0.1 & 20000 &  25.703 ( 1.149) &   1.032 ( 0.046) &   4.954 ( 0.222) &   4.954 ( 0.222) &   1.031 ( 0.046) &   1.031 ( 0.046) \\ 
  2 & 1.0 & 0.0 & 0.5 & 0.1 & 30000 &  24.924 ( 1.115) &   1.070 ( 0.048) &   5.001 ( 0.224) &   5.001 ( 0.224) &   1.070 ( 0.048) &   1.070 ( 0.048) \\ 
  2 & 1.0 & 0.0 & 0.5 & 0.1 & 40000 &  26.654 ( 1.192) &   1.038 ( 0.046) &   5.710 ( 0.255) &   5.710 ( 0.255) &   1.038 ( 0.046) &   1.038 ( 0.046) \\ 
  2 & 1.0 & 0.0 & 0.5 & 0.1 & 50000 &  26.874 ( 1.202) &   1.010 ( 0.045) &   5.403 ( 0.242) &   5.403 ( 0.242) &   1.010 ( 0.045) &   1.010 ( 0.045) \\ 
  3 & 0.0 & 1.5 & 0.5 & 0.1 & 100 &   5.322 ( 0.238) &   8.884 ( 0.397) &   8.638 ( 0.386) &   2.820 ( 0.126) &   8.521 ( 0.381) &   2.733 ( 0.122) \\ 
  3 & 0.0 & 1.5 & 0.5 & 0.1 & 500 &   5.129 ( 0.229) &   9.835 ( 0.440) &   9.426 ( 0.422) &   2.448 ( 0.109) &   9.413 ( 0.421) &   2.432 ( 0.109) \\ 
  3 & 0.0 & 1.5 & 0.5 & 0.1 & 1000 &   5.076 ( 0.227) &   8.854 ( 0.396) &   8.641 ( 0.386) &   2.323 ( 0.104) &   8.601 ( 0.385) &   2.306 ( 0.103) \\ 
  3 & 0.0 & 1.5 & 0.5 & 0.1 & 5000 &   5.186 ( 0.232) &   9.535 ( 0.426) &   9.349 ( 0.418) &   2.507 ( 0.112) &   9.346 ( 0.418) &   2.503 ( 0.112) \\ 
  3 & 0.0 & 1.5 & 0.5 & 0.1 & 10000 &   4.923 ( 0.220) &   9.650 ( 0.432) &   9.301 ( 0.416) &   2.621 ( 0.117) &   9.308 ( 0.416) &   2.619 ( 0.117) \\ 
  3 & 0.0 & 1.5 & 0.5 & 0.1 & 20000 &   4.925 ( 0.220) &   9.396 ( 0.420) &   8.939 ( 0.400) &   2.515 ( 0.112) &   8.942 ( 0.400) &   2.515 ( 0.112) \\ 
  3 & 0.0 & 1.5 & 0.5 & 0.1 & 30000 &   4.918 ( 0.220) &  10.201 ( 0.456) &   9.918 ( 0.444) &   2.336 ( 0.104) &   9.918 ( 0.444) &   2.336 ( 0.104) \\ 
  3 & 0.0 & 1.5 & 0.5 & 0.1 & 40000 &   4.827 ( 0.216) &   9.442 ( 0.422) &   9.054 ( 0.405) &   2.466 ( 0.110) &   9.050 ( 0.405) &   2.465 ( 0.110) \\ 
  3 & 0.0 & 1.5 & 0.5 & 0.1 & 50000 &   4.971 ( 0.222) &   9.887 ( 0.442) &   9.522 ( 0.426) &   2.429 ( 0.109) &   9.524 ( 0.426) &   2.429 ( 0.109) \\ 
  4 & 1.0 & 1.5 & 0.5 & 0.1& 100 &  27.476 ( 1.229) &   7.768 ( 0.347) &  11.463 ( 0.513) &   7.321 ( 0.327) &   7.425 ( 0.332) &   3.838 ( 0.172) \\ 
  4 & 1.0 & 1.5 & 0.5 & 0.1& 500 &  26.663 ( 1.192) &  12.802 ( 0.573) &  15.898 ( 0.711) &   9.623 ( 0.430) &  10.391 ( 0.465) &   3.625 ( 0.162) \\ 
  4 & 1.0 & 1.5 & 0.5 & 0.1& 1000 &  28.438 ( 1.272) &  10.170 ( 0.455) &  14.143 ( 0.632) &   8.190 ( 0.366) &  10.013 ( 0.448) &   3.360 ( 0.150) \\ 
  4 & 1.0 & 1.5 & 0.5 & 0.1& 5000 &  25.361 ( 1.134) &  10.802 ( 0.483) &  14.430 ( 0.645) &   8.090 ( 0.362) &  10.545 ( 0.472) &   3.960 ( 0.177) \\ 
  4 & 1.0 & 1.5 & 0.5 & 0.1& 10000 &  29.269 ( 1.309) &  10.149 ( 0.454) &  14.558 ( 0.651) &   8.315 ( 0.372) &  10.075 ( 0.451) &   3.693 ( 0.165) \\ 
  4 & 1.0 & 1.5 & 0.5 & 0.1& 20000 &  24.081 ( 1.077) &   9.555 ( 0.427) &  13.032 ( 0.583) &   7.211 ( 0.322) &   9.303 ( 0.416) &   3.645 ( 0.163) \\ 
  4 & 1.0 & 1.5 & 0.5 & 0.1& 30000 &  25.390 ( 1.135) &   9.072 ( 0.406) &  12.748 ( 0.570) &   7.184 ( 0.321) &   8.960 ( 0.401) &   3.412 ( 0.153) \\ 
  4 & 1.0 & 1.5 & 0.5 & 0.1& 40000 &  23.328 ( 1.043) &   9.997 ( 0.447) &  13.090 ( 0.585) &   6.494 ( 0.290) &   9.883 ( 0.442) &   3.256 ( 0.146) \\ 
  4 & 1.0 & 1.5 & 0.5 & 0.1& 50000 &  25.066 ( 1.121) &   9.628 ( 0.431) &  12.794 ( 0.572) &   7.198 ( 0.322) &   9.426 ( 0.422) &   3.740 ( 0.167) \\ 
  5 & 0.0 & 0.0 & 1.5 & 0.1& 100 &  14.476 ( 0.647) &   8.746 ( 0.391) &   9.374 ( 0.419) &   9.410 ( 0.421) &   8.739 ( 0.391) &   8.776 ( 0.392) \\ 
  5 & 0.0 & 0.0 & 1.5 & 0.1& 500 &  13.978 ( 0.625) &   9.434 ( 0.422) &   9.492 ( 0.425) &   9.489 ( 0.424) &   9.433 ( 0.422) &   9.430 ( 0.422) \\ 
  5 & 0.0 & 0.0 & 1.5 & 0.1& 1000 &  13.094 ( 0.586) &   9.789 ( 0.438) &   9.817 ( 0.439) &   9.812 ( 0.439) &   9.791 ( 0.438) &   9.787 ( 0.438) \\ 
  5 & 0.0 & 0.0 & 1.5 & 0.1& 5000 &  13.018 ( 0.582) &   8.821 ( 0.394) &   8.827 ( 0.395) &   8.827 ( 0.395) &   8.820 ( 0.394) &   8.820 ( 0.394) \\ 
  5 & 0.0 & 0.0 & 1.5 & 0.1& 10000 &  12.134 ( 0.543) &   8.812 ( 0.394) &   8.826 ( 0.395) &   8.826 ( 0.395) &   8.813 ( 0.394) &   8.813 ( 0.394) \\ 
  5 & 0.0 & 0.0 & 1.5 & 0.1& 20000 &  14.071 ( 0.629) &   9.208 ( 0.412) &   9.226 ( 0.413) &   9.226 ( 0.413) &   9.208 ( 0.412) &   9.208 ( 0.412) \\ 
  5 & 0.0 & 0.0 & 1.5 & 0.1& 30000 &  12.663 ( 0.566) &   9.046 ( 0.405) &   9.048 ( 0.405) &   9.048 ( 0.405) &   9.046 ( 0.405) &   9.046 ( 0.405) \\ 
  5 & 0.0 & 0.0 & 1.5 & 0.1& 40000 &  13.027 ( 0.583) &   8.492 ( 0.380) &   8.485 ( 0.379) &   8.485 ( 0.379) &   8.493 ( 0.380) &   8.493 ( 0.380) \\ 
  5 & 0.0 & 0.0 & 1.5 & 0.1& 50000 &  12.250 ( 0.548) &   8.755 ( 0.392) &   8.767 ( 0.392) &   8.767 ( 0.392) &   8.757 ( 0.392) &   8.757 ( 0.392) \\ 
  6 & 1.0 & 0.0 & 1.5 & 0.1& 100 & 110.994 ( 4.964) &   9.044 ( 0.404) & 122.328 ( 5.471) & 122.320 ( 5.470) &   9.054 ( 0.405) &   9.036 ( 0.404) \\ 
  6 & 1.0 & 0.0 & 1.5 & 0.1& 500 &  49.493 ( 2.213) &   9.470 ( 0.423) &  36.351 ( 1.626) &  36.353 ( 1.626) &   9.476 ( 0.424) &   9.477 ( 0.424) \\ 
  6 & 1.0 & 0.0 & 1.5 & 0.1& 1000 &  69.967 ( 3.129) &   9.698 ( 0.434) &  48.141 ( 2.153) &  48.140 ( 2.153) &   9.694 ( 0.434) &   9.693 ( 0.433) \\ 
  6 & 1.0 & 0.0 & 1.5 & 0.1& 5000 &  65.897 ( 2.947) &   9.896 ( 0.443) &  41.541 ( 1.858) &  41.541 ( 1.858) &   9.896 ( 0.443) &   9.896 ( 0.443) \\ 
  6 & 1.0 & 0.0 & 1.5 & 0.1& 10000 &  64.108 ( 2.867) &   9.666 ( 0.432) &  45.373 ( 2.029) &  45.373 ( 2.029) &   9.665 ( 0.432) &   9.665 ( 0.432) \\ 
  6 & 1.0 & 0.0 & 1.5 & 0.1& 20000 &  70.295 ( 3.144) &   8.975 ( 0.401) &  45.055 ( 2.015) &  45.056 ( 2.015) &   8.974 ( 0.401) &   8.974 ( 0.401) \\ 
  6 & 1.0 & 0.0 & 1.5 & 0.1& 30000 &  64.432 ( 2.881) &   9.894 ( 0.442) &  47.285 ( 2.115) &  47.285 ( 2.115) &   9.893 ( 0.442) &   9.893 ( 0.442) \\ 
  6 & 1.0 & 0.0 & 1.5 & 0.1& 40000 &  78.064 ( 3.491) &   9.684 ( 0.433) &  45.049 ( 2.015) &  45.049 ( 2.015) &   9.684 ( 0.433) &   9.684 ( 0.433) \\ 
  6 & 1.0 & 0.0 & 1.5 & 0.1& 50000 &  86.437 ( 3.866) &   9.019 ( 0.403) &  59.158 ( 2.646) &  59.158 ( 2.646) &   9.019 ( 0.403) &   9.019 ( 0.403) \\ 
  7 & 0.0 & 1.5 & 1.5 & 0.1& 100 &  14.032 ( 0.628) &  19.080 ( 0.853) &  17.919 ( 0.801) &  11.010 ( 0.492) &  17.255 ( 0.772) &  10.514 ( 0.470) \\ 
  7 & 0.0 & 1.5 & 1.5 & 0.1& 500 &  13.478 ( 0.603) &  17.312 ( 0.774) &  16.864 ( 0.754) &  10.844 ( 0.485) &  16.867 ( 0.754) &  10.777 ( 0.482) \\ 
  7 & 0.0 & 1.5 & 1.5 & 0.1& 1000 &  13.069 ( 0.584) &  17.869 ( 0.799) &  17.208 ( 0.770) &  10.353 ( 0.463) &  17.185 ( 0.769) &  10.276 ( 0.460) \\ 
  7 & 0.0 & 1.5 & 1.5 & 0.1& 5000 &  13.245 ( 0.592) &  18.767 ( 0.839) &  18.513 ( 0.828) &  10.940 ( 0.489) &  18.443 ( 0.825) &  10.890 ( 0.487) \\ 
  7 & 0.0 & 1.5 & 1.5 & 0.1& 10000 &  12.759 ( 0.571) &  16.924 ( 0.757) &  16.238 ( 0.726) &  10.328 ( 0.462) &  16.217 ( 0.725) &  10.294 ( 0.460) \\ 
  7 & 0.0 & 1.5 & 1.5 & 0.1& 20000 &  12.729 ( 0.569) &  17.390 ( 0.778) &  17.080 ( 0.764) &  10.114 ( 0.452) &  17.079 ( 0.764) &  10.108 ( 0.452) \\ 
  7 & 0.0 & 1.5 & 1.5 & 0.1& 30000 &  13.465 ( 0.602) &  18.032 ( 0.806) &  17.713 ( 0.792) &  10.517 ( 0.470) &  17.715 ( 0.792) &  10.519 ( 0.470) \\ 
  7 & 0.0 & 1.5 & 1.5 & 0.1& 40000 &  13.603 ( 0.608) &  18.636 ( 0.833) &  18.392 ( 0.823) &  10.979 ( 0.491) &  18.392 ( 0.823) &  10.981 ( 0.491) \\ 
  7 & 0.0 & 1.5 & 1.5 & 0.1& 50000 &  12.863 ( 0.575) &  16.925 ( 0.757) &  16.723 ( 0.748) &  10.868 ( 0.486) &  16.718 ( 0.748) &  10.858 ( 0.486) \\ 
  8 & 1.0 & 1.5 & 1.5 & 0.1& 100 &  68.441 ( 3.061) &  14.864 ( 0.665) &  40.062 ( 1.792) &  36.377 ( 1.627) &  14.756 ( 0.660) &  10.982 ( 0.491) \\ 
  8 & 1.0 & 1.5 & 1.5 & 0.1& 500 &  60.633 ( 2.712) &  24.658 ( 1.103) &  56.216 ( 2.514) &  44.763 ( 2.002) &  24.247 ( 1.084) &  12.405 ( 0.555) \\ 
  8 & 1.0 & 1.5 & 1.5 & 0.1& 1000 & 161.693 ( 7.231) &  19.547 ( 0.874) &  87.635 ( 3.919) &  78.136 ( 3.494) &  19.382 ( 0.867) &  11.489 ( 0.514) \\ 
  8& 1.0 & 1.5 & 1.5 & 0.1 & 5000 &  60.515 ( 2.706) &  17.717 ( 0.792) &  49.715 ( 2.223) &  43.973 ( 1.967) &  17.437 ( 0.780) &  11.529 ( 0.516) \\ 
  8 & 1.0 & 1.5 & 1.5 & 0.1& 10000 &  63.620 ( 2.845) &  16.518 ( 0.739) &  52.928 ( 2.367) &  47.845 ( 2.140) &  16.192 ( 0.724) &  11.435 ( 0.511) \\ 
  8 & 1.0 & 1.5 & 1.5 & 0.1& 20000 &  65.244 ( 2.918) &  18.120 ( 0.810) &  53.155 ( 2.377) &  48.255 ( 2.158) &  17.839 ( 0.798) &  12.244 ( 0.548) \\ 
  8 & 1.0 & 1.5 & 1.5 & 0.1& 30000 &  67.921 ( 3.038) &  17.838 ( 0.798) &  60.753 ( 2.717) &  52.532 ( 2.349) &  17.679 ( 0.791) &  10.964 ( 0.490) \\ 
  8 & 1.0 & 1.5 & 1.5 & 0.1& 40000 &  72.604 ( 3.247) &  17.502 ( 0.783) &  51.462 ( 2.301) &  46.380 ( 2.074) &  17.246 ( 0.771) &  11.382 ( 0.509) \\ 
  8 & 1.0 & 1.5 & 1.5 & 0.1 & 50000 &  63.512 ( 2.840) &  17.825 ( 0.797) &  50.326 ( 2.251) &  44.108 ( 1.973) &  17.525 ( 0.784) &  11.956 ( 0.535) \\ 
   \hline
\end{tabular}
\caption{Scaled empirical variance (Monte Carlo SEs in parentheses) of estimators, part 1. *DGP - data generating process}
\end{table}

\newpage
\begin{table}[ht]
\centering
\tiny
    \setlength{\tabcolsep}{2pt}
\begin{tabular}{rrrrrrrrrrrrr}
  \hline
DGP* &  $\alpha$ & $\beta$ & $\gamma_1$ & $\gamma_2$& n & BD & FD & TD & BD TD & FD TD & BD FD TD \\ 
  \hline 
  9& 0.0 & 0.0 & 0.5 & 1.0 & 100 &   5.337 ( 0.239) &   1.333 ( 0.060) &   1.181 ( 0.053) &   1.179 ( 0.053) &   1.083 ( 0.048) &   1.080 ( 0.048) \\ 
  9 & 0.0 & 0.0 & 0.5 & 1.0& 500 &   5.313 ( 0.238) &   1.238 ( 0.055) &   1.182 ( 0.053) &   1.184 ( 0.053) &   1.166 ( 0.052) &   1.167 ( 0.052) \\ 
  9& 0.0 & 0.0 & 0.5 & 1.0 & 1000 &   5.073 ( 0.227) &   1.030 ( 0.046) &   1.016 ( 0.045) &   1.015 ( 0.045) &   1.014 ( 0.045) &   1.013 ( 0.045) \\ 
  9 & 0.0 & 0.0 & 0.5 & 1.0& 5000 &   4.981 ( 0.223) &   0.982 ( 0.044) &   0.979 ( 0.044) &   0.979 ( 0.044) &   0.979 ( 0.044) &   0.979 ( 0.044) \\ 
  9& 0.0 & 0.0 & 0.5 & 1.0 & 10000 &   5.184 ( 0.232) &   0.974 ( 0.044) &   0.971 ( 0.043) &   0.971 ( 0.043) &   0.971 ( 0.043) &   0.971 ( 0.043) \\ 
  9& 0.0 & 0.0 & 0.5 & 1.0 & 20000 &   4.875 ( 0.218) &   0.982 ( 0.044) &   0.982 ( 0.044) &   0.982 ( 0.044) &   0.982 ( 0.044) &   0.982 ( 0.044) \\ 
  9& 0.0 & 0.0 & 0.5 & 1.0 & 30000 &   5.364 ( 0.240) &   1.053 ( 0.047) &   1.050 ( 0.047) &   1.050 ( 0.047) &   1.051 ( 0.047) &   1.051 ( 0.047) \\ 
  9 & 0.0 & 0.0 & 0.5 & 1.0& 40000 &   4.978 ( 0.223) &   1.055 ( 0.047) &   1.051 ( 0.047) &   1.051 ( 0.047) &   1.051 ( 0.047) &   1.051 ( 0.047) \\ 
  9& 0.0 & 0.0 & 0.5 & 1.0 & 50000 &   5.449 ( 0.244) &   1.034 ( 0.046) &   1.034 ( 0.046) &   1.034 ( 0.046) &   1.034 ( 0.046) &   1.034 ( 0.046) \\ 
  10 & 1.0 & 0.0 & 0.5 & 1.0& 100 &  25.579 ( 1.144) &   1.105 ( 0.049) &   6.070 ( 0.271) &   6.070 ( 0.271) &   1.000 ( 0.045) &   1.004 ( 0.045) \\ 
  10& 1.0 & 0.0 & 0.5 & 1.0 & 500 &  23.455 ( 1.049) &   1.121 ( 0.050) &   4.947 ( 0.221) &   4.945 ( 0.221) &   1.109 ( 0.050) &   1.109 ( 0.050) \\ 
  10& 1.0 & 0.0 & 0.5 & 1.0 & 1000 &  30.338 ( 1.357) &   1.194 ( 0.053) &   6.736 ( 0.301) &   6.737 ( 0.301) &   1.156 ( 0.052) &   1.156 ( 0.052) \\ 
  10& 1.0 & 0.0 & 0.5 & 1.0 & 5000 &  26.216 ( 1.172) &   0.964 ( 0.043) &   5.568 ( 0.249) &   5.569 ( 0.249) &   0.956 ( 0.043) &   0.956 ( 0.043) \\ 
  10& 1.0 & 0.0 & 0.5 & 1.0 & 10000 &  30.665 ( 1.371) &   1.035 ( 0.046) &   6.202 ( 0.277) &   6.202 ( 0.277) &   1.035 ( 0.046) &   1.035 ( 0.046) \\ 
  10& 1.0 & 0.0 & 0.5 & 1.0 & 20000 &  25.751 ( 1.152) &   1.026 ( 0.046) &   5.772 ( 0.258) &   5.772 ( 0.258) &   1.022 ( 0.046) &   1.022 ( 0.046) \\ 
  10& 1.0 & 0.0 & 0.5 & 1.0 & 30000 &  30.961 ( 1.385) &   1.065 ( 0.048) &   4.723 ( 0.211) &   4.723 ( 0.211) &   1.069 ( 0.048) &   1.069 ( 0.048) \\ 
  10& 1.0 & 0.0 & 0.5 & 1.0 & 40000 &  32.451 ( 1.451) &   0.987 ( 0.044) &   5.149 ( 0.230) &   5.149 ( 0.230) &   0.987 ( 0.044) &   0.987 ( 0.044) \\ 
  10& 1.0 & 0.0 & 0.5 & 1.0 & 50000 &  26.334 ( 1.178) &   0.979 ( 0.044) &   4.652 ( 0.208) &   4.652 ( 0.208) &   0.979 ( 0.044) &   0.979 ( 0.044) \\ 
  11& 0.0 & 1.5 & 0.5 & 1.0 & 100 &   5.260 ( 0.235) &  44.473 ( 1.989) &   8.672 ( 0.388) &   2.908 ( 0.130) &   8.579 ( 0.384) &   2.846 ( 0.127) \\ 
  11& 0.0 & 1.5 & 0.5 & 1.0 & 500 &   5.040 ( 0.225) &  41.215 ( 1.843) &   8.480 ( 0.379) &   2.366 ( 0.106) &   8.476 ( 0.379) &   2.344 ( 0.105) \\ 
  11 & 0.0 & 1.5 & 0.5 & 1.0& 1000 &   4.747 ( 0.212) &  54.978 ( 2.459) &   9.183 ( 0.411) &   2.413 ( 0.108) &   9.176 ( 0.410) &   2.411 ( 0.108) \\ 
  11& 0.0 & 1.5 & 0.5 & 1.0 & 5000 &   5.261 ( 0.235) &  43.385 ( 1.940) &   8.823 ( 0.395) &   2.481 ( 0.111) &   8.821 ( 0.394) &   2.482 ( 0.111) \\ 
  11& 0.0 & 1.5 & 0.5 & 1.0 & 10000 &   4.656 ( 0.208) &  46.606 ( 2.084) &   9.351 ( 0.418) &   2.588 ( 0.116) &   9.351 ( 0.418) &   2.587 ( 0.116) \\ 
  11 & 0.0 & 1.5 & 0.5 & 1.0& 20000 &   5.043 ( 0.226) &  46.175 ( 2.065) &   9.127 ( 0.408) &   2.559 ( 0.114) &   9.128 ( 0.408) &   2.558 ( 0.114) \\ 
  11 & 0.0 & 1.5 & 0.5 & 1.0& 30000 &   4.824 ( 0.216) &  47.409 ( 2.120) &  10.094 ( 0.451) &   2.457 ( 0.110) &  10.094 ( 0.451) &   2.456 ( 0.110) \\ 
  11& 0.0 & 1.5 & 0.5 & 1.0 & 40000 &   5.036 ( 0.225) &  43.515 ( 1.946) &   8.971 ( 0.401) &   2.566 ( 0.115) &   8.967 ( 0.401) &   2.565 ( 0.115) \\ 
  11& 0.0 & 1.5 & 0.5 & 1.0 & 50000 &   4.671 ( 0.209) &  46.933 ( 2.099) &  10.121 ( 0.453) &   2.343 ( 0.105) &  10.119 ( 0.453) &   2.343 ( 0.105) \\ 
  12& 1.0 & 1.5 & 0.5 & 1.0 & 100 &  27.565 ( 1.233) &  24.418 ( 1.092) &  12.041 ( 0.538) &   8.612 ( 0.385) &   7.211 ( 0.322) &   4.106 ( 0.184) \\ 
  12& 1.0 & 1.5 & 0.5 & 1.0 & 500 &  26.830 ( 1.200) &  28.556 ( 1.277) &  13.640 ( 0.610) &   8.088 ( 0.362) &   9.366 ( 0.419) &   3.823 ( 0.171) \\ 
  12& 1.0 & 1.5 & 0.5 & 1.0 & 1000 &  22.104 ( 0.989) &  29.971 ( 1.340) &  13.199 ( 0.590) &   7.698 ( 0.344) &   9.094 ( 0.407) &   3.688 ( 0.165) \\ 
  12& 1.0 & 1.5 & 0.5 & 1.0 & 5000 &  24.909 ( 1.114) &  32.123 ( 1.437) &  13.037 ( 0.583) &   7.083 ( 0.317) &   9.337 ( 0.418) &   3.466 ( 0.155) \\ 
  12& 1.0 & 1.5 & 0.5 & 1.0 & 10000 &  24.016 ( 1.074) &  30.976 ( 1.385) &  14.693 ( 0.657) &   8.009 ( 0.358) &   9.963 ( 0.446) &   3.824 ( 0.171) \\ 
  12 & 1.0 & 1.5 & 0.5 & 1.0& 20000 &  27.107 ( 1.212) &  31.631 ( 1.415) &  14.931 ( 0.668) &   8.664 ( 0.387) &   9.365 ( 0.419) &   3.574 ( 0.160) \\ 
  12& 1.0 & 1.5 & 0.5 & 1.0 & 30000 &  33.908 ( 1.516) &  31.587 ( 1.413) &  17.604 ( 0.787) &  11.920 ( 0.533) &   9.605 ( 0.430) &   3.625 ( 0.162) \\ 
  12& 1.0 & 1.5 & 0.5 & 1.0 & 40000 &  25.191 ( 1.127) &  35.220 ( 1.575) &  13.116 ( 0.587) &   7.864 ( 0.352) &   8.665 ( 0.387) &   3.447 ( 0.154) \\ 
  12& 1.0 & 1.5 & 0.5 & 1.0 & 50000 &  27.827 ( 1.244) &  32.483 ( 1.453) &  13.549 ( 0.606) &   8.138 ( 0.364) &   9.252 ( 0.414) &   3.700 ( 0.165) \\ 
  13& 0.0 & 0.0 & 1.5 & 1.0 & 100 &  13.732 ( 0.614) &   8.425 ( 0.377) &   8.829 ( 0.395) &   8.834 ( 0.395) &   8.187 ( 0.366) &   8.192 ( 0.366) \\ 
  13 & 0.0 & 0.0 & 1.5 & 1.0& 500 &  13.303 ( 0.595) &   8.970 ( 0.401) &   9.105 ( 0.407) &   9.106 ( 0.407) &   8.961 ( 0.401) &   8.962 ( 0.401) \\ 
  13& 0.0 & 0.0 & 1.5 & 1.0 & 1000 &  11.552 ( 0.517) &   8.689 ( 0.389) &   8.590 ( 0.384) &   8.588 ( 0.384) &   8.648 ( 0.387) &   8.646 ( 0.387) \\ 
  13 & 0.0 & 0.0 & 1.5 & 1.0& 5000 &  11.971 ( 0.535) &   8.823 ( 0.395) &   8.816 ( 0.394) &   8.816 ( 0.394) &   8.798 ( 0.393) &   8.798 ( 0.393) \\ 
  13 & 0.0 & 0.0 & 1.5 & 1.0& 10000 &  12.715 ( 0.569) &   8.395 ( 0.375) &   8.400 ( 0.376) &   8.400 ( 0.376) &   8.389 ( 0.375) &   8.389 ( 0.375) \\ 
  13& 0.0 & 0.0 & 1.5 & 1.0 & 20000 &  12.865 ( 0.575) &   8.820 ( 0.394) &   8.835 ( 0.395) &   8.835 ( 0.395) &   8.814 ( 0.394) &   8.814 ( 0.394) \\ 
  13& 0.0 & 0.0 & 1.5 & 1.0 & 30000 &  12.827 ( 0.574) &   9.004 ( 0.403) &   9.004 ( 0.403) &   9.004 ( 0.403) &   9.005 ( 0.403) &   9.005 ( 0.403) \\ 
  13& 0.0 & 0.0 & 1.5 & 1.0 & 40000 &  12.252 ( 0.548) &   8.735 ( 0.391) &   8.728 ( 0.390) &   8.728 ( 0.390) &   8.731 ( 0.390) &   8.731 ( 0.390) \\ 
  13 & 0.0 & 0.0 & 1.5 & 1.0& 50000 &  13.486 ( 0.603) &   9.014 ( 0.403) &   9.000 ( 0.403) &   9.000 ( 0.403) &   9.010 ( 0.403) &   9.010 ( 0.403) \\ 
  14& 1.0 & 0.0 & 1.5 & 1.0 & 100 &  63.020 ( 2.818) &   9.351 ( 0.418) &  66.100 ( 2.956) &  66.105 ( 2.956) &   9.383 ( 0.420) &   9.362 ( 0.419) \\ 
  14& 1.0 & 0.0 & 1.5 & 1.0 & 500 &  56.290 ( 2.517) &   9.230 ( 0.413) &  37.402 ( 1.673) &  37.407 ( 1.673) &   9.189 ( 0.411) &   9.187 ( 0.411) \\ 
  14 & 1.0 & 0.0 & 1.5 & 1.0& 1000 &  64.273 ( 2.874) &   9.321 ( 0.417) &  49.792 ( 2.227) &  49.789 ( 2.227) &   9.442 ( 0.422) &   9.441 ( 0.422) \\ 
  14& 1.0 & 0.0 & 1.5 & 1.0 & 5000 &  65.704 ( 2.938) &   9.252 ( 0.414) &  46.026 ( 2.058) &  46.026 ( 2.058) &   9.265 ( 0.414) &   9.265 ( 0.414) \\ 
  14& 1.0 & 0.0 & 1.5 & 1.0 & 10000 &  62.914 ( 2.814) &   8.873 ( 0.397) &  41.801 ( 1.869) &  41.801 ( 1.869) &   8.863 ( 0.396) &   8.863 ( 0.396) \\ 
  14 & 1.0 & 0.0 & 1.5 & 1.0& 20000 &  70.381 ( 3.148) &   9.259 ( 0.414) &  45.507 ( 2.035) &  45.507 ( 2.035) &   9.260 ( 0.414) &   9.260 ( 0.414) \\ 
  14& 1.0 & 0.0 & 1.5 & 1.0 & 30000 &  70.430 ( 3.150) &   8.969 ( 0.401) &  46.123 ( 2.063) &  46.124 ( 2.063) &   8.969 ( 0.401) &   8.969 ( 0.401) \\ 
  14& 1.0 & 0.0 & 1.5 & 1.0 & 40000 &  70.721 ( 3.163) &   9.592 ( 0.429) &  47.366 ( 2.118) &  47.366 ( 2.118) &   9.590 ( 0.429) &   9.590 ( 0.429) \\ 
  14& 1.0 & 0.0 & 1.5 & 1.0 & 50000 &  65.161 ( 2.914) &   9.677 ( 0.433) &  47.915 ( 2.143) &  47.915 ( 2.143) &   9.669 ( 0.432) &   9.669 ( 0.432) \\ 
  15& 0.0 & 1.5 & 1.5 & 1.0 & 100 &  14.203 ( 0.635) &  45.077 ( 2.016) &  16.556 ( 0.740) &  11.467 ( 0.513) &  15.872 ( 0.710) &  10.687 ( 0.478) \\ 
  15 & 0.0 & 1.5 & 1.5 & 1.0& 500 &  12.935 ( 0.578) &  49.839 ( 2.229) &  17.138 ( 0.766) &  10.694 ( 0.478) &  17.232 ( 0.771) &  10.708 ( 0.479) \\ 
  15 & 0.0 & 1.5 & 1.5 & 1.0& 1000 &  14.049 ( 0.628) &  50.325 ( 2.251) &  18.188 ( 0.813) &  11.811 ( 0.528) &  18.104 ( 0.810) &  11.699 ( 0.523) \\ 
  15& 0.0 & 1.5 & 1.5 & 1.0 & 5000 &  13.016 ( 0.582) &  51.914 ( 2.322) &  18.138 ( 0.811) &  10.950 ( 0.490) &  18.133 ( 0.811) &  10.954 ( 0.490) \\ 
  15& 0.0 & 1.5 & 1.5 & 1.0 & 10000 &  12.281 ( 0.549) &  53.562 ( 2.395) &  18.294 ( 0.818) &  10.279 ( 0.460) &  18.278 ( 0.817) &  10.271 ( 0.459) \\ 
  15& 0.0 & 1.5 & 1.5 & 1.0 & 20000 &  13.148 ( 0.588) &  51.415 ( 2.299) &  16.030 ( 0.717) &  10.493 ( 0.469) &  16.015 ( 0.716) &  10.481 ( 0.469) \\ 
  15& 0.0 & 1.5 & 1.5 & 1.0 & 30000 &  13.003 ( 0.582) &  53.297 ( 2.384) &  17.891 ( 0.800) &  10.897 ( 0.487) &  17.901 ( 0.801) &  10.901 ( 0.487) \\ 
  15& 0.0 & 1.5 & 1.5 & 1.0 & 40000 &  12.621 ( 0.564) &  52.419 ( 2.344) &  16.790 ( 0.751) &  10.014 ( 0.448) &  16.795 ( 0.751) &  10.005 ( 0.447) \\ 
  15& 0.0 & 1.5 & 1.5 & 1.0 & 50000 &  13.342 ( 0.597) &  52.433 ( 2.345) &  16.448 ( 0.736) &  10.182 ( 0.455) &  16.458 ( 0.736) &  10.187 ( 0.456) \\ 
  16& 1.0 & 1.5 & 1.5 & 1.0 & 100 &  87.358 ( 3.907) &  32.432 ( 1.450) &  66.371 ( 2.968) &  62.728 ( 2.805) &  16.566 ( 0.741) &  12.061 ( 0.539) \\ 
  16 & 1.0 & 1.5 & 1.5 & 1.0 & 500 & 107.124 ( 4.791) &  38.888 ( 1.739) &  72.425 ( 3.239) &  65.172 ( 2.915) &  19.232 ( 0.860) &  12.253 ( 0.548) \\ 
  16 & 1.0 & 1.5 & 1.5 & 1.0 & 1000 & 153.531 ( 6.866) &  39.391 ( 1.762) & 266.236 (11.906) & 258.586 (11.564) &  17.238 ( 0.771) &  11.181 ( 0.500) \\ 
  16 & 1.0 & 1.5 & 1.5 & 1.0& 5000 &  67.995 ( 3.041) &  44.140 ( 1.974) &  50.934 ( 2.278) &  47.373 ( 2.119) &  17.267 ( 0.772) &  11.462 ( 0.513) \\ 
  16& 1.0 & 1.5 & 1.5 & 1.0 & 10000 &  67.592 ( 3.023) &  41.062 ( 1.836) &  61.603 ( 2.755) &  55.286 ( 2.472) &  17.973 ( 0.804) &  11.648 ( 0.521) \\ 
  16 & 1.0 & 1.5 & 1.5 & 1.0& 20000 &  59.580 ( 2.665) &  43.753 ( 1.957) &  49.953 ( 2.234) &  42.924 ( 1.920) &  17.548 ( 0.785) &  11.242 ( 0.503) \\ 
  16 & 1.0 & 1.5 & 1.5 & 1.0& 30000 &  59.543 ( 2.663) &  40.627 ( 1.817) &  54.854 ( 2.453) &  46.793 ( 2.093) &  16.706 ( 0.747) &  10.391 ( 0.465) \\ 
  16& 1.0 & 1.5 & 1.5 & 1.0 & 40000 &  56.936 ( 2.546) &  42.033 ( 1.880) &  51.886 ( 2.320) &  45.059 ( 2.015) &  18.020 ( 0.806) &  12.244 ( 0.548) \\ 
  16& 1.0 & 1.5 & 1.5 & 1.0 & 50000 &  72.134 ( 3.226) &  38.512 ( 1.722) &  60.447 ( 2.703) &  54.527 ( 2.439) &  18.434 ( 0.824) &  12.997 ( 0.581) \\ 
 \hline
\end{tabular}
\caption{Scaled empirical variance (Monte Carlo SEs in parentheses) of estimators, part 2. *DGP - data generating process}
\end{table}

\bibliography{bibliography}
\end{document}